\documentclass[journal]{IEEEtran}

\ifCLASSOPTIONcompsoc
  \usepackage[nocompress]{cite}
\else
  \usepackage{cite}
\fi

\ifCLASSINFOpdf

\else

\fi

\usepackage{array}
\usepackage{graphicx}
\usepackage{amsmath,amssymb,bm,amsthm,amsfonts}
\usepackage{subfig}
\usepackage{color}
\usepackage{float}
\newcommand{\code}[1]{\texttt{#1}}

\begin{document}

\title{Rapid, User-Transparent, and Trustworthy Device Pairing for D2D-Enabled Mobile Crowdsourcing}

\author{\IEEEauthorblockN{Cong Zhao\IEEEauthorrefmark{1},
Shusen Yang\IEEEauthorrefmark{1},
Xinyu Yang\IEEEauthorrefmark{1}, and
Julie McCann\IEEEauthorrefmark{2}}\\
\IEEEauthorblockA{\IEEEauthorrefmark{1} Xi'an Jiaotong University, P.R.China}\\
\IEEEauthorblockA{\IEEEauthorrefmark{2} Imperial College London, UK}}

\maketitle

\begin{abstract}
Mobile Crowdsourcing is a promising service paradigm utilizing ubiquitous mobile devices to facilitate large-scale crowdsourcing tasks (\emph{e.g.} urban sensing and collaborative computing).
Many applications in this domain require Device-to-Device (D2D) communications between participating devices for interactive operations such as task collaborations and file transmissions.
Considering the private participating devices and their opportunistic encountering behaviors, it is highly desired to establish secure and trustworthy D2D connections in a fast and autonomous way, which is vital for implementing practical Mobile Crowdsourcing Systems (MCSs).
In this paper, we develop an efficient scheme, Trustworthy Device Pairing (TDP), which achieves user-transparent secure D2D connections and reliable peer device selections for trustworthy D2D communications.
Through rigorous analysis, we demonstrate the effectiveness and security intensity of TDP in theory.
The performance of TDP is evaluated based on both real-world prototype experiments  and extensive trace-driven simulations.
Evaluation results verify our theoretical analysis and show that TDP significantly outperforms existing approaches in terms of pairing speed, stability, and security.
\end{abstract}


\IEEEpeerreviewmaketitle

\section{Introduction}\label{sec1}

Smart mobile devices have promoted the proliferation of Mobile Crowdsourcing Systems (MCSs), which facilitates large-scale crowdsouring tasks by exploiting the capacities of ubiquitous mobile users in a collaborative way~\cite{ren2015exploiting}. Typical MCS applications include mobile crowdsensing~\cite{Zhang:2014:CSP:2632048.2632059,Gao:2014:JIF:2639108.2639134,yangbackpressure,nawaz2013parksense}, information searching~\cite{shin2014cosmic}, crowdsourced content caching and sharing~\cite{yangbackpressure,nawaz2013parksense}, and collaborative mobile could computing~\cite{li2014multiple,murray2010case}, \emph{etc.} Many of these applications require exploiting the collaborative interactions among individual participants in geographical proximity. For instance, in collaborative searching applications (\emph{e.g.} finding a missing child~\cite{shin2014cosmic}), local searching results are required to be shared among surrounding participators.

Such geographically proximal interactions can be achieved by using Device-to-Device (D2D) communication (\emph{e.g.} WiFi direct and Bluetooth), which is much more agile and cost-effective compared with using traditional cellular networks, due to its shorter transmission delay, lower power consumption, and ignorable financial cost~\cite{yangbackpressure,karaliopoulos2015userinfocom}. For instance, in D2D-enabled mobile crowdsensing systems~\cite{yangbackpressure,Yang2013Selfish,Han2016Competition,adeel2014self}, a participating mobile user can report his or her sensory data to the server via other participating phones through opportunistic D2D transmissions rather than through cellular communications directly, which significantly improves the network throughput and reduces the system financial cost. It has also been shown that D2D communication is promising in cellular traffic offloading in crowdsourced video streaming~\cite{ren2015exploiting,Zhang:2014:CSP:2632048.2632059} and  content sharing in mobile social networking~\cite{nawaz2013parksense}. In addition, D2D communications are considered as the fundamental communication infrastructure in more and more emerging mobile could computing architectures~\cite{shin2014cosmic,li2014multiple}.

There exist an increasing number of efforts from both academia and industry to develop D2D related communication techniques, including efficient neighbor discovery~\cite{bakht2012searchlight}, secure device pairing~\cite{Varshavsky:2007:APA:1771592.1771607,Miettinen:2014:CZP:2660267.2660334,truong2014comparing}, D2D enabled cellular networks~\cite{liu2015stochasticD2D}, incentivisation for D2D communications~\cite{liincentivizingD2D,yangbackpressure,karaliopoulos2015userinfocom}, and D2D standards such as WiFi direct, Bluetooth Smart, and LTE direct. However, these approaches cannot  be directly applied to achieve D2D-enabled MCSs, because they cannot simultaneously satisfy the following three requirements in practice:

1. \textbf{Secure D2D Connection.} Since participating devices are privately held and temporarily recruited, there is no prior trust relation between them. Therefore, establishing a shared secret key for the firstly encountered devices (\emph{i.e.} device pairing) is vital for secure collaborative interactions.

2. \textbf{Rapid and User-Transparent Connection.} Due to the opportunistic human contact behaviors, it is highly desired to achieve rapid, autonomous, and user-transparent device pairing processes. However, existing device pairing approaches are based on either physical interactions (\emph{e.g.} button clicks) between firstly encountered devices or common contextual information gathered gradually~\cite{Miettinen:2014:CZP:2660267.2660334,truong2014comparing}, and are therefore unsuitable for practical MCSs.

3. \textbf{Addressing Peer Diversity.} In a MCS, a participating device could have multiple potential D2D peer devices and need to choose the optimal one (or a set of optimal ones). This optimization decision should depend on the trustworthiness and other metrics (\emph{e.g.} transmission data rate) of each candidate. As a result, a mechanism estimating the trustworthiness of a certain device as a D2D peer is necessary to guarantee the quality of D2D communications.
\vspace{-1em}
\subsection{Contribution} \label{subsec11}
In this paper, we present both theoretical and practical studies to simultaneously achieve above three requirements. Our contributions are summarized as follows:
\begin{itemize}
\item We develop Trustworthy Device Pairing (TDP), the first device pairing scheme for trustworthy opportunistic D2D communications in MCSs. TDP achieves rapid and user-transparent device paring, which is not supported by the off-the-shelf protocols or state-of-art user-transparent approaches, without introducing extra pairing labors. In addition, TDP introduces a new trust management method for reliable peer device selections based on online trustworthiness estimation of D2D peers.

\item Through rigorous security analysis, we demonstrate that TDP is immune to six potential security threats, including \textit{Passive Eavesdropping attacks}, \textit{Impersonating attacks}, \textit{Man-in-the-Middle attacks}, \textit{Trust Forging attacks}, \textit{Independent Negative Attacks}, and \textit{Collusive attacks}. Particularly, besides guaranteeing a comparable security intensity to the off-the-shelf protocols, TDP can also resist the \textit{Traffic-Oriented attacks} that are detrimental to D2D communications in MCSs.

\item We implement TDP in the Android operating system and the OMNeT++ simulator, and conduct real-world experiments and extensive trace-driven simulations to evaluate the performance of TDP. Results demonstrate that  TDP significantly outperforms existing approaches in terms of pairing speed and stability, and it can effectively eliminate the impact of \textit{Traffic-Oriented attacks} in D2D-enabled MCSs.
\end{itemize}
\vspace{-1em}
\subsection{Paper Organization} \label{subsec12}
The next section presents related work. Section \ref{sec3} presents the system model. The device trust estimation method in D2D-enabled MCSs is presented in Section \ref{sec4}. Section \ref{sec5} presents the Trustworthy Device Pairing scheme. Real-world experiments and extensive simulations are discussed in Section \ref{sec6}. And we conclude this paper in Section \ref{sec7}. Discussions on parameter determinations and security proofs are placed in Appendices A and B respectively, which can be found in the supplemental material.

\section{Related Work}\label{sec2}

\textbf{D2D Communications:}
There exist a large body of D2D communication based schemes, including efficient neighbor discovery~\cite{bakht2012searchlight}, D2D-enabled cellular networks~\cite{liu2015stochasticD2D}, incentivisation for D2D~\cite{liincentivizingD2D,yangbackpressure,karaliopoulos2015userinfocom},  network protocols~\cite{yang2009cross,yang2012hlls}, etc. Among those, the work~\cite{yangbackpressure,karaliopoulos2015userinfocom,xiao2015multi} specifically consider MCSs with D2D communications. However, none of these work considers security issues.

\textbf{Device Pairing:}
Device pairing focuses on establishing shared keys between freshly encountered devices. Some approaches~\cite{Li2013Secure,aditya2014encore,WiFiDirect,BL} rely on user-interactions to distribute initial trust credentials. However, in D2D-enabled MCSs, mobile users are normally strangers who are unlikely to conduct physical interactions. Since user-transparency is more preferable for spontaneous pairing, the proximity-based approach ~\cite{Varshavsky:2007:APA:1771592.1771607,Miettinen:2014:CZP:2660267.2660334,truong2014comparing} attracts significant attentions, which authenticates adjacent devices leveraging locally sensed contextual information. However, this approach commonly assumes that adversaries have no sustained access to legal contextual information, which is not acceptable for highly open MCSs. One of the most relevant work is the non-interactive device pairing approach in~\cite{184395}, which, however, requires asynchronous broadcasting of public credentials that may lead to unpredictable pairing time fluctuations. In summary, none of No existing device pairing approach can achieve both secure and user-transparent D2D connections, as our TDP.

\textbf{Trust Management:}
For participatory distributed systems, trust management is an effective approach to enhance system performance by determining the potential gain of self-interested entities according to their trustworthiness~\cite{Mousa2015Trust,Selvaraj2012Survey}. In traditional P2P systems, the trust framework is usually applied for either enabling access control~\cite{Blaze1996Decentralized} or encouraging benign resource sharing~\cite{Damiani2002A}. SepRep~\cite{Chu2010Reputation} establishes a general Quality-of-Service (QoS) based trust framework for heterogeneous P2P networks. For mobile P2P networks,~\cite{Qureshi2012A} proposes M-trust for accurate, robust and light-weight trust rating aggregation. In proliferating MCS applications, existing trust frameworks~\cite{Huang2010Are,Wang2014Enabling} mainly focus on evaluating the participator contribution from the data quality perspective. In~\cite{Talasila2015Collaborative}, a reputation-based trust framework is used to estimate the trustworthiness of a mobile device in location authentications, which is collaboratively conducted by proximal devices using Bluetooth communications. Overall, existing approaches maintain application-specific trustworthiness for self-interested entities, which is usually estimated using upper-level metrics without considering link-layer constrains. They cannot be directly used to evaluate the trustworthiness of D2D communications in MCSs.

The workshop abstract~\cite{Zhao:2015:TDP:2801694.2801705} presents the very initial idea of our TDP, while this paper establishes a much more complete and systematic trust management scheme with comprehensive theoretical analysis. Furthermore,  real-world experiments and extensive trace-driven simulations are also conducted to study the practical performance of TDP in realistic D2D-enabled MCSs.

\section{System Model}\label{sec3}

In this section, we present the model of general D2D-enabled MCSs and potential attacks to opportunistic D2D communications.
\vspace{-1em}
\subsection{D2D-enabled MCS Architecture} \label{subsec31}

\begin{figure}
\setlength{\belowcaptionskip}{-1em}
\centering
\includegraphics[width=3in]{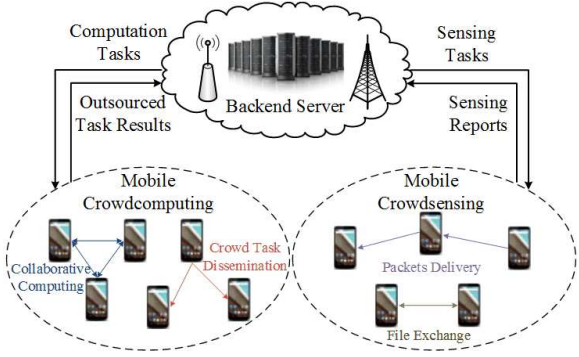}
\caption{Architecture of D2D-Enabled MCSs.} \label{fig1}
\end{figure}

The typical architecture of a D2D-enabled MCS is shown in Fig.\ref{fig1}: multiple participating personal mobile devices communicate with the Backend Server (BS) through the Internet via WiFi or cellular access points, and they communicate with devices opportunistically encountered through outband D2D radios such as Bluetooth and WiFi direct.

Generally, the BS is responsible for the maintenance of the entire MCS by establishing system regulations, publishing crowdsourcing applications, publishing tasks and recording device profiles. Let $\mathcal{N}$ be the set of all regulated MCS task types (\emph{e.g.} mobile social networking, crowdsensing, collaborative computing).

To participate in crowdsourcing tasks, personal mobile devices need to install the crowdsourcing application and register to the BS. Let $\mathcal{D}$ be the set of all registered devices. Driven by a specific crowdsourcing task, any pair of registered devices can establish a secure D2D connection for further communications, which is denoted as a D2D transaction for the rest of the paper. A device $a\in\mathcal{D}$ can have a sequence of D2D transactions $i=\{1,2,...\}$ with any other device in $\mathcal{D}$.

For the establishment of secure D2D connections, each device $a \in \mathcal{D}$ possesses a unique device pairing credential, which is verifiable to both the BS and any other registered device encountered. Meanwhile, a `trustvalue' is maintained for each device according to its behavior in previous D2D transactions, which is treats as its trustworthiness to fulfill future transaction requests. Here, the `trustvalue' is the predominant concern of the peer selection when there are multiple candidates.

\vspace{-1em}
\subsection{Security Threats and Attack Models} \label{subsec32}

Considering the sensitivity of personal mobile devices, \textit{neither devices opportunistically encountered nor the BS should be fully trusted in MCSs}.

Intuitively, potential attacks launched by malicious devices within the D2D-enabled MCS can be divided into two categories: Connection-Oriented (CO) attacks and Traffic-Oriented (TO) attacks. On one hand, CO attackers intend to compromise the pairing process between encountered devices for illegal communication contents or MCS services. In fact, the prevention of CO attacks is the predominant concentration of security mechanism of off-the-shelf D2D communication protocols (\emph{e.g} Bluetooth~\cite{BL} and WiFi direct~\cite{WiFiDirect}). On the other hand, the TO attackers are more interested in manipulating D2D traffics in MCSs for simply unfair rewards (\emph{e.g.} popularity, credits) or future advanced attacks (\emph{e.g.} false data injection attacks). Unlike CO attacks, as far as we know, there is still no systematic discussion on the impact and prevention of TO attacks in MCSs from the perspective of opportunistic D2D communications.

We formally model these potential attacks as follows (let there be devices $a,b,m \in \mathcal{D}$):

\textbf{1. Passvie-Eavesdropping Attack (\textbf{CO$_1$})}: any third party adversary $m$, who is not the intended end $a$ or $b$ of a D2D transaction, tries to decrypt eavesdropped data;

\textbf{2. Impersonating Attack (CO$_2$)}: any adversary $m$, except the device impersonated $a$, tries to use $a$'s credential to establish D2D connections with others;

\textbf{3. Man-in-the-Middle Attack (CO$_3$)}: any third party adversary $m$, except both ends of a D2D transaction $a$ and $b$, tries to intercept and replace $a$ ($b$)'s credential to establish D2D connections with $b$ ($a$) in middle without being noticed;

\textbf{4. Trust Forging Attack (TO$_1$)}: any adversary $m$ forges its trustvalue to attract unfair D2D transaction requests;

\textbf{5. Independent Negative Attack (TO$_2$)}: any adversary $m$ intentionally replies negative ratings (continuously or intermittently) to downgrade others' trustvalue and consequently attract unfair D2D transaction requests;

\textbf{6. Collusive Attack (TO$_3$)}: any set of adversaries $\mathcal{M}$ intentionally reply positive ratings to members within $\mathcal{M}$ and negative ratings to others (continuously or intermittently) to attract unfair D2D transaction requests.

Moreover, we should treat the BS as `benign but curious': generally, it will jeopardize neither the device pairing nor trust management process for public credibility; meanwhile, it is curious about the content of D2D transactions among devices. In this case, the BS may try to launch CO$_1$ attacks as a third party adversary.

\section{Device Trust Estimation in D2D-Enabled MCSs} \label{sec4}

In D2D-enabled MCSs, the reliability of a device to fulfill future D2D requests can be reflected by its behavior in historical transactions. In this paper, we denote this reliability as the trustworthiness of a device, which is treated as the basic metric for the peer selection in device pairing process (will be discussed latter).

In this section, we develop an online method to: (1) accurately estimate the trustworthiness of participating devices according to their historical behaviors; (2) selectively profile the device behavior during D2D transactions; (3) adjust the device trustworthiness using behavior profiles at real-time.
\vspace{-1em}
\subsection{Device Trustworthiness} \label{subsec41}

We use the term of `trustvalue' to quantitatively represent the trustworthiness of each device, and, for the authenticity concern, it can only be adjusted by the BS after each transaction according to the device behavior.

Mathematically, let the trustvalue of each device $a\in \mathcal{D}$ after its $i^{\rm th}$ D2D transaction be a $|\mathcal{N}|$-dimensional vector $\bm{t}_a(i)$. Here, the value of the $n^{\rm th}$ $(1\leqslant n \leqslant |\mathcal{N}|)$ entry of $\bm{t}_a(i)$ is a value within the range of $[0,~1]$, which represents the trustworthiness of $a$ fulfilling the $n^{\rm th}$ type of D2D request (for the corresponding type of MCS task).

Since the trustworthiness of a device should be gradually built over a period of benign behaviors, we use the widely-adopted Gompertz function~\cite{kenney1954mathematics} in reputation mapping to compute the device trustvalue:
\begin{equation}\label{eq1}
\bm{t}_a(i)=G(G^{-1}(\bm{t}_a(i-1))+\Delta\bm{t}_a(i)),
\end{equation}
where, for a given vector $\bm{x}$, $G(\bm{x})$ denotes calculating the value of the Gompertz function:
\begin{equation}\label{eq2}
g(x)=e^{-e^{-c_gx}}
\end{equation}
for each entry of vector $\bm{x}$, and the parameter $c_g$ determines the sensitivity of trustvalue variation along time. $\Delta\bm{t}_a(i)$ is the estimation of $a$'s behavior at its $i^{\rm th}$ transaction, which is computed by the BS according to behavior profiles uploaded by participating devices (will be explained later).
\vspace{-1em}
\subsection{Device Behavior Profiling during D2D Transactions} \label{subsec42}

For each D2D transaction, both ends of the transaction profile necessary information that can objectively reflect each other's behavior, which is then recorded in a D2D receipt respectively for future trustvalue adjustment at the BS.
Intuitively, for a reasonable profiling of the $i^{\rm th}$ D2D transaction of a device $a\in\mathcal{D}$ (with its peer device $b$), the following four factors should be recorded in $a$'s receipt (omitting index $i$ for concision): the Quality-of-Service (QoS) provided by the D2D link $q_{ab}$, the credibility of $b$ from $a$'s perspective $c_{ab}$, the rating from $b$ on $a$'s behavior $r_{ba}$, and the type of the current transaction $\boldsymbol{\lambda}_{a}$. Following is their specific definitions.

\textbf{QoS}: We use the link-layer QoS value $q_{ab}$ to denote the quality of the D2D link constructed by a given pair of encountered devices $a$ and $b$. Different QoS metrics are used according to  different D2D transactions, such as transmission delay, data rate, or packet loss rate. For example, image-based applications would require high data rate, while event-monitoring tasks need small delay.
In our system, $q_{ab}$ is a normalized value that can be obtained by both $a$ and $b$ during their transactions.

We use link-layer QoS  because it can precisely reflect the  application-layer QoS (due to the one-hop nature of D2D transactions), and it is much easier to be estimated in practice.
For instance, if the minimal possible delay for transmitting a given-size packet over a D2D link is 10 ms (computed using datasheet), and the actual per-packet delay is 40 ms (measured online), then the QoS value can be normalized as $q_{ab}=10/40=0.25$.

\textbf{Credibility}: The credibility estimation between encountered devices is important for the evaluation of current behavior of each other, considering their historical knowledge.
In practice, it is reasonable for a person to consider a stranger credible if their judges to any common third party are similar. Inspired by this, we estimate $c_{ab}$ based on the comparability of contact histories of both device $a$ and device $b$. Without loss of generality, let device $a$'s contact history $\mathcal{E}_a$ be a set of devices that $a$ has paired with during a trust management cycle regulated by the BS. For each device $x\in \mathcal{E}_a$, a value $e_{ax}$ that denotes the percentage of positive ratings from $a$ to $x$ is also maintained by $a$, which is recorded in a vector $\bm{\theta}_a$ according to the encountering order of device $x$. $e_{ax}$ will be refreshed according to the rating from device $a$ to device $x$ whenever they finish a D2D transaction.
Specifically, $c_{ab}$ is defined as:
\begin{equation} \label{eq3}
c_{ab}=(\alpha s_{ab}^2+(1-\alpha)d_{ab}^2)w_{ab},
\end{equation}
where $s_{ab}$ is the similarity between percentages of positive ratings separately replied by $a$ and $b$ to other devices, $d_{ab}$ is the relative diversity of percentages of positive ratings separately replied by $a$ and $b$ to other devices, $\alpha$ within the range of $[0,~1]$ is used to adjust the focus of the credibility estimation, and $w_{ab}$ is the damping factor that is inversely proportion of the intimacy between $a$ and $b$.

In Eq.(\ref{eq3}), $s_{ab}$ and $d_{ab}$ are defined as:
\begin{equation} \label{eq4}
s_{ab}=1-\sqrt{\displaystyle{\frac{\sum_{x\in \mathcal{E}_{ab}}(e_{ax}-e_{bx})^2}{|\mathcal{E}_{ab}|}}},
\end{equation}
\begin{equation} \label{eq5}
d_{ab}=1-\left|H(\bm{\theta}_a)-H(\bm{\theta}_b)\right|,
\end{equation}
where $\mathcal{E}_{ab}=\mathcal{E}_{a}\cap \mathcal{E}_{b}$, and $H(\bm{\theta})$ is calculated as:
\begin{equation} \label{eq6}
H(\bm{\theta})=\sqrt{\displaystyle{\frac{\sum_{1\leqslant y\leqslant dim(\bm{\theta})}(\theta_y-\theta_{(y+1)\text{mod}dim(\bm{\theta})})^2}{dim(\bm{\theta})}}},
\end{equation}
where $dim(\bm{\theta})$ is $\bm{\theta}$'s dimensions, and $\theta_y$ is $\bm{\theta}$'s $y^{\rm th}$ entry.
$w_{ab}$ is used to restrict the credibility estimation between devices that contact more frequently than normal, which is defined as:
\begin{equation} \label{eq7}
w_{ab}=
\begin{cases}
1 &\mbox{if $\sigma_{ab} = 0$}\\
-c_w(\sigma_{ab}/\bar{\sigma})^{2}+ 1 &\mbox{if $0 < \sigma_{ab} \leqslant \lfloor \sqrt{1/c_w} \bar{\sigma} \rfloor$}\\
0 &\mbox{if $\sigma_{ab} > \lfloor \sqrt{1/c_w} \bar{\sigma} \rfloor$}
\end{cases},
\end{equation}
where parameter $c_w$ determines the sensitivity of credibility damping, $\sigma_{ab}$ is the transaction count between $a$ and $b$, and $\bar{\sigma}$ is the average transaction count of all devices within a trust management cycle regulated by the BS.

\textbf{Rating}: The rating indicates the satisfactory of a device on the behavior of its peer device in the current D2D transaction, which is determined by the comparison of the QoS and credibility of the current transaction with that of historical transactions with any device.
For the transaction between devices $a$ and $b$:
\begin{equation} \label{eq8}
r_{ab}=
\begin{cases}
1 &\mbox{if $\beta q_{ab}^2+(1-\beta)c_{ab}^2 \geqslant \beta\bar{q}_a^2+(1-\beta)\bar{c}_a^2$}\\
-1 &\mbox{if $\beta q_{ab}^2+(1-\beta)c_{ab}^2 < \beta\bar{q}_a^2+(1-\beta)\bar{c}_a^2$}
\end{cases},
\end{equation}
where $\bar{q}_a$ and $\bar{c}_a$ are the EWMAs of historical QoS and credibility that are locally updated by device $a$ after each transaction, and $\beta$ within the range of $[0,~1]$ is used to adjust the focus of the rating decision.

\textbf{Transaction Type}: The type of the current transaction is profiled to estimate the trustworthiness of a device fulfilling D2D requests from different type of MCS tasks respectively.

For a normalized and comparable profiling, we define $\boldsymbol{\lambda}_{a}$ as a $|\mathcal{N}|$-dimensional vector to represent a D2D transaction that is initiated from device $a$. If the current transaction is a type-$n$ transaction, the $n^{\rm th}$ entry of $\boldsymbol{\lambda}_{a}$ is a positive value within the range of $(0,~1]$, while all other entries are $0$. The value of the $n^{\rm th}$ entry indicates resources (computing/communicating) that are contributed by such a transaction, \emph{e.g.} bandwidth, which is determined according to the criteria regulated by the BS.
\vspace{-1em}
\subsection{Device Trustworthiness Adjustment at the BS} \label{subsec43}

According to the uploaded receipts, the BS can estimate the behavior of a device in a D2D transaction. For instance, for the $i^{\rm th}$ transaction of a device $a\in\mathcal{D}$ (with its peer device $b$), the BS can obtain $q_{ab}$, $c_{ab}$, $r_{ba}$ and $\boldsymbol{\lambda}_{a}$ from $a$'s receipt.

From the BS's perspective, one more thing should be considered is that D2D transactions required by different crowdsourcing tasks lead to different device trustvalue adjustment patterns (\emph{i.e.} which end of the transaction should have a trustvalue adjustment). For instance, for an outsourced collaborative computing task, both ends of the D2D transaction should have trustvalue adjustments because of their equal contributions to task offloading.

Without loss of generality, for $\bm{\lambda}_{a}$, the trustvalue adjustment pattern $\bm{p}_a$ is defined as an element of set $\mathcal{P}=\{(0,0),(0,1),(1,0),(1,1)\}$ that indicates whether the initiator (indicated by the $1^{\rm st}$ entry, role $\bm{r}_a=(1,0)$) or the peer (indicated by the $2^{\rm nd}$ entry, role $\bm{r}_a=(0,1)$) should have a trustvalue adjustment. Let $\Lambda$ be the whole set of the transaction types ($\bm{\lambda}$). It is viable for the BS to establish a mapping:
\begin{equation}\label{eq9}
f: \Lambda \rightarrow \mathcal{P}
\end{equation}
that indicates which end of the transaction should have a trustvalue adjustment according to the transaction type.

In this case, for the BS, the behavior estimation of device $a$ in its $i^{\rm th}$ transaction $\Delta\bm{t}_a(i)$ in Eq.(\ref{eq1}) is computed as (omitting index $i$ for concision):
\begin{equation}\label{eq10}
\Delta\bm{t}_a=q_{ab}c_{ab}r_{ba}f(\boldsymbol{\lambda}_{a})\bm{r}_a\boldsymbol{\lambda}_{a}.
\end{equation}
\vspace{-1em}
\subsection{Trustvalue Bootstrapping} \label{subsec44}

For the bootstrapping of the device trustvalue, an initial value of $e^{-1}$ ($g(0)$) is issued when the device joins the MCS. Besides, for two devices freshly encountered (\emph{e.g.} $a$ and $b$, where $|\mathcal{E}_a|$, $|\mathcal{E}_b|$ or $|\mathcal{E}_{ab}|$ is 0), the weight factor of rating $\beta$ is set as $1$ to neglect the credibility factor. Therefore, any device that freshly joins the MCS will not be starved and is guaranteed a fair probability to be involved as long as behaving well in D2D transactions.
\vspace{-1em}
\subsection{Discussion on the Trust Estimation Method} \label{subsec45}

In this subsection, we firstly provide justifications for the trustvalue definition and device behavior profiling, then we conduct comparisons between our method and the Josang model~\cite{J2002The} in the D2D-enabled MCS scenario.

\subsubsection{Device Trustvalue Accumulation} \label{subsubsec451}

Intuitively, we estimate the trustworthiness of a device performing D2D transactions according to its historical behaviors. The trustvalue of a device is determined by the accumulation of its quantified behavior profiles (Eq.(\ref{eq1})).

Considering that any freshly joint device with continuously positive (negative) behaviors deserves a rapid trustvalue enhancement (decrement), and that the trustvalue itself should be normalized for effective comparisons, we use a type of sigmoid function, \emph{i.e.} the Gompertz Function~\cite{kenney1954mathematics} $g(x)=e^{-e^{-c_gx}}$ (see Fig.A1 (a) in Appendix A.1), to map the behavior accumulation to the device trustvalue. Any other function with similar properties, \emph{e.g.} the Logistic Function~\cite{Jones1975Probability} $l(x)=1/(1+e^{-c_lx})$ (see Fig.A1 (a) in Appendix A.1), is also applicable in trustworthiness mapping.

One thing should be noted is that the value of $c_g$ determines the sensitivity of trustvalue variations. We provide an instance for $c_g$ determination in Appendix A.1.

\subsubsection{Device Credibility Estimation} \label{subsubsec452}

Intuitively, to stimulate more D2D transactions, devices with benign and stable behaviors in extensive active scopes should be guaranteed higher credibilities.

The first metric of credibility estimation is the comparability of the behavior profiles separatively maintained by both ends of the transaction. We assume that devices strictly following MCS regulations receive relatively similar ratings from different devices, then the comparability is quantified by the similarity (Eq.(\ref{eq4})) and the relative diversity (Eq.(\ref{eq5})) of their contact histories. Specifically, the similarity estimation encourages devices to objectively reflect other's behavior through justifiable ratings, while that the diversity estimation restricts devices from downgrading other's trustvalue with extreme or discriminative ratings.

Besides, to prevent the potential trustvalue boost caused by massive meaningless D2D transactions between collusive devices, the intimacy between both ends of a transaction is treated as a damping factor of their credibility (Eq.(\ref{eq7})), \emph{i.e.} suspiciously frequent contacts lead to rapid credibility decreases. We use a type of monotonic descending function, \emph{i.e.} a quadratic function $w(\sigma)=-c_w(\sigma/\bar{\sigma})^{2}+ 1$, to represent the relation between the device contact number $\sigma$ and the damping factor $w$ (see Fig.A2 (a) in Appendix A.2). Any other function with similar properties, \emph{e.g.} a cubic function $w(\sigma)=-c_w(\sigma/\bar{\sigma})^{3}+ 1$ (see Fig.A2 (a) in Appendix A.2), is also applicable.

One thing should be noted is that the value of $c_w$ determines the sensitivity of credibility damping on the device intimacy. We provide an instance for $c_w$ determination in Appendix A.2.

\begin{figure}
\setlength{\belowcaptionskip}{-1em}
\centering
\subfloat[Impact of Positive Ratings]{\includegraphics[width=1.5in]{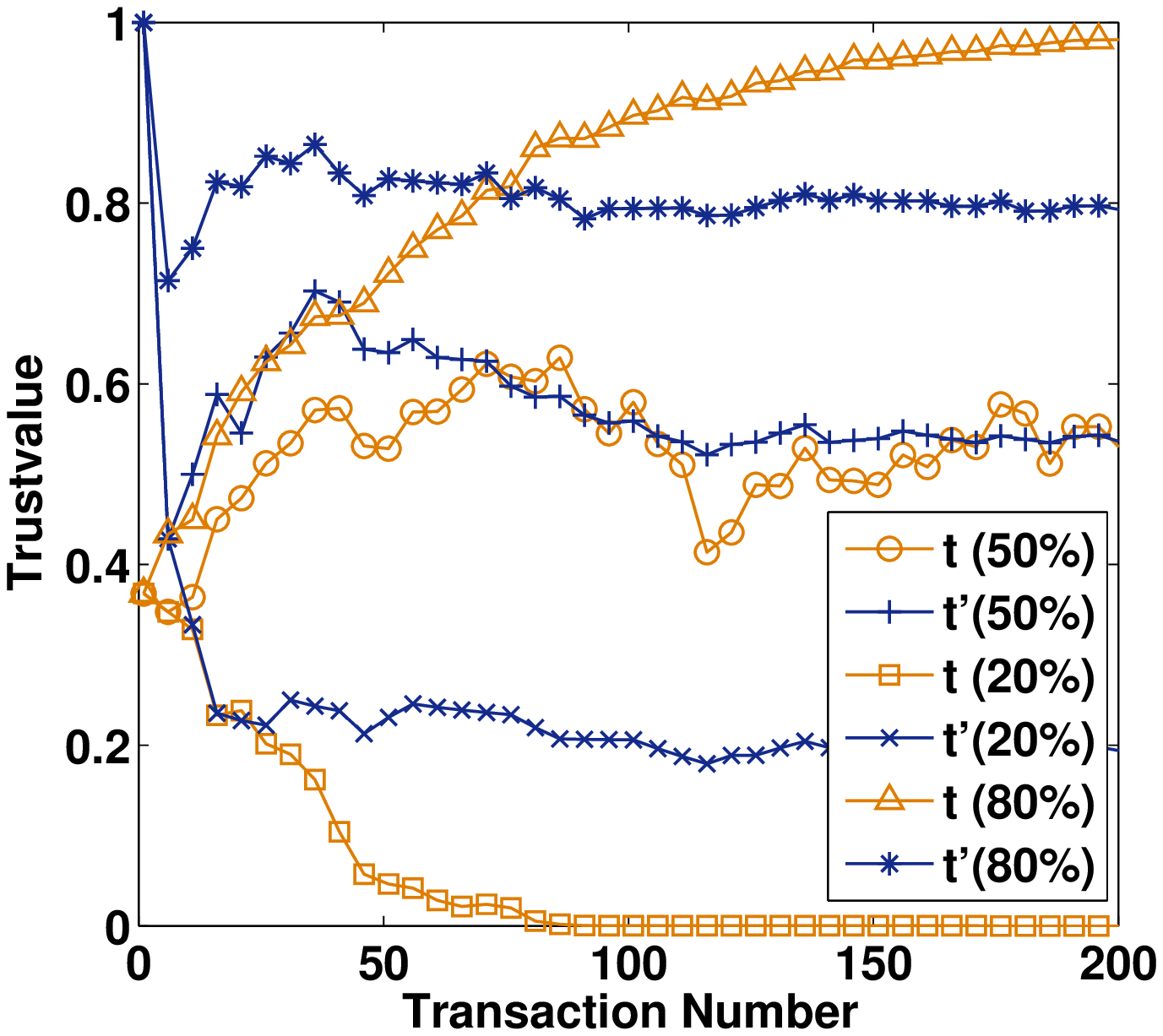}%
\label{fig2a}}
\hfil
\subfloat[Impact of D2D Profiles]{\includegraphics[width=1.5in]{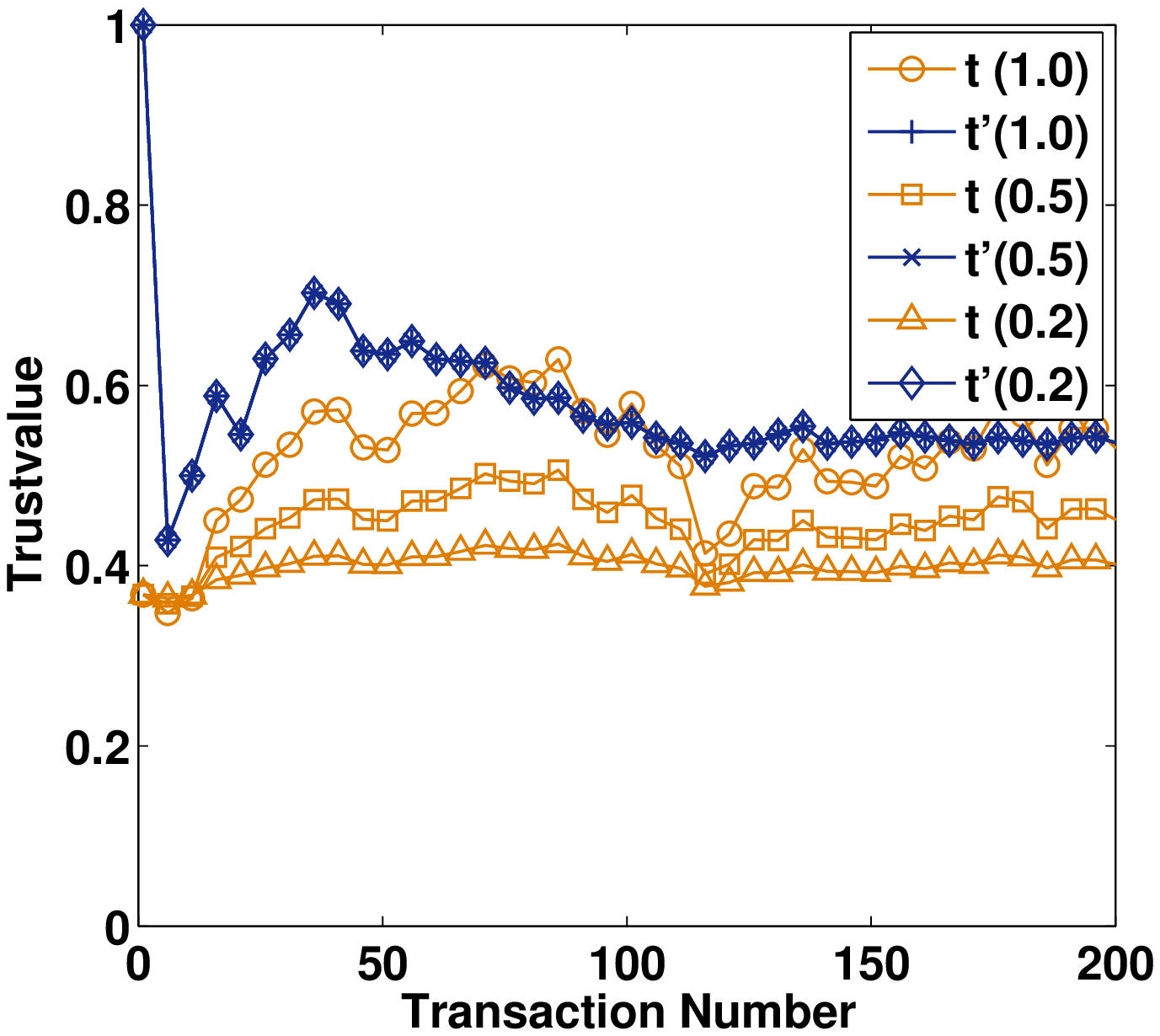}%
\label{fig2b}}
\caption{Comparisons between TDP and the Josang Model.}
\label{fig2}
\end{figure}

\subsubsection{Comparison with the Josang Trust Model} \label{subsubsec453}
In this part, we compare the effectiveness of our trust estimation method and that of the Josang trust model~\cite{J2002The}, one of the most representative trust estimation methods, in D2D-enabled MCS scenarios.

Specifically, the Josang model~\cite{J2002The} leverages the beta probability density function to map the reputation of self-interested entities, whose expectation is treated as entity trustworthiness on fulfilling future transactions. According to~\cite{J2002The}, we calculate the trustvalue based on the Josang model of each device $a$ after its $i^{\rm th}$ transaction as:
\begin{equation}\label{eq11}
t'_a(i)=\displaystyle{\frac{Count_{p}(i)+1}{Count_{p}(i)+Count_{n}(i)+2}},
\end{equation}
where $Count_{p}(i)$ and $Count_{n}(i)$ denote the number of positive and negative ratings that $a$ receives after its $i^{\rm th}$ transaction, respectively.

For comparison, we conducted two sets of numerical simulations to study the impact of the positive rating percentage and the D2D profile on the accuracy and sensitivity of trustworthiness mapping of our method and the Josang model. For each round of simulation, we recorded the trustvalue variation ($t$ for our method, and $t'$ for the Josang model) of a single device $a$ in 200 transactions. For our method, we set $t_0=e^{-1}$, and $t$ was updated after each transaction based on the rating and D2D profiles (\emph{i.e.} QoS and credibility); for the Josang model, we set $t'_0=1$, and $t'$ was updated after each transaction based on the rating only.

In the first set of simulations, we set that device $a$ got a positive rating at different probabilities (\emph{i.e.} 50\%, 20\% and 80\%) and a random D2D profile within the range of $[0,1]$ in a transaction. As shown in Fig.\ref{fig2} (a), both models can maintain a distinguishable trustvalue with a similar adjusting trend. However, their sensitivities are obviously different: for more extreme cases (\emph{i.e.} 20\% and 80\%), our method guarantees a gradual accumulation of the device trustworthiness, whereas the Josang model just presents a rapid trust convergence; for the neutral case (\emph{i.e.} 50\%), our device trustworthiness is adjusted more sensitively because of more objective device behavior estimations.

In the second set of simulations, we set that device $a$ got a positive rating with a 50\% probability and a random D2D profile within different ranges (\emph{i.e.} $[0,1]$, $[0,0.5]$ and $[0,0.2]$) in a transaction. As shown in Fig.\ref{fig2} (b), TDP manages to maintain a more distinguishable and sensitive trustvalue compared to the Josang model. In D2D-enabled MCSs, it is not objective to estimate the trustworthiness of a device only based on ratings from other peers. Counting more detailed information into trustworthiness estimation not only guarantees a better objectivity but also provides devices a fair chance to attract more transactions with better behaviors.

\section{The Trustworthy Device Pairing (TDP) Scheme} \label{sec5}
With the accurate trustworthiness estimated by the method in Section \ref{sec4}, in this section, we develop TDP to realize rapid, user-transparent, and trustworthy device pairing for D2D-enabled MCSs.
\vspace{-1em}
\subsection{Overview} \label{subsec51}

Considering the security threats in Subsection \ref{subsec32}, TDP establishes a Certificate-Less Public Key Cryptography (CL-PKC)~\cite{Al2003Certificateless} framework to issue a unique private-public key pair to each registered device as the D2D credential. Since the credential is collaboratively generated by the BS and the registering device, it is verifiable to any other registered device as well as the BS.

\begin{figure}
\setlength{\belowcaptionskip}{-1em}
    \centering
    \includegraphics[width=2.5in]{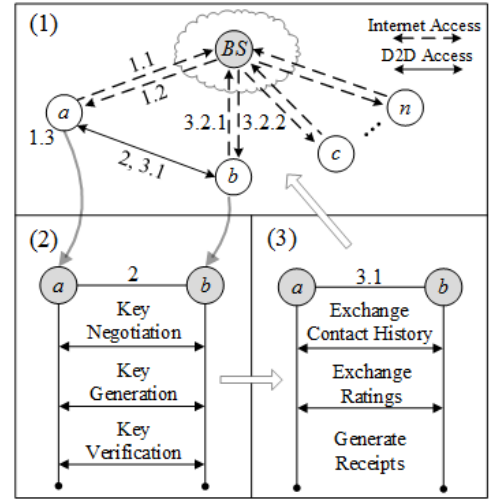}
    \caption{Architecture of TDP.} \label{fig3}
\end{figure}

Based on the credential, any pair of registered devices encountered can negotiate a shared key spontaneously without contacting the BS, which introduces no extra labor compared with existing device pairing methods. Besides, because a part of the credential is privately held by the device only, the `benign but curious' BS cannot restore such shared keys to decrypt the transaction contents. Moreover, according to the D2D receipts authenticated by the credential, participating devices of D2D transactions can have justifiable trustvalue adjustments from the BS based on their behaviors, which will guarantee a well-behaving device a fair advantage in peer selections in future D2D transactions.

Specifically, TDP consists of three components: \emph{device registration}, \emph{device pairing}, and \emph{device trustvalue management}, whose basic procedure is illustrated in Fig.\ref{fig3}. Specific operations are presented as follows.
\vspace{-1em}
\subsection{Device Registration} \label{subsec52}

In this process, the BS generates and issues a D2D credential to each registering device.

For the system initialization, the BS determines all system parameters in Table \ref{tab1}. Then, the parameter set $\Omega=\{F_q, G_q, P, P_{pub}, h_0, h_1, h_2\}$ and the bootstrapping trustvalue $\bm{t}_0$ are deployed in the crowdsourcing application.
\begin{table}[!t]
\caption{System Parameters of TDP.}
\label{tab1}
\centering
\begin{tabular}{|m{35pt}<{\centering}|m{70pt}<{\centering}|m{85pt}<{\centering}|}
\hline
\textbf{Parameter} & \textbf{Meaning} & \textbf{Note}\\
\hline
$E$ & an elliptic curve on a finite filed $F_q$ & treated as a cyclic addition group $G_q$; $q$: a large prime as the group order; $P$: the generator of $G_q$\\
\hline
$x$ & a random positive integer $<q$ ($x \in_R Z_q^*$) & the master private key, only possessed by the BS \\
\hline
$P_{pub}$ & a point on $E$ ($P_{pub}=xP$) & the master public key; $xP$: the point multiplication on $E$ \\
\hline
$h_{0,1,2}$ & one-way hash functions & $h_0:\{0,1\}^* \rightarrow Z^*$; $h_1:\{0,1\}^* \rightarrow \{0,1\}^*$; $h_2:\{0,1\}^* \rightarrow Z^*$ \\
\hline
\end{tabular}
\end{table}

\begin{figure}[h!]
\setlength{\belowcaptionskip}{-1em}
    \centering
    \includegraphics[width=3in]{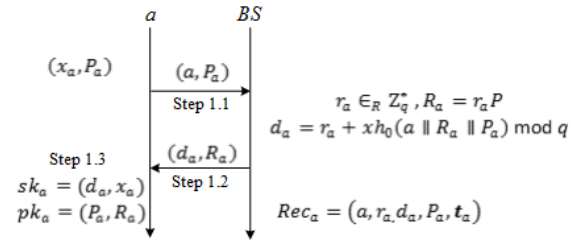}
    \caption{Device Registration (Step 1 in Fig.\ref{fig3}).} \label{fig4}
\end{figure}

To join the MCS, personal mobile devices need to install the crowdsourcing application and register to the BS. Without loss of generality, let device $a$ register to the BS. Specific operations (Step 1.1-1.3 in Fig.\ref{fig3}) are demonstrated in Fig.\ref{fig4}, where the operator`$\Vert$' denotes the concatenation operation.

Here, device $a$ determines its local partial key pair $(x_a,P_a)$, where $P_a=x_aP$. Then the BS extracts an authenticated partial key pair $(d_a, R_a)$ according to $P_a$. Finally, $(sk_a, pk_a)$, where $sk_a=(d_a, x_a)$ and $pk_a=(P_a, R_a)$, is treated as $a$'s credential. Meanwhile, the BS records $(a,r_a,d_a,P_a,\bm{t}_a)$.
\vspace{-1em}
\subsection{Device Pairing} \label{subsec53}

In this process, any pair of registered devices encountered use their credentials to negotiate a shared key for a secure D2D connection.

\begin{figure}
\setlength{\belowcaptionskip}{-1em}
    \centering
    \includegraphics[width=3in]{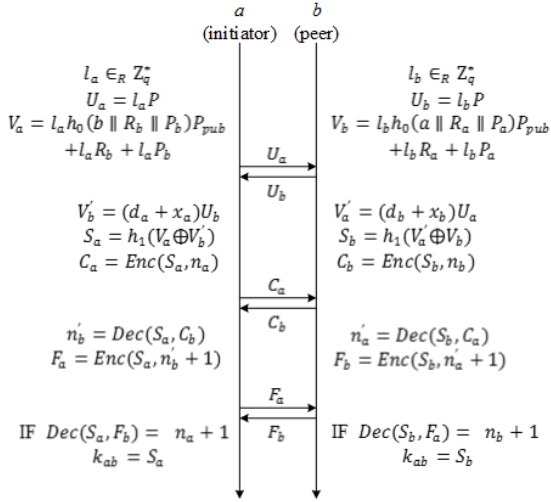}
    \caption{Device Pairing (Step 2 in Fig.\ref{fig3}).} \label{fig5}
\end{figure}

To accomplish a specific kind of D2D transaction, the initiator $a$ selects a peer $b$ among multiple available candidates considering their device trustvalues and other states (\emph{e.g.} D2D channel signal strength). Then $a$ and $b$ start the device pairing. Specific operations (Step 2 in Fig.\ref{fig3}) are demonstrated in Fig.\ref{fig5}, where `$\oplus$' denotes the $xor$ operation, and $Enc(x,y)$ and $Dec(x,y)$ denote the symmetric encryption and decryption with $x$ as the key and $y$ as the plaintext. The shared key $k_{ab}$ is generated according to $a$ and $b$'s credentials. $l_{a(b)}$ and $n_{a(b)}$ are random numbers guaranteeing the freshness of $k_{ab}$ and the Challenge-Response messages $C$ and $F$, respectively.

\vspace{-1em}
\subsection{Device Trustvalue Management} \label{subsec54}

In this process, each end of a finished D2D transaction generates a verifiable receipt to profile the current transaction, which is uploaded to the BS for trustvalue adjustment.

\textbf{D2D Transaction Receipt Generation:}
After the $i^{\rm th}$ D2D transaction of device $a$ (the $j^{\rm th}$ of device $b$), $a$ and $b$ generate receipts that record both the transaction information and the profiles of each other's behavior.

\begin{figure}
\setlength{\belowcaptionskip}{-1em}
    \centering
    \includegraphics[width=3.3in]{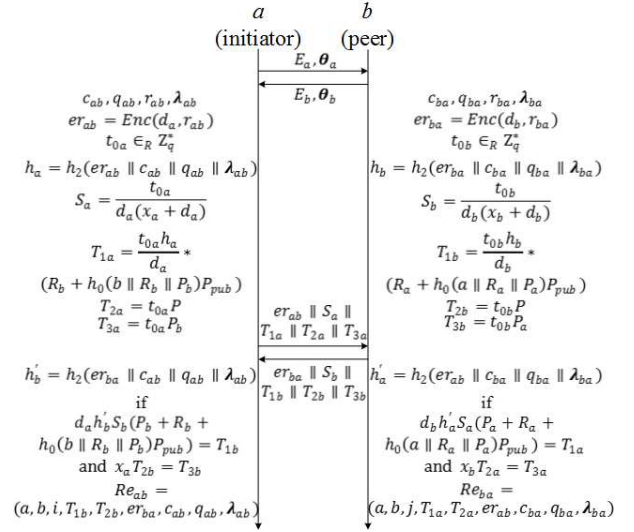}
    \caption{D2D Receipt Generation (Step 3 in Fig.\ref{fig3}).} \label{fig6}
\end{figure}

Specific operations (Step 3.1 in Fig.\ref{fig3}) are demonstrated in Fig.\ref{fig6}. Note that all interactions between $a$ and $b$ are encrypted by $k_{ab}$ (the symmetric cryptography operation ($Enc,Dec$) and the transaction indexes ($i,j$) are omitted for conciseness). Here, $a$ and $b$ exchange the contact histories $\mathcal{E}_a$ and $\mathcal{E}_b$ (together with $\bm{\theta}_a$ and $\bm{\theta}_b$) to calculate each other's credibility $c_{ab}=c_{ba}$. Then, considering the QoS of the D2D link $q_{ab}=q_{ba}$ and the knowledge of historical transactions $\bar{q}_{a(b)}$ and $\bar{c}_{a(b)}$, device $a(b)$ generates a rating $r_{ab(ba)}$ to evaluate $b(a)$'s behavior. For the integrity and accountability concerns, the rating is encrypted before exchanged that only the generator and the BS can decrypt. $T_{1,2,3}$ are signatures that are generated using the credentials of $a$, $b$, and the BS combined, while $t_0$ is used to guarantee the signature freshness. $Re_{ab(ba)}$ is the receipt of device $a(b)$.

\textbf{Trustvalue Adjustment:}
Since D2D receipts will expire after a certain period of time, participating devices will upload locally verified receipts to the BS (Step 3.2.1 in Fig.\ref{fig3}) when there is a cost-effective Internet access (\emph{e.g.} free WiFi).

When device $a$ uploads $Re_{ab}$, the BS can get the transaction information and behavior profiles including $a,b,c_{ab}, q_{ab}, \boldsymbol{\lambda}_{ab}$ and $r_{ba} = Dec(d_b,er_{ba})$. To verify the validity of $Re_{ab}$, the BS computes
\begin{equation}\label{eq12}
h_b'=h_2(er_{ba}\Vert c_{ab}\Vert q_{ab}\Vert \boldsymbol{\lambda}_{ab}),
\end{equation}then, according to the partial key pairs of $a$ and $b$, the BS verifies whether
\begin{equation}\label{eq13}
\displaystyle\frac{d_ah_b'}{d_b}T_{2b}=T_{1b}
\end{equation}
holds. If Eq.(\ref{eq13}) holds, the receipt is valid.

With a verified receipt, according to Eq.(\ref{eq1}) and (\ref{eq10}), the trustvalue of device $a$ can be adjusted.

For the integrity and authenticity concerns, the BS generates a signature for each adjusted device trustvalue. For device $a$ discussed above, the BS generates
\begin{equation} \label{eq14}
\begin{split}
T_{4bs}&=\displaystyle{\frac{l_{bs}\bm{t}_a\bm{u}}{x+r_a}},\\
&T_{5bs}=l_{bs}P,
\end{split}
\end{equation} where $l_{bs} \in_R Z_q^*$ is used to guarantee the signature freshness, $\bm{t}_a$ is the adjusted trustvalue of device $a$, and $\bm{u}$ is a $|\mathcal{N}|$-dimensional unit vector.

Then $\bm{t}_a\Vert T_{4bs}\Vert T_{5bs}$ is replied to device $a$ (Step 3.2.2 in Fig.\ref{fig3}). Any device that encounters $a$ can verify whether
\begin{equation} \label{eq15}
T_{4bs}(P_{pub}+R_a)=\bm{t}_a\bm{u}T_{5bs}
\end{equation}holds. If Eq.(\ref{eq15}) holds, the trustvalue is valid.

Until now, TDP realizes the secure device pairing and the online trust management in D2D-enabled MCSs.
\vspace{-1em}
\subsection{Security Analysis} \label{subsec55}
In this subsection, we conduct theoretical analysis to demonstrate the security intensity of TDP confronting potential threats in D2D-enabled MCSs (Subsection \ref{subsec32}). Note that we consider the Computational Diffie-Hellman Problem on Elliptic Curves (ECDHP)~\cite{joux2003separating} as intractable for any adversary with finite resources, which is clarified as follows:

Let $E$ be an elliptic curve on a finite field $F_q$, whose generator is $P$: given $P$, $aP$ and $bP$, where $a,b \in Z_q^*$, the computation of $abP$ is intractable.

\subsubsection{Confronting CO Attacks} \label{subsubsec551}

Adversaries launch CO attacks to compromise the device pairing process between registered devices. Without loss of generality, let an adversary $m \in \mathcal{M}$ ($\mathcal{M} \subset \mathcal{D}$) be a registered device, who has a verifiable credential $(sk_m, pk_m)$, that tries to compromise the device pairing process between benign devices $a,b \in \mathcal{D}$. In Appendixes B.1 to B.3, we demonstrate TDP's immunity to all CO attacks in Subsection \ref{subsec32}, which indicates its comparable security intensity to the off-the-shelf protocols~\cite{BL,WiFiDirect}.

\subsubsection{Confronting TO Attacks} \label{subsubsec552}

Adversaries launch TO attacks to manipulate D2D traffics. Without loss of generality, let a set of adversaries $\mathcal{M} \subset \mathcal{D}$ be multiple registered devices with verifiable credentials. They try to, either independently or collusively, attract unfair D2D transaction requests than benign devices in $\mathcal{D}$ do. In Appendixes B.4 to B.6, we demonstrate TDP's resistance to all TO attacks in Subsection \ref{subsec32}, which can effectively prevent traffic manipulations in D2D-enabled MCSs.
\vspace{-1em}
\subsection{Case Studies} \label{subsec56}
In this part, we conduct two case studies to demonstrate how to apply TDP to mobile collaborative computing tasks~\cite{li2014multiple,murray2010case} and mobile crowdsensing tasks~\cite{yangbackpressure,xiao2015multi,karaliopoulos2015userinfocom}.

\textbf{Collaborative Computing Tasks:}
In a mobile crowdcomputing task, the BS outsources computing requests to registered devices through the crowdsourcing application. Let device $a$ accept the task and recruit peer devices for collaborative computing. By investigating the status of nearby devices, including the trustvalue in terms of collaborative computing, $a$ determines its peer devices and then conducts device pairing with each one of them.

During the pairing process between $a$ and one of the recruited devices $b$, both of them use registered credentials $(sk,pk)$ to negotiate, generate and verify a shared key $k_{ab}$, which is used to secure D2D communications during the task. After the computing task, recruited devices reply ratings to demonstrate their satisfactory. To generate the receipt, $a$ and $b$ exchange $\mathcal{E}_a$ and $\mathcal{E}_b$ using the established secure connection. Then receipts from $a$ and $b$, \emph{i.e.} $Re_{ab}$ and $Re_{ba}$, are generated using the credentials of $a$, $b$ and the BS.	

When there is a cost-effective access to the Internet (\emph{e.g.} free WiFi at work, home or a cafe), $a$ and $b$ separately upload the receipts to the BS. After authenticating the receipts, the BS will adjust their trustvalues according to Eq.(\ref{eq1}) and (\ref{eq10}). For the collaborative computing, since both ends of the transaction equally contribute, both ends will have a trustvalue adjustment (\emph{i.e.} $f(\boldsymbol{\lambda}_{ab})=(1,1)$). Then the BS generates signatures for their adjusted trustvalues. Finally, the adjusted trustvalues and signatures are replied to $a$ and $b$ for future D2D transactions.

\textbf{Mobile Crowdsensing Tasks:}
In a mobile crowdsensing task, the BS outsources a sensing request to a participating device $a$ to collect and report sensing data. Perhaps there are Internet-access or cost issues and $a$ requires a peer device to deliver its data packets in a D2D manner. By investigating the status of nearby devices, including the trustvalue in terms of packet delivery, $a$ selects a peer device $b$ and conducts device pairing with it.

The device pairing and D2D receipt generation in such a mobile crowdsensing task are similar to that in the last case, therefore we make no further discussion for conciseness.

For the packet delivery, only the device that initiates the data transmission, which is device $a$ in this case, will have an `actual' trustvalue adjustment (\emph{i.e.} $f(\boldsymbol{\lambda}_{ab})=(1,0)$). Therefore the peer device that receives packets will try further deliveries to get a trustvalue adjustment. Then the BS generates a signature for $a$'s adjusted trustvalue. Finally, the adjusted trustvalue and signatures are replied to $a$ and $b$ (no change) for future D2D transactions.

\section{Evaluation} \label{sec6}
In this section, we evaluate the performance of TDP by both prototype experiments and trace-driven simulations.

\begin{figure}
    \centering
    \includegraphics[width=2.8in]{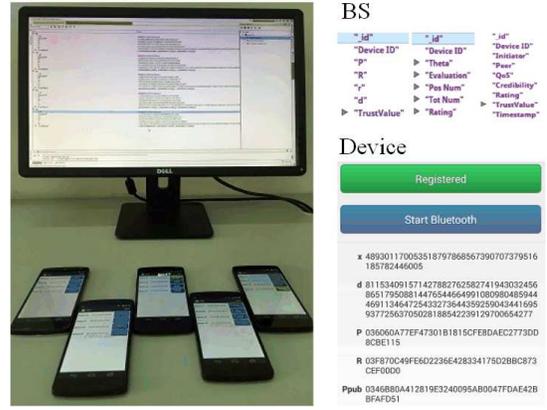}
    \caption{D2D-Enabled MCS Prototype Illustration.} \label{fig7}
\end{figure}

\begin{table}[!t]
  \centering
  \caption{Implementation Details of TDP.}
  \begin{tabular}{|m{0.20\columnwidth}<{\centering}|m{0.31\columnwidth}<{\centering}|m{0.31\columnwidth}<{\centering}|}
    \hline
    \textbf{Schemes} & \textbf{BTDP} & \textbf{WTDP}\\
    \hline
    \textbf{D2D Channel} & Bluetooth v3.0~\cite{BL} & WiFi direct v1.5~\cite{WiFiDirect}\\
    \hline
    \textbf{ROM} & 4.20 MB & 7.47 MB\\
    \hline
    \textbf{RAM} & 19.29 MB & 23.54 MB\\
    \hline
    \textbf{Cryptography Primitives} & ECC-secp160r1, AES-128~\cite{bouncy2014,certicomsec2000} & ECC-secp160r1, AES-128~\cite{bouncy2014,certicomsec2000}\\
    \hline
  \end{tabular} \label{tab2}
\end{table}
\vspace{-1em}
\subsection{Real-world Experiments} \label{subsec61}

As shown in Fig.\ref{fig7}, we constructed a MSC prototype consisting of:

 \begin{itemize}
 \item \textbf{A PC based BS}: A Dell OPTIPLEX 9010 with Linux Mint 17.3-64bit, Intel Core i5-3470 3.2GHz CPU, and 8GB RAM.
 \item \textbf{Five Android devices}: Google Nexus5 with Android 4.4.4, Qualcomm Snapdragon 800 2.3GHz CPU, 2GB RAM, and 16GB ROM.
\end{itemize}

Devices can communicate with the BS through WiFi and communicate with each other through D2D channels.

As shown in Table~\ref{tab2}, we implemented two versions of TDP as independent security services upon \textit{user-transparent but insecure} D2D connections on Android mobile devices: BTDP is based on a Bluetooth connection in the \code{JustWork} mode~\cite{BL}, and WTDP is based on a WiFi direct connection in the \code{ServiceDiscovery} mode~\cite{WiFiDirect} whose mandatory manual authentication of WiFi Protected Setup (WPS) is blocked. Our implementation adopted the cryptography primitives in~\cite{bouncy2014,certicomsec2000}.

In prototype experiments, we evaluated the performance of device pairing and trust management processes.

\begin{figure}
\setlength{\belowcaptionskip}{-1em}
\centering
\subfloat[Bluetooth Pairing]{\includegraphics[width=1.5in]{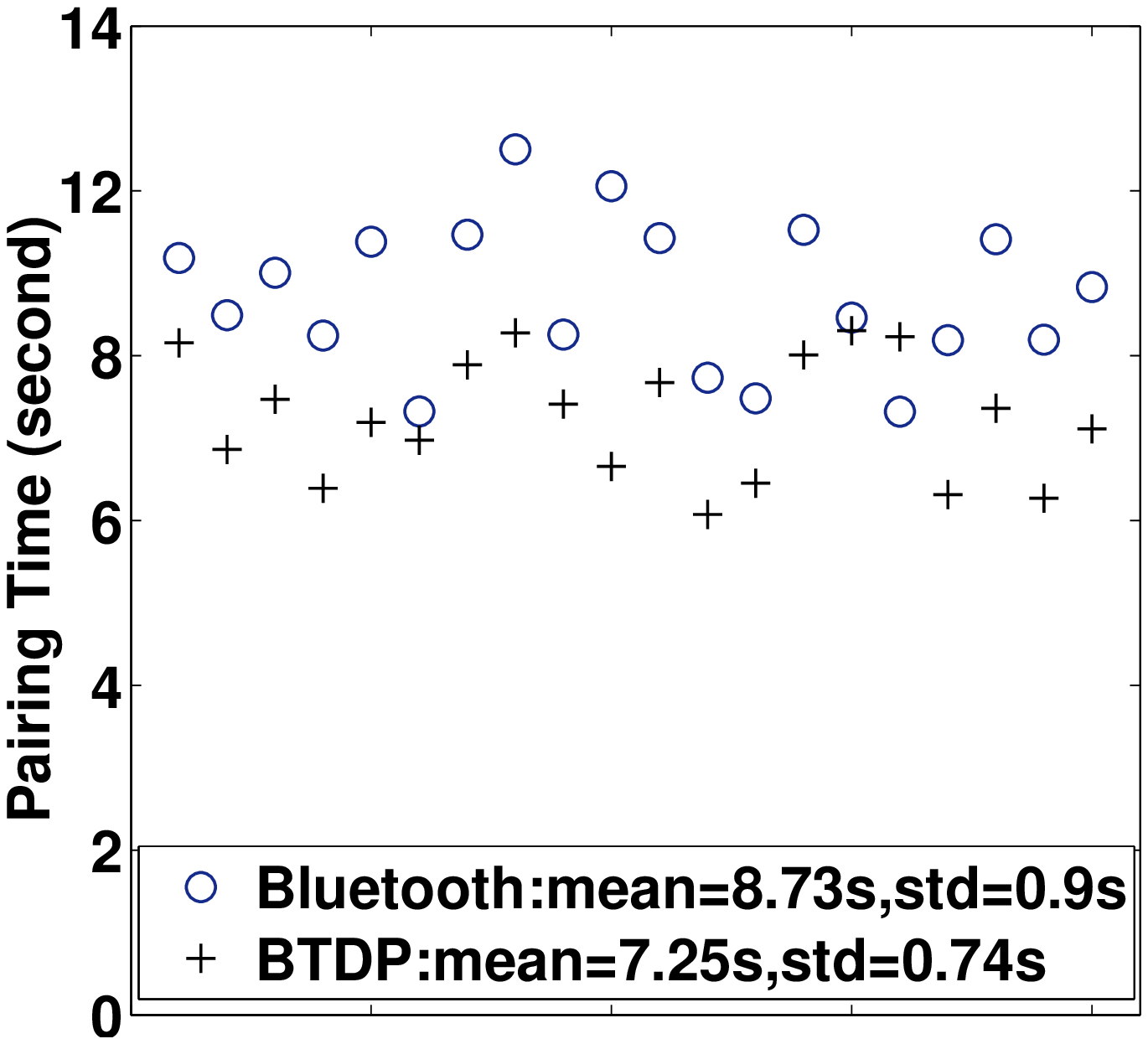}%
\label{fig8a}}
\hfil
\subfloat[WiFi direct Pairing]{\includegraphics[width=1.5in]{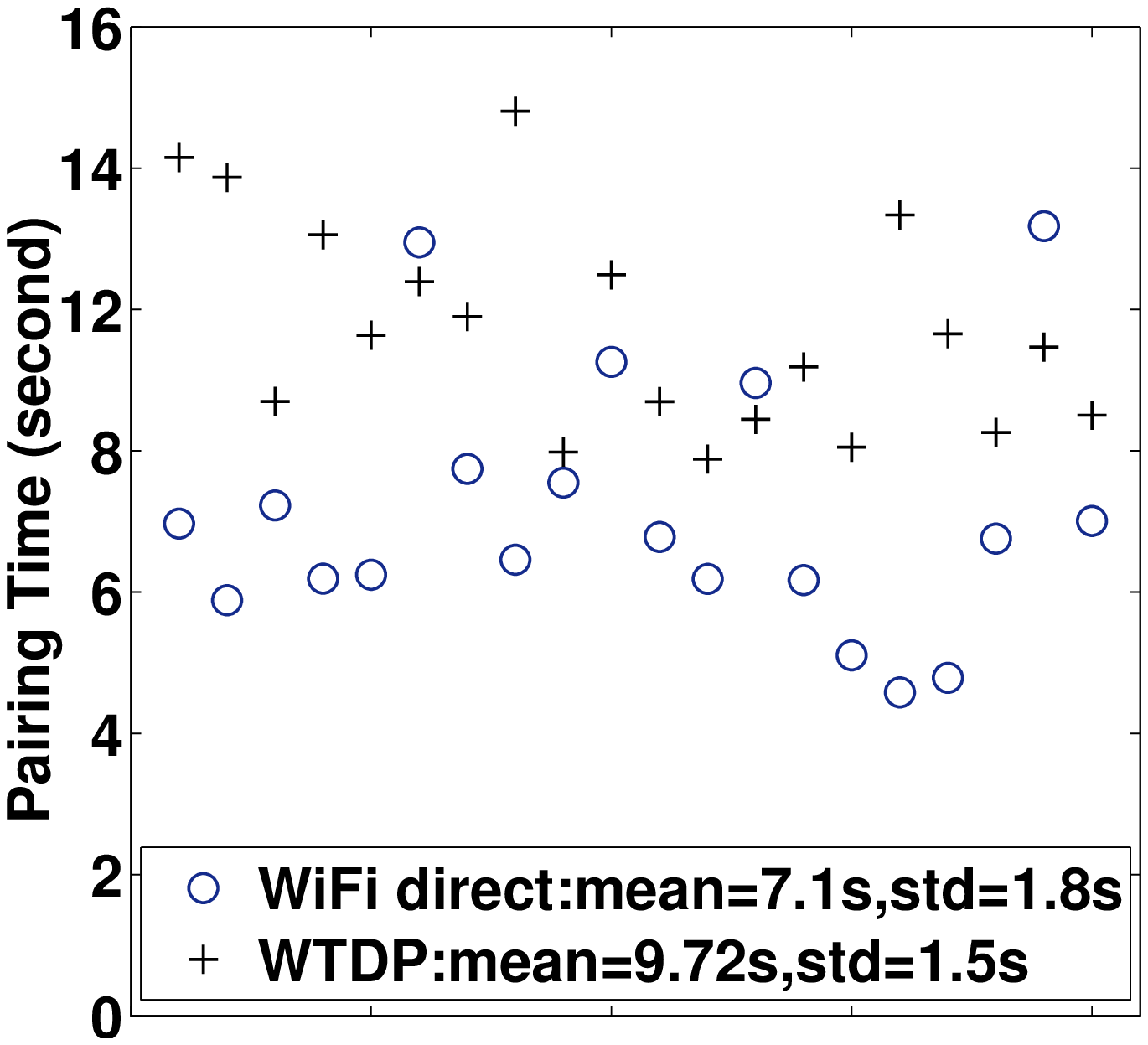}%
\label{fig8b}}
\caption{Off-the-Shelf Pairing Time Comparison.}
\label{fig8}
\end{figure}

\subsubsection{Device Pairing Study} \label{subsubsec611}

In this set of experiments, we evaluated the performance of TDP device pairing process. Specifically, we considered \textit{pairing time}, \emph{i.e.} the time consumed from the device discovery to the establishment of a shared key, as the performance metric to measure the pairing speed. We compared BTDP and WTDP with original Bluetooth and WiFi direct with default settings~\footnote{Note that we did our best to shorten the reactive time whenever there was a physical interaction (\emph{i.e.} manual confirmations for pairing requests: PIN comparison, pairing button click).}. As shown in Fig.\ref{fig8} (a) and (b), 20 pairing time samples were collected for BTDP (and original Bluetooth) and WTDP (and original WiFi direct) respectively.

Fig.\ref{fig8} (a) shows that both the mean and standard deviation (std) of the pairing time of BTDP are smaller than that of original Bluetooth pairing. This demonstrates that the pairing process of BTDP is not only faster, but also more stable, in comparison with Bluetooth. The autonomous key negotiation of BTDP avoids undetermined delays caused by manual confirmations in original Bluetooth pairing. It shortens the pairing time and limits any potential fluctuations other than that caused by Bluetooth device discovery itself.

From Fig.\ref{fig8} (b), we can see that WTDP outperforms WiFi Direct in terms of pairing stability (by avoiding potential manual operation delays), but its average pairing speed is slower. Different from the BTDP implemented on the \textit{insecure but autonomous} Bluetooth connections (\emph{i.e.} the \code{JustWork} mode of Bluetooth~\cite{BL}), there exists no such WiFi direct connecting mode available for WTDP~\footnote{The default WiFi direct pairing is in the \code{ActiveScan} mode~\cite{WiFiDirect}, which requires both users to scan simultaneously for device discovery.}.
Therefore, we had to implement WTDP in the \code{ServiceDiscovery} mode~\cite{WiFiDirect} of WiFi direct for autonomous device discovery, and block the mandatory manual confirmation in the Android Wifip2p service. This inefficient implementation somehow emulated \textit{insecure but autonomous} WiFi direct connections, but leading to significant performance degradation of our approach, \emph{i.e.} each recorded pairing time of WTDP consists of both an entire round of original WiFi direct pairing and our upper-layer security service.

Table~\ref{tab3} compares the pairing time of TDP and state-of-the-art user-transparent device pairing approaches that are promising in implementing D2D-enabled MCSs. The proximity-based approaches in~\cite{Miettinen:2014:CZP:2660267.2660334} and~\cite{truong2014comparing} collect contextual data for 120s and 10s respectively to generate key materials, and the time for key negotiation and authentication is not considered. The non-interactive approach in~\cite{184395} pairs up two devices based on a series of asynchronous broadcasts of device public key materials with a 60s or 120s interval. it is possible that a pairing time fluctuation up to 100\% will be introduced. Compared with them, TDP has no requirement on contextual information collection or device proximity, and its pairing fluctuation is well restricted without the synchronization concern.

\begin{table}[!t]
  \centering
  \caption{User-Transparent Pairing Time.}
  \begin{tabular}{|m{0.18\columnwidth}<{\centering}|m{0.14\columnwidth}<{\centering}|m{0.14\columnwidth}<{\centering}|m{0.13\columnwidth}<{\centering}|m{0.13\columnwidth}<{\centering}|}
    \hline
    \textbf{Related Schemes} & \textbf{Miettinen \emph{et al.}\cite{Miettinen:2014:CZP:2660267.2660334}} & \textbf{Truong \emph{et al.}\cite{truong2014comparing}} & \textbf{Lentz \emph{et al.}\cite{184395}} & \textbf{TDP (B/W)}\\
    \hline
    \textbf{Secret Extraction/s} & 120 & 10 & No need & No need\\
    \hline
    \textbf{Key Negotiation/s} & Not considered & Not considered & Not specified & 7.25/9.72\\
    \hline
    \textbf{Fluctuation/\%} & Not specified & Not specified & 0-100 & 10.2/15.4 \\
    \hline
  \end{tabular} \label{tab3}
\end{table}

\subsubsection{Trust Management Study} \label{subsubsec612}

This set of experiments for trust management evaluation were based on BTDP pairing. Specifically, we adopted the average RSSI of the Bluetooth channel measured by both ends at the start of device pairing process as the metric for QoS estimation. The RSSI values were normalized, \emph{i.e.} proportionally mapped from the range of [-100 dBm,~0dBm] to [0,~1]. According to Eq.(\ref{eq8}), the rating that reflects the device behavior during a D2D transaction can be either QoS-driven or Credibility-Driven by adjusting the value of $\beta$. Therefore, we conducted two sets of experiments that separately focused on QoS ($\beta=0.8$) and Credibility ($\beta=0.2$), meanwhile we monitored the variation of device trustvalue and the direction of D2D traffic within the system.

Each set of experiments had 40 D2D transactions. At an experiment round, registered devices (Devices 1-5) took turns to post a transaction request and autonomously paired with the most trustworthy device that was visible. One of the devices (in turn) was excluded from candidates for a single round. During the experiments, only the trustvalue of transactions that lead to trust variations at the peer side was considered.
We set the average transaction count $\bar{\sigma}=30$. As a result, the trust accumulation sensitivity $c_g=0.5$ and the damping sensitivity $c_w=0.5$ according to Eq.(A1) and (A2) respectively. The credibility sensitivity $\alpha=0.5$.

\begin{figure}[!t]
\setlength{\belowcaptionskip}{-1em}
\centering
\subfloat[Trustvalue]{\includegraphics[width=1.5in]{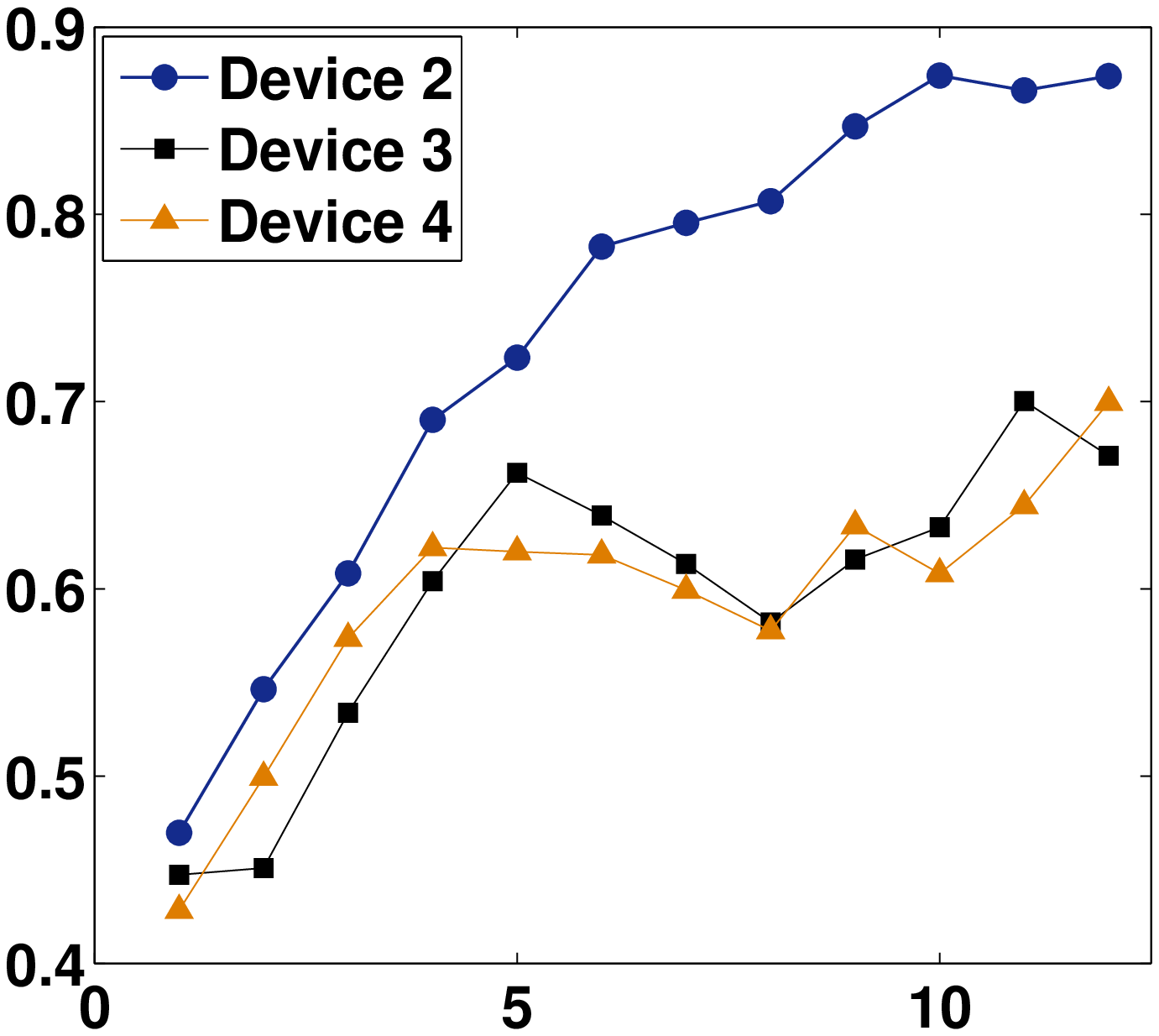}%
\label{fig9a}}
\hfil
\subfloat[QoS]{\includegraphics[width=1.5in]{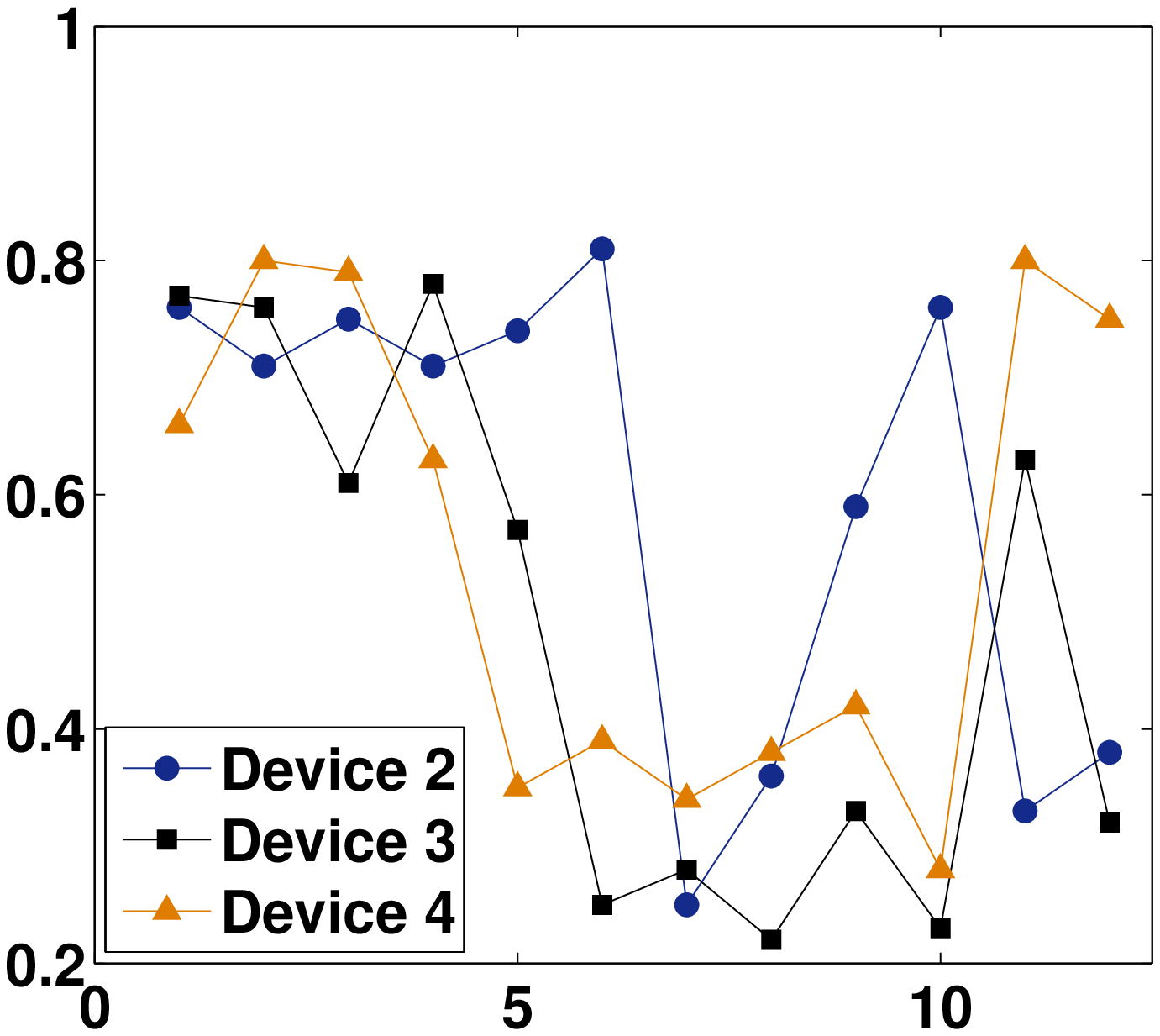}%
\label{fig9b}}
\caption{QoS-Driven Trust Management.}
\label{fig9}
\end{figure}

Fig.\ref{fig9} shows the results of the QoS-driven experiments. Here, each device was carried by a student roaming around the laboratory to introduce QoS variations. The majority of D2D requests were attracted by Devices 2-4. For Devices 3 and 4, a higher QoS boosts their trustvalue and vice versa. Moreover, since Device 2 has a trustvalue boost at the beginning due to our device encountering setting, it stays in a dominant position all the time compared with others. Even it provides lower QoS for a while, its trustvalue has no severely but only a slightly drop since almost all of contact histories of others are with Device 2. To avoid such kind of monopoly, $\beta$ should be well adjusted to offer devices with inherently lower QoS a chance for a trustvalue boost with higher credibilities.

\begin{figure}[!t]
\setlength{\belowcaptionskip}{-1em}
\centering
\subfloat[Device 2 Profiles]{\includegraphics[width=1.5in]{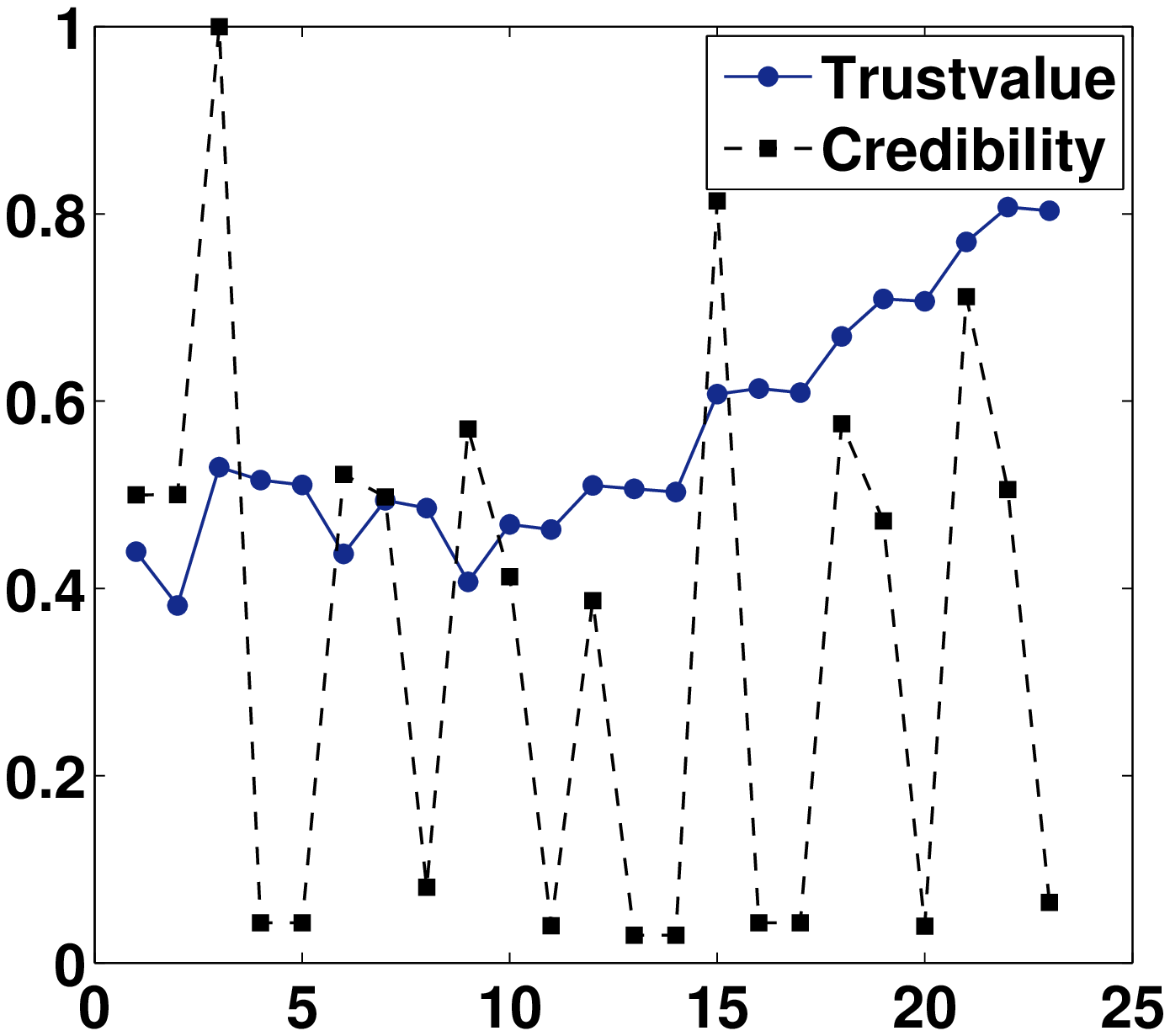}%
\label{fig10a}}
\hfil
\subfloat[Device 3 Profiles]{\includegraphics[width=1.5in]{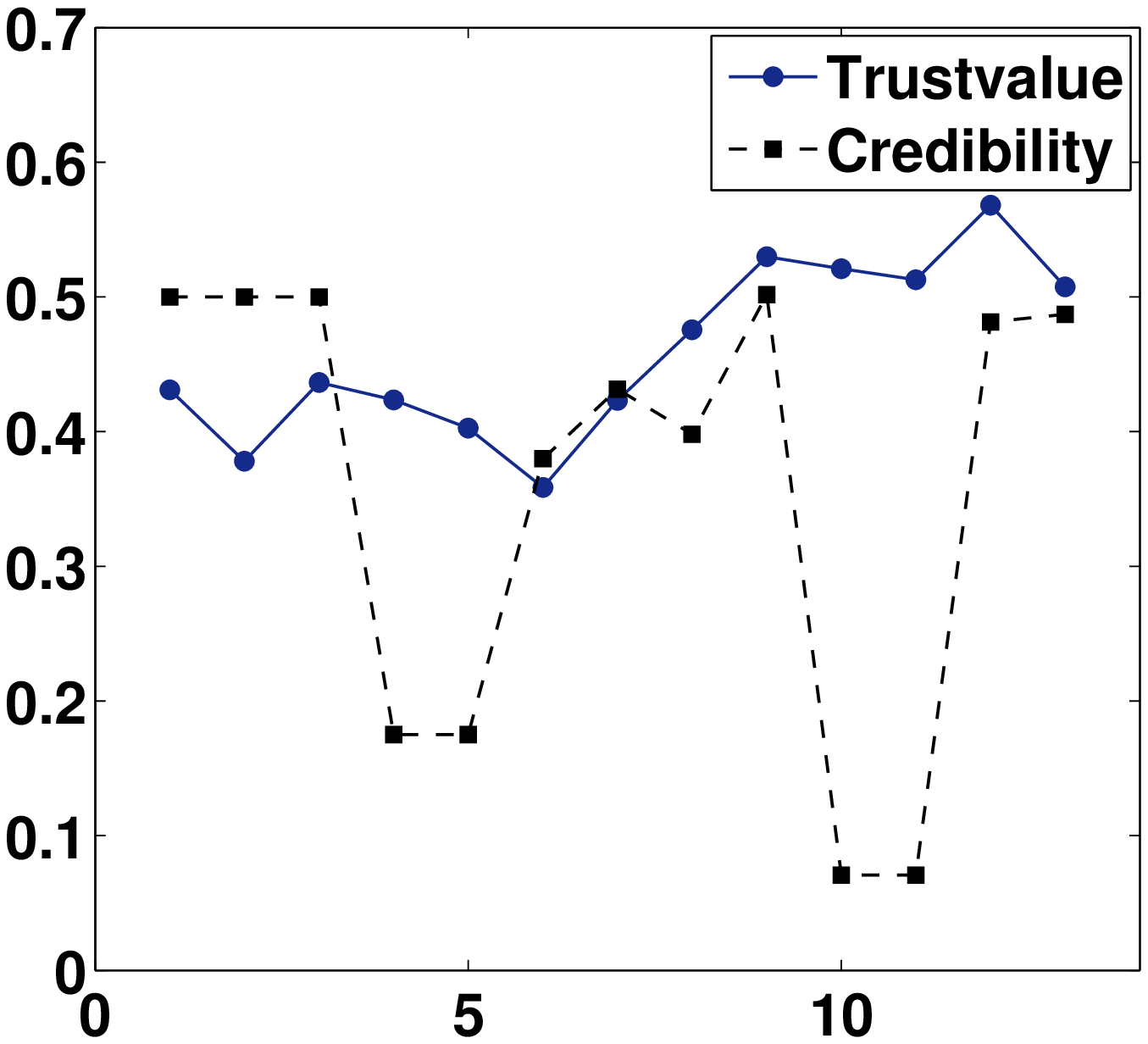}%
\label{fig10b}}
\caption{Credibility-Driven Trust Management.}
\label{fig10}
\end{figure}

Fig.\ref{fig10} shows the results of Credibility-driven experiments. Here, Devices 4 and 5 were set to reply negative ratings in all D2D transactions to introduce credibility differences. The majority of D2D requests were attracted by Devices 2 and 3. Devices 4 and 5 only attracted two requests because of low credibilities. For Devices 2 and 3, transactions with a credibility around $0.1$ were conducted with Device 4 or 5. The influence of ratings with abnormal credibilities on the trustvalue of Devices 2 and 3 is nearly negligible. When they conduct transactions with other devices, a higher credibility boosts their trustvalue and vice versa. Actually, to encourage users refining their QoS, $\beta$ should be well adjusted to offer devices with lower credibilities a chance for a trustvalue boost with higher QoS.

\vspace{-1em}
\subsection{Trace-Driven Simulations} \label{subsec62}

To study the practical performance of TDP in larger-scale MCSs, we constructed extensive trace-driven simulations based on OMNeT++ 4.6~\cite{Omnet}, by using real human mobility data collected from the IEEE Infocom'05 conference~\cite{cambridge-haggle-20090529} and the ACM Sigcomm'09 conference~\cite{thlab-sigcomm2009-20120715}. The settings of these two sets of simulations are as follows:

\begin{itemize}
\item \textbf{The Infocom'05 Trace} was used for constructing a D2D-enabled MCS with 41 registered devices. Here, the connecting time of D2D channels between encountered devices was used to estimate the D2D QoS. According to the simulation result with no attacker, we set the average transaction count $\bar{\sigma}=110$. Therefore sensitivities $c_g=0.14$ and $c_w=0.5$ according to Eq.(A1) and (A2) respectively. For Eq.(\ref{eq3}), we set $\alpha=0.5$ to equally treat the similarity and diversity, and $\beta=0.1$ for Eq.(\ref{eq8}) to compensate the inherent unfairness on the device QoS. According to duration of the Infocom'05 trace, each round of simulations lasted for 30000 simulation seconds, which was treated as a single trustvalue management cycle.

\item \textbf{The Sigcomm'09 Trace} has nearly twice node records than the Infocom'05 trace. By using this trace, a D2D-enabled MCS with 76 devices was constructed to evaluate the scalability of TDP. Here, the volume of data transmitted through D2D channels was used to estimate the D2D QoS. In this set of simulations, we set $\bar{\sigma}=31$, $c_g=0.5$, $c_w=0.5$, $\alpha=0.5$ and $\beta=0.1$. Each round of simulations lasted for 15000 simulation seconds.
\end{itemize}

In all simulations, devices can spontaneously register to the BS and publish or accept D2D requests. The trustvalue was treated as the unique metric for peer selections. Meanwhile, since the type of D2D transactions has no impact on the performance of TDP, we only considered the trustvalue of transactions that lead to trust variations at the peer side.

\subsubsection{Performance Metrics} \label{subsubsec621}

Since the goal of TO attackers in D2D-enabled MCSs is to get unfair advantages in peer selections over benign devices, we collected the following two metrics to evaluate the effectiveness of TDP against potential traffic manipulations:

\begin{itemize}
\item \textbf{Device Trustvalue}, the time-varying device trustvalue, whose cumulative distribution indicates whether malicious behaviors of TO attackers downgrade the trustworthiness of benign devices. Ideally, TDP should be able to automatically downgrade the trustvalue of TO attackers without disturbing (downgrading) the trustvalue of benign devices;
\item \textbf{False Positive Transactions Numbers}, the number of transactions attracted by TO attackers from benign devices, which quantifies the impact of traffic manipulations of TO attackers. Ideally, TDP should be able to automatically limit the number of transactions attracted by TO attackers.
\end{itemize}

In our simulations, participating devices in the MCS were divided into two categories: attackers and benign devices, whose trustvalue and false positive transactions were recorded separately for analysis. For comparison, we ran two simulations with no attacker for both the Infocom and Sigcomm traces (\emph{i.e.} the original scenario). For the rest of the paper, we use the Cumulative Distribution Figure (CDF) to present the trustvalue of both benign devices and TO attackers, and we use the Time Variance Figure (TVF) to present the number of false positive transactions attracted by TO attackers.

\subsubsection{Impact of Attacker Percentage} \label{subsubsec622}

\begin{figure}[!t]
\setlength{\belowcaptionskip}{-1em}
\centering
\subfloat[Trustvalue (5\%)]{\includegraphics[width=1.7in]{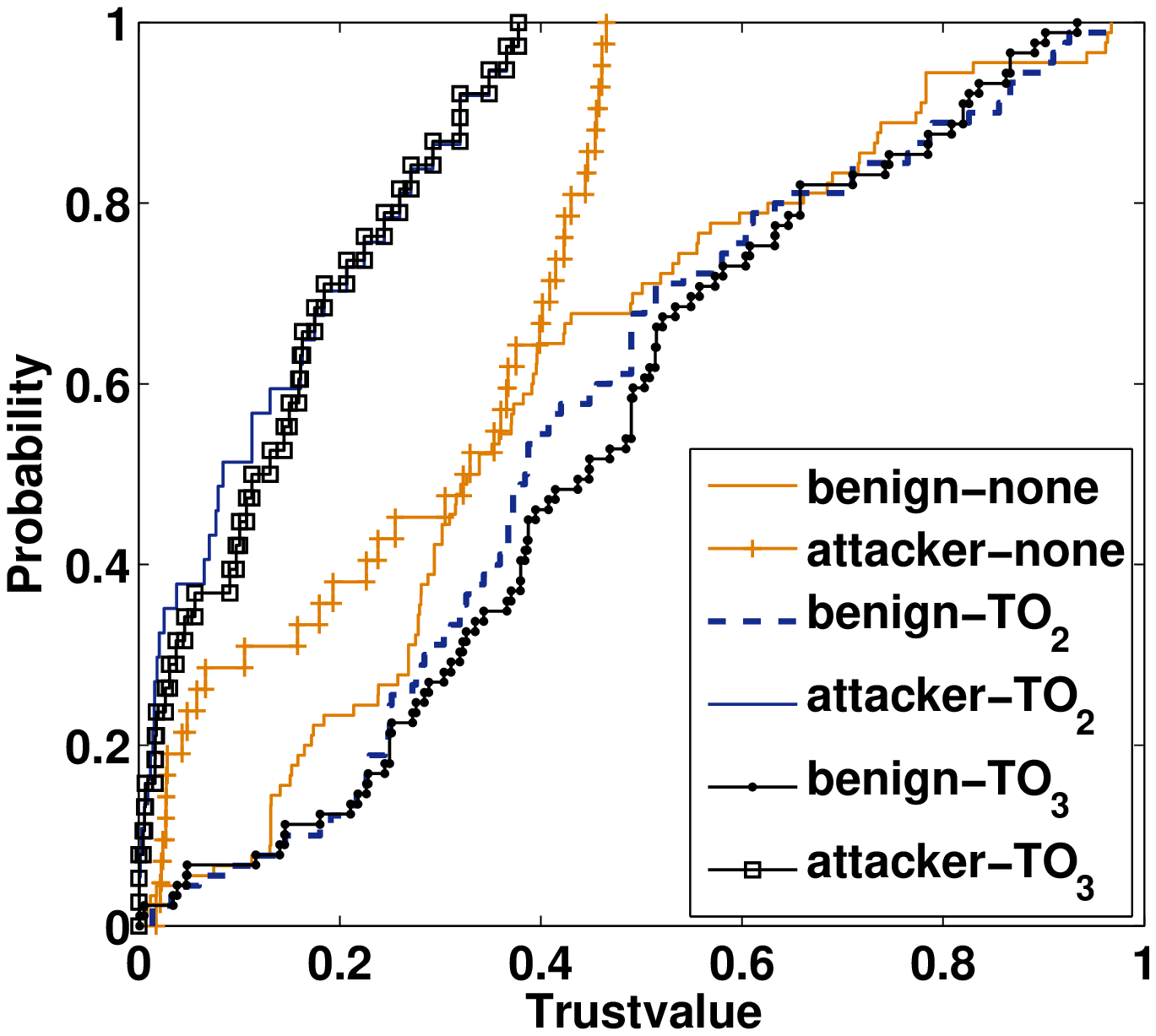}%
\label{fig11a}}
\hfil
\subfloat[False Positive Count (5\%)]{\includegraphics[width=1.7in]{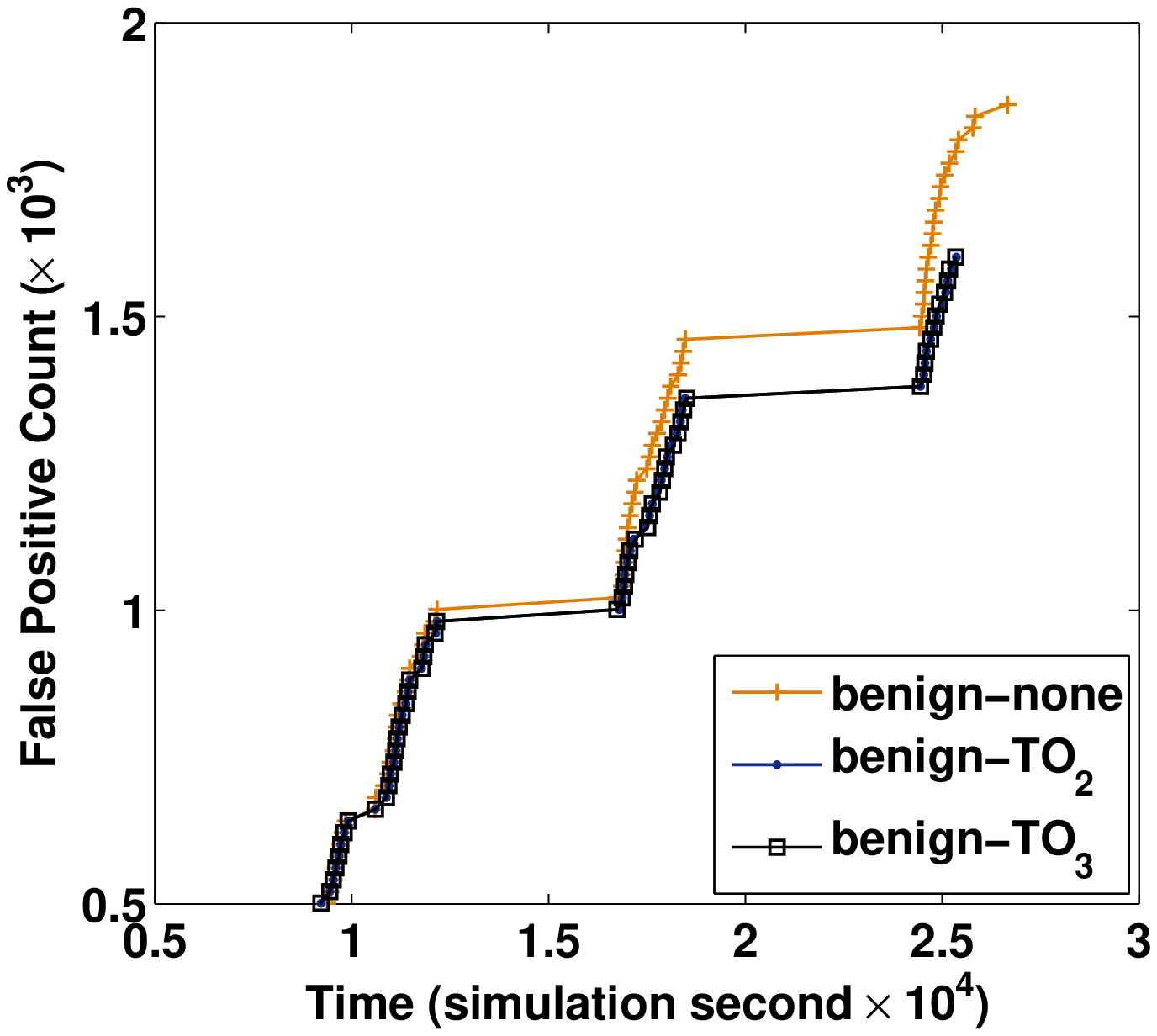}%
\label{fig11b}}
\hfil
\subfloat[Trustvalue (10\%)]{\includegraphics[width=1.7in]{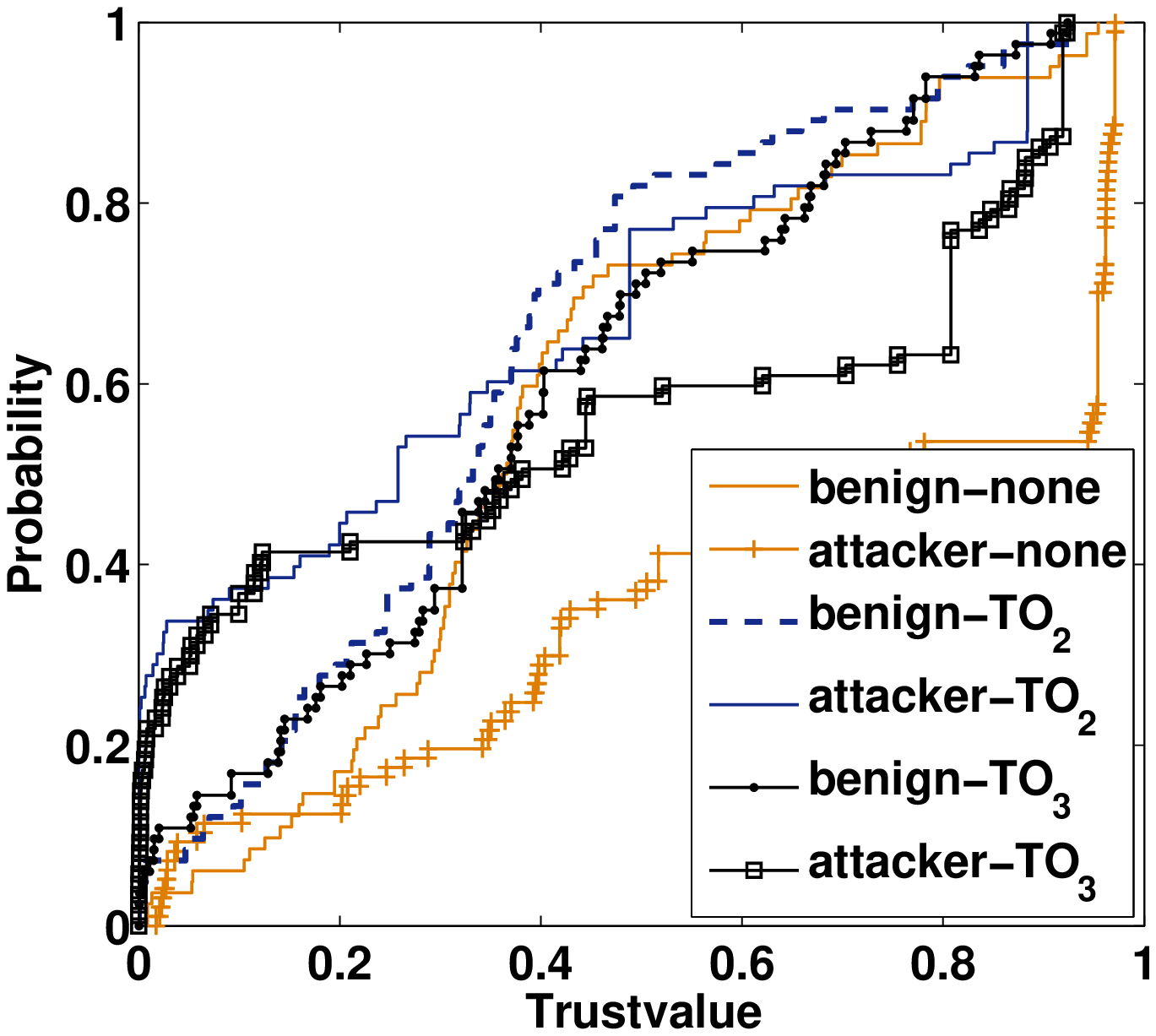}%
\label{fig11c}}
\hfil
\subfloat[False Positive Count (10\%)]{\includegraphics[width=1.7in]{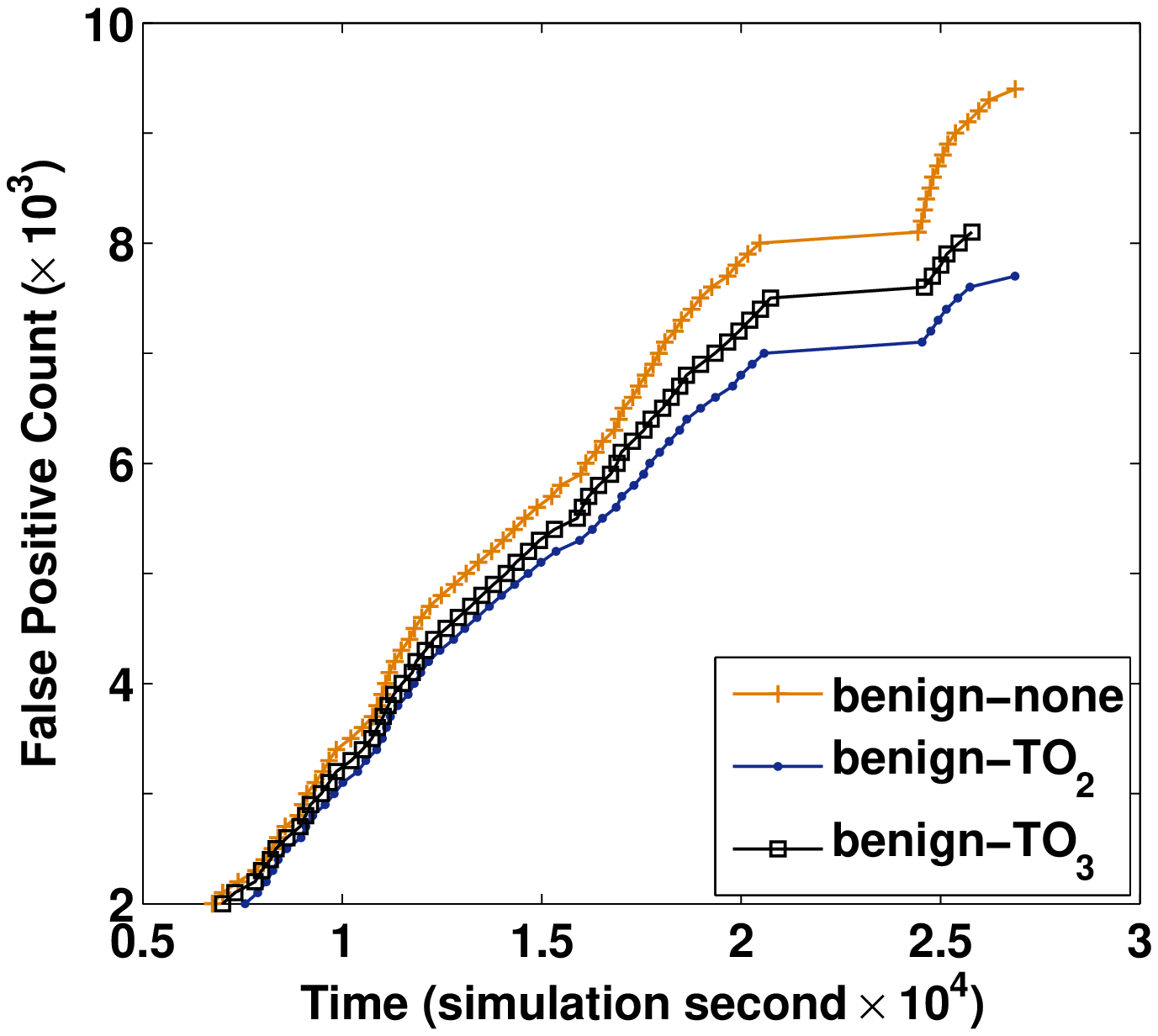}%
\label{fig11d}}
\hfil
\subfloat[Trustvalue (15\%)]{\includegraphics[width=1.7in]{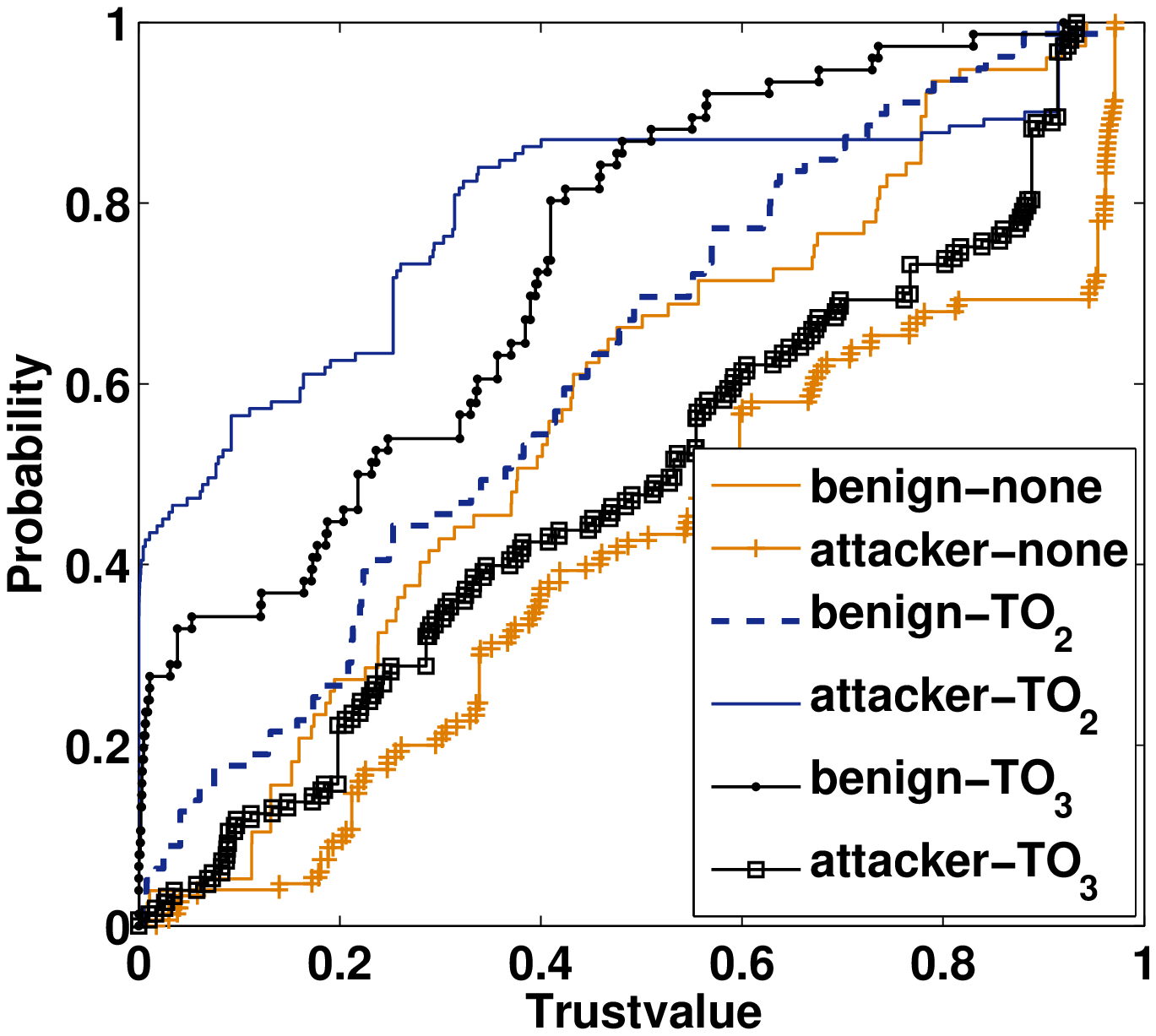}%
\label{fig11e}}
\hfil
\subfloat[False Positive Count (15\%)]{\includegraphics[width=1.7in]{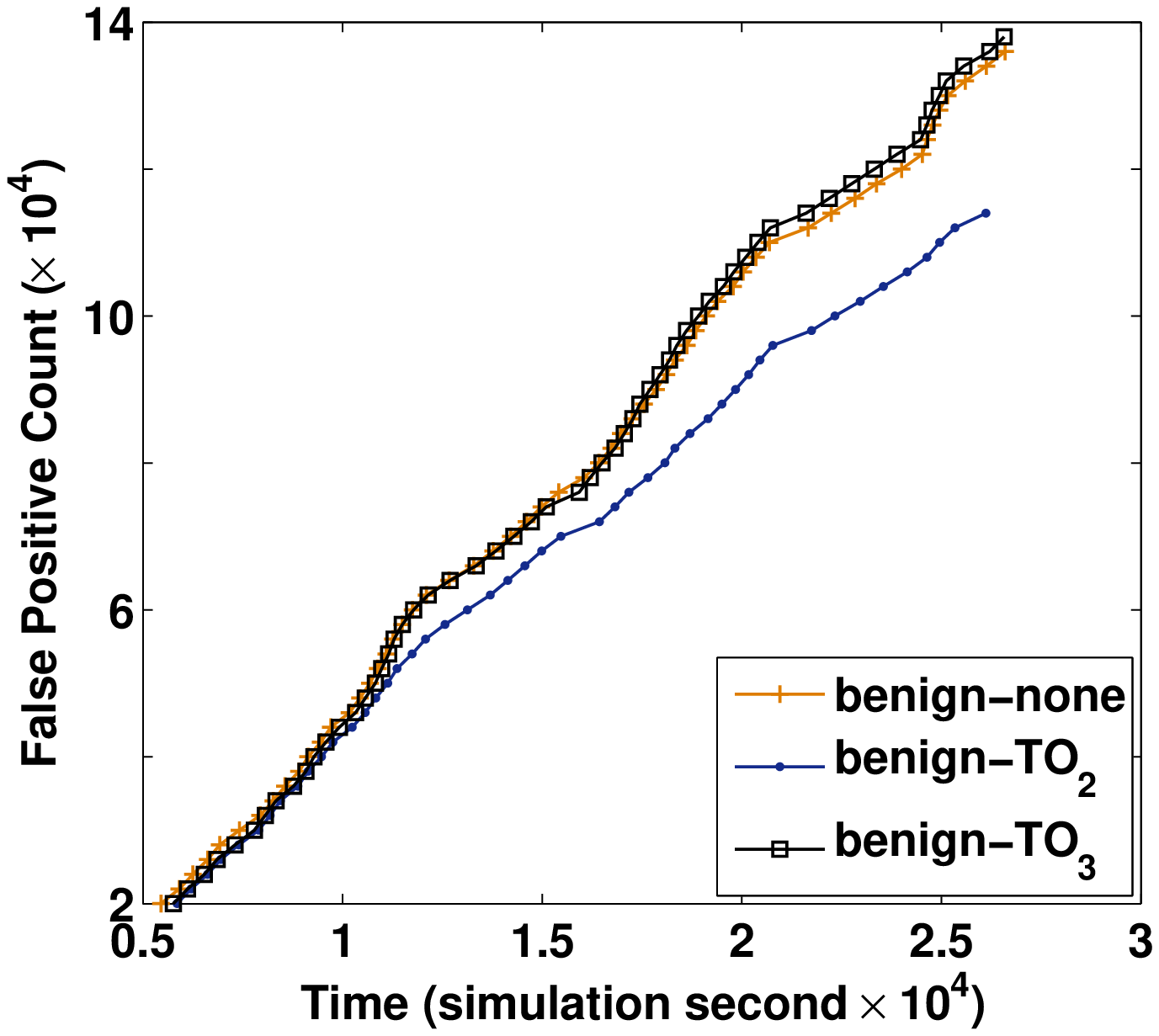}%
\label{fig11f}}
\caption{Impact of Different Attacker Percentages (Infocom).}
\label{fig11}
\end{figure}

\begin{figure}[!t]
\setlength{\belowcaptionskip}{-1em}
\centering
\subfloat[Trustvalue (5\%)]{\includegraphics[width=1.7in]{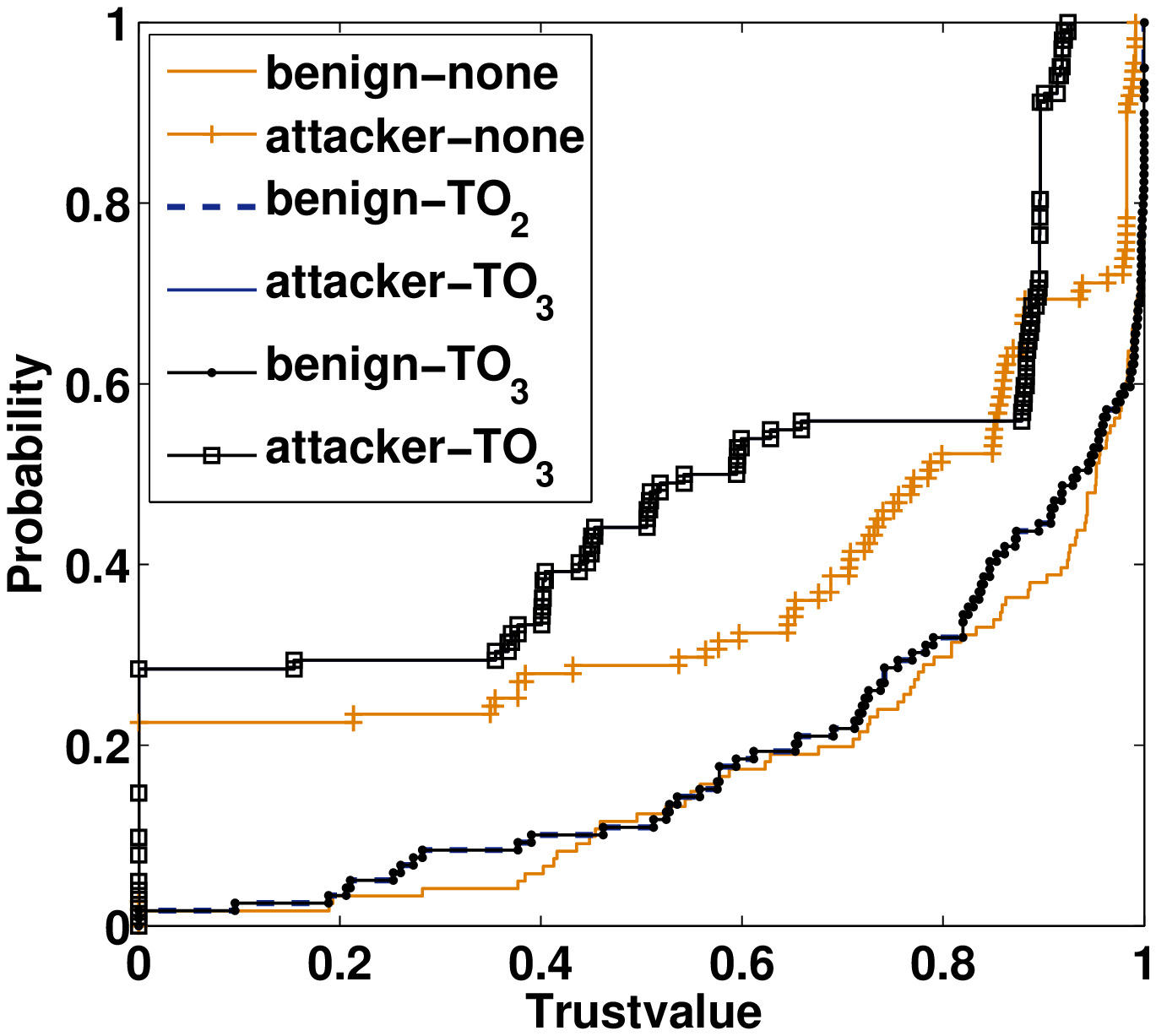}%
\label{fig12a}}
\hfil
\subfloat[False Positive Count (5\%)]{\includegraphics[width=1.7in]{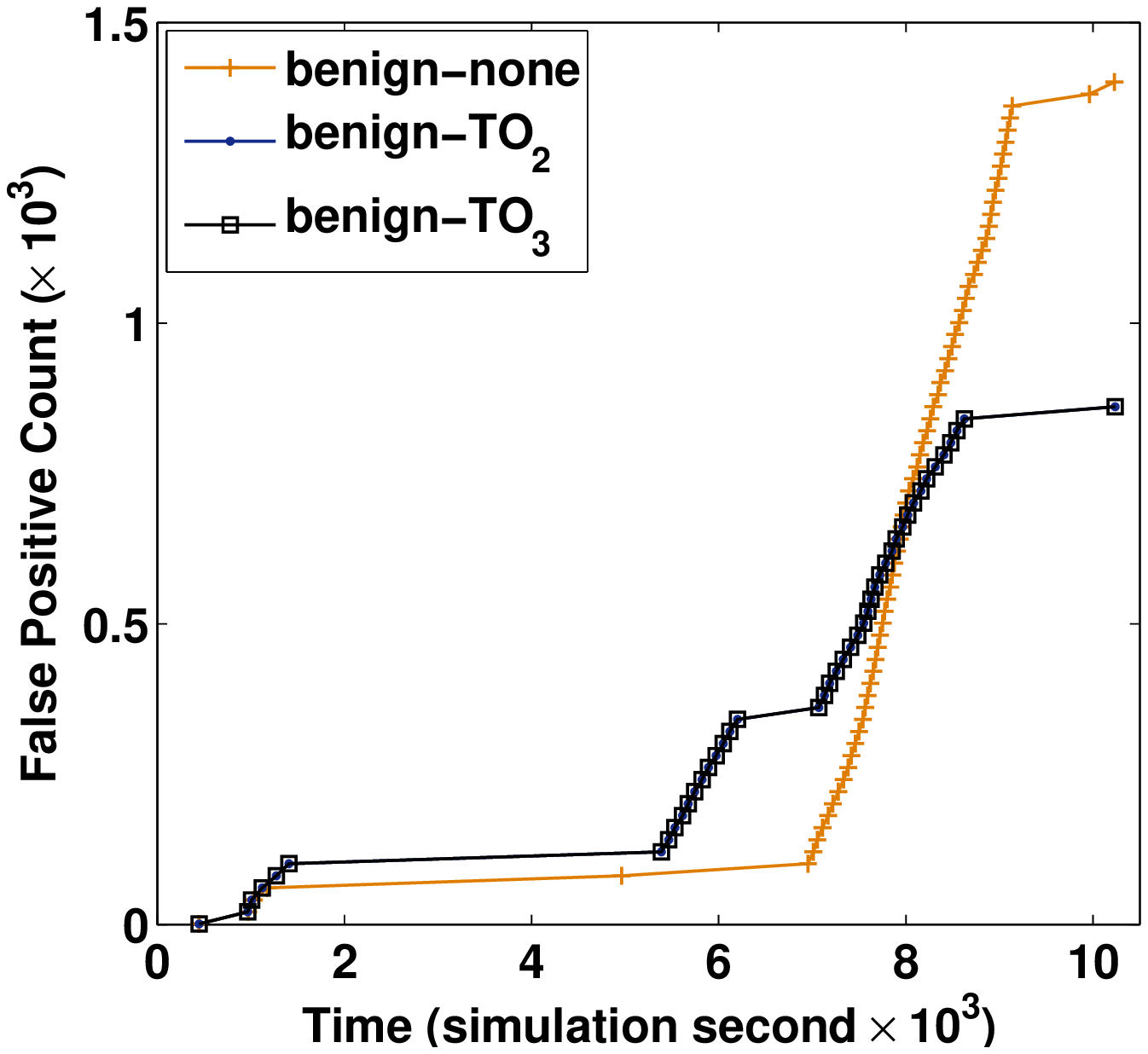}%
\label{fig12b}}
\hfil
\subfloat[Trustvalue (10\%)]{\includegraphics[width=1.7in]{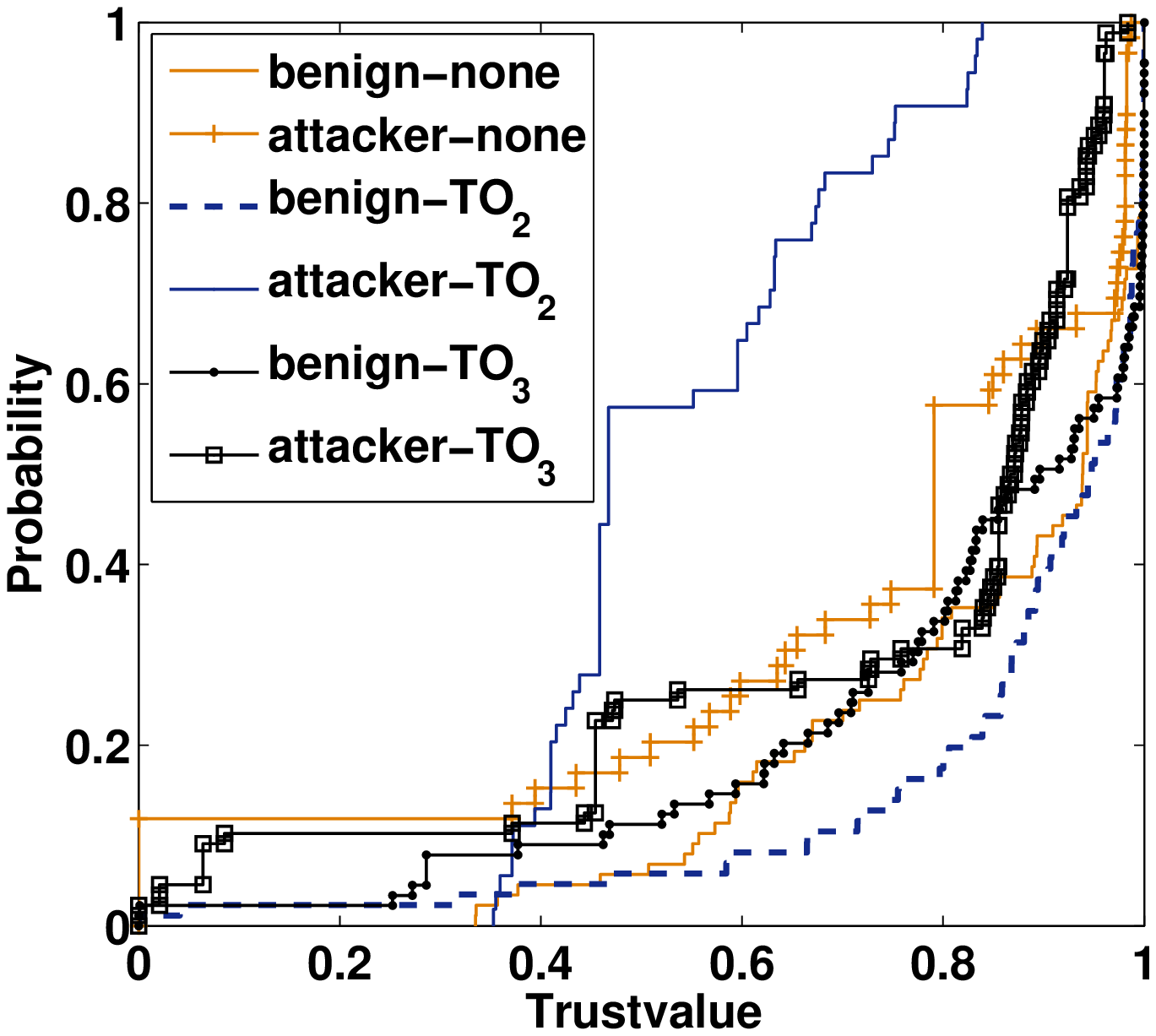}%
\label{fig12c}}
\hfil
\subfloat[False Positive Count (10\%)]{\includegraphics[width=1.7in]{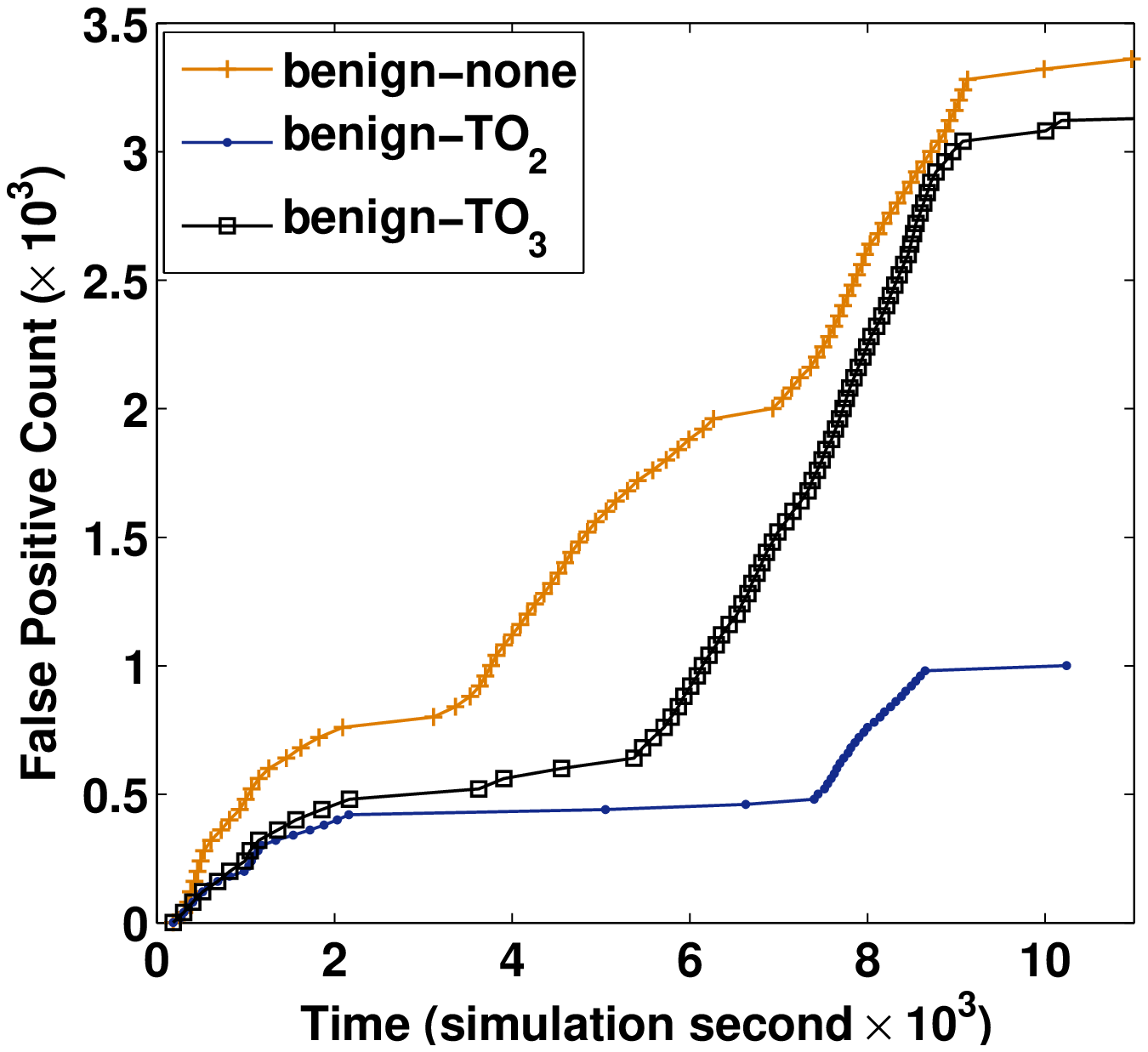}%
\label{fig12d}}
\hfil
\subfloat[Trustvalue (15\%)]{\includegraphics[width=1.7in]{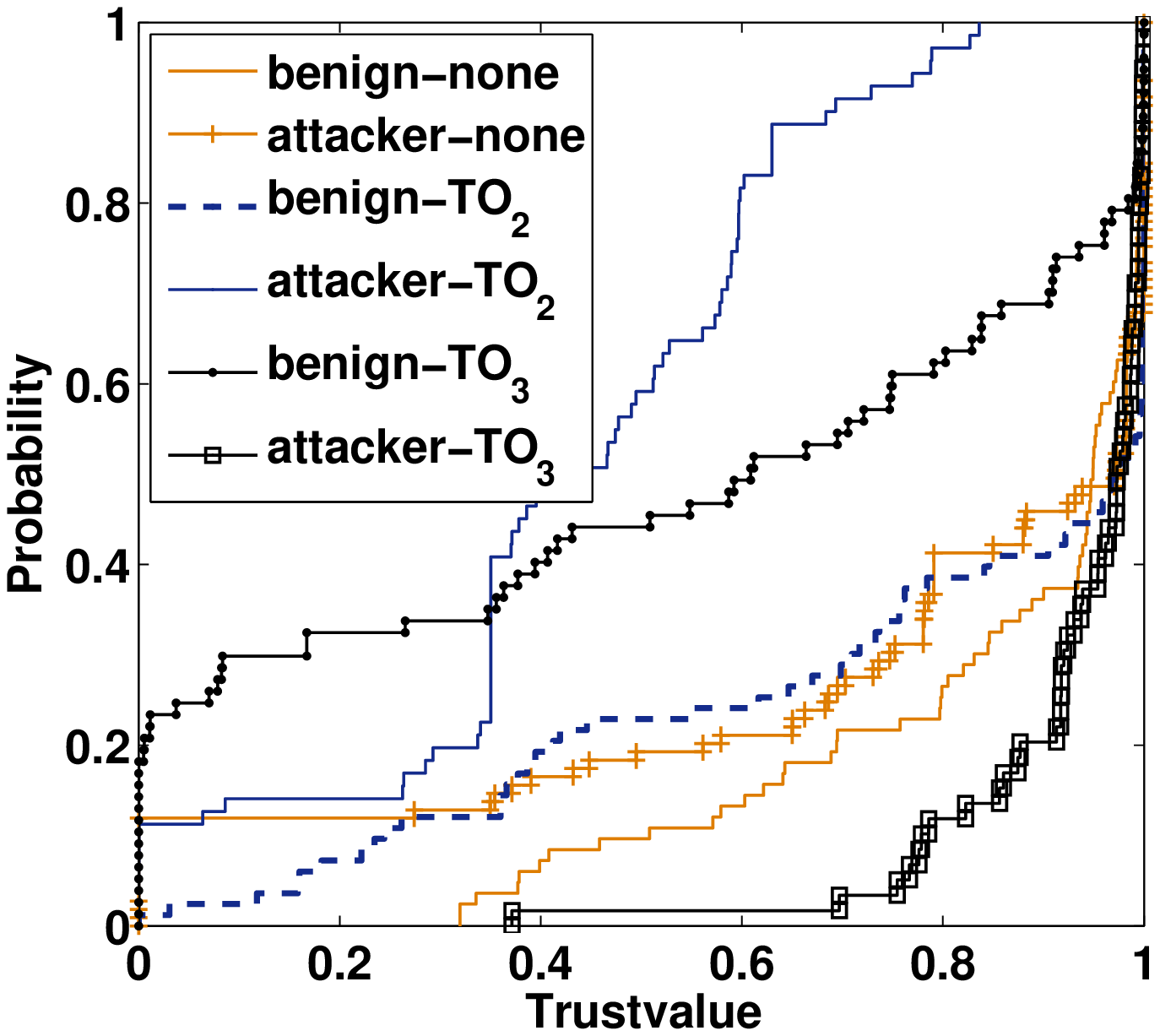}%
\label{fig12e}}
\hfil
\subfloat[False Positive Count (15\%)]{\includegraphics[width=1.7in]{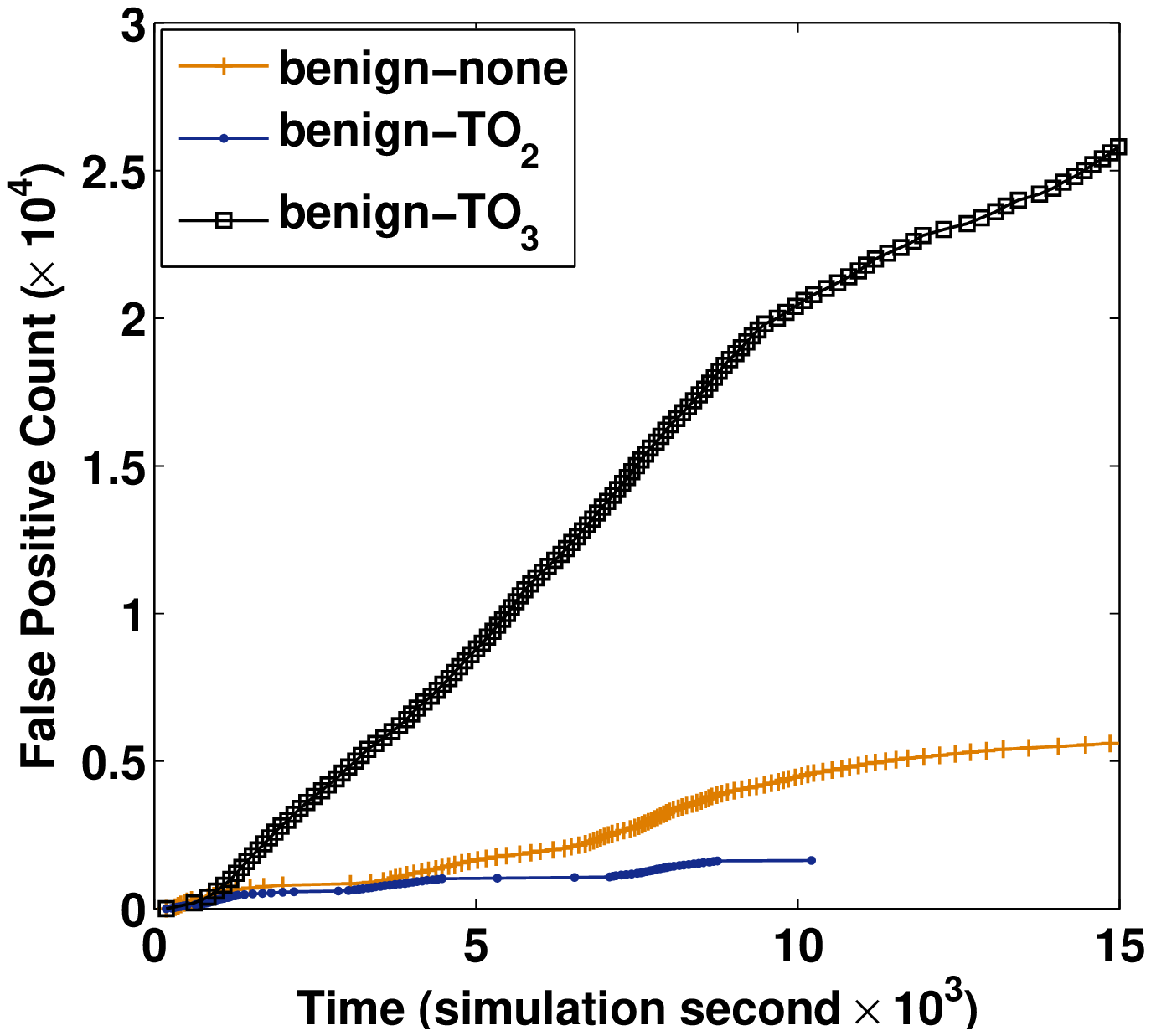}%
\label{fig12f}}
\caption{Impact of Different Attacker Percentages (Sigcomm).}
\label{fig12}
\end{figure}

In this set of simulations, different percentages (5\%, 10\%, and 15\%) of registered devices were randomly selected as attackers that were continuously launching TO attacks. The results of the Infocom and Sigcomm simulations are shown in Fig.\ref{fig11} and Fig.\ref{fig12} respectively.

For both sets of simulations, the impact of TO$_2$ attacks is well eliminated in all three scenarios: the average trustvalue of benign devices is not obviously affected compared with the original scenario (\emph{e.g.} benign-TO$_2$ ($0.46/9.5\%^\uparrow$) vs benign-none ($0.42$) in Fig.\ref{fig11} (a)). However, the average trustvalue of attackers is severely downgraded because of their TO$_2$ attacks (\emph{e.g.} attacker-TO$_2$ ($0.13/50\%^\downarrow$) vs attacker-none ($0.26$) in Fig.\ref{fig11} (a)). Meanwhile, the false positive transaction number also decreases when there are TO$_2$ attacks (\emph{e.g.} benign-TO$_2$ ($1612/13.5\%^\downarrow$) vs benign-none ($1864$) in Fig.\ref{fig11} (b)).

For the Infocom simulations, the impact of TO$_3$ attacks is well eliminated in all three scenarios. For the Sigcomm simulations, when the attacker percentage is 15\%, the impact of TO$_3$ attacks cannot be neglected. However, in general, it is well accepted that devices in the MCS should form a relatively good community with small percentage of malicious devices~(\emph{e.g.} 4\% in~\cite{Shen2015Enhancing} and 10\% in~\cite{Li2012Scalable}). For the scenario with the attacker percentage up to 10\%, TDP manages to effectively eliminate the impact of TO$_3$ attacks.

\subsubsection{Impact of Attackers with Different Properties} \label{subsubsec623}

\begin{figure*}[!t]
\setlength{\belowcaptionskip}{-1em}
\centering
\subfloat[Trustvalue (TopTV)]{\includegraphics[width=1.7in]{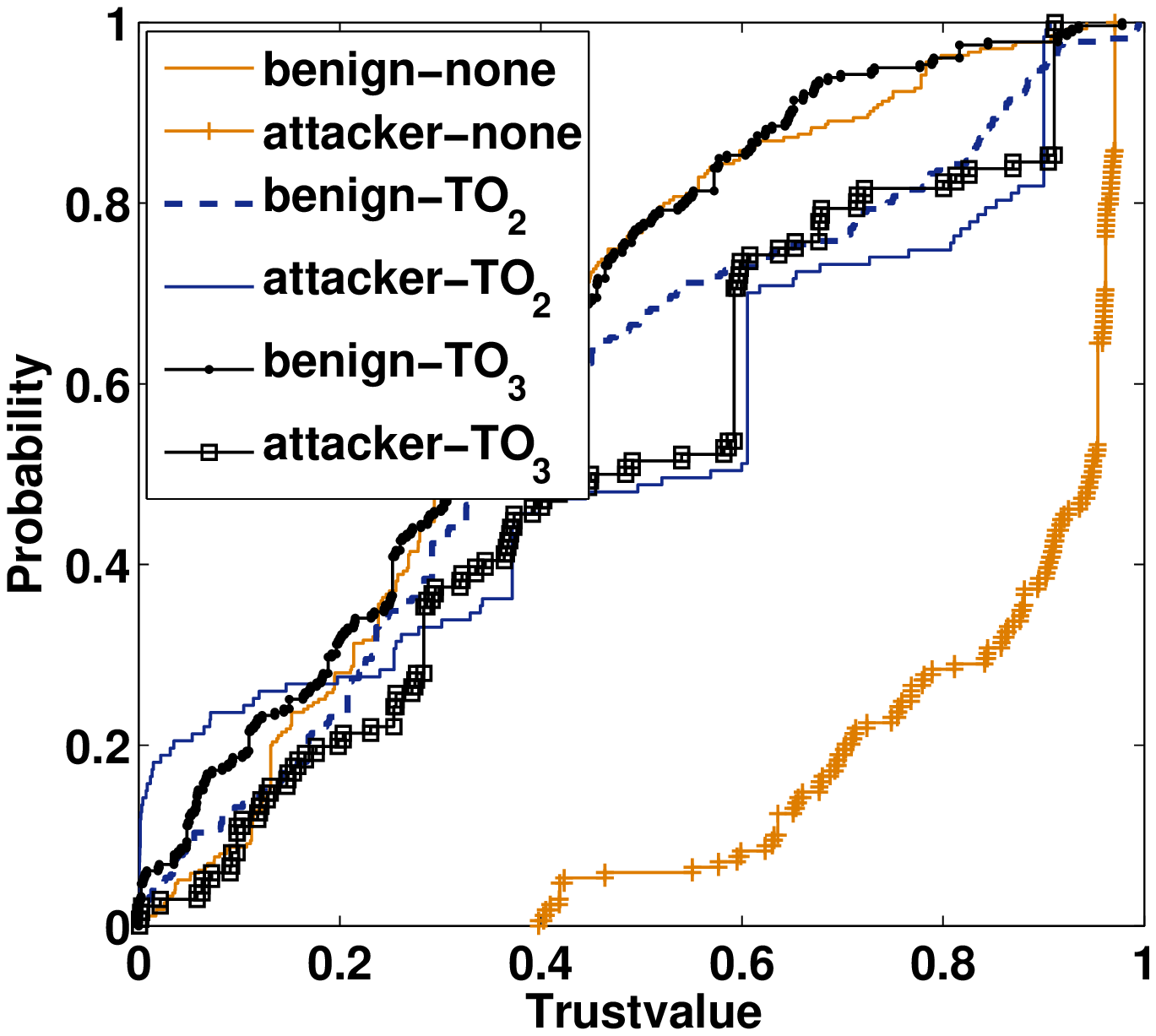}%
\label{fig13a}}
\hfil
\subfloat[False Positive Count (TopTV)]{\includegraphics[width=1.7in]{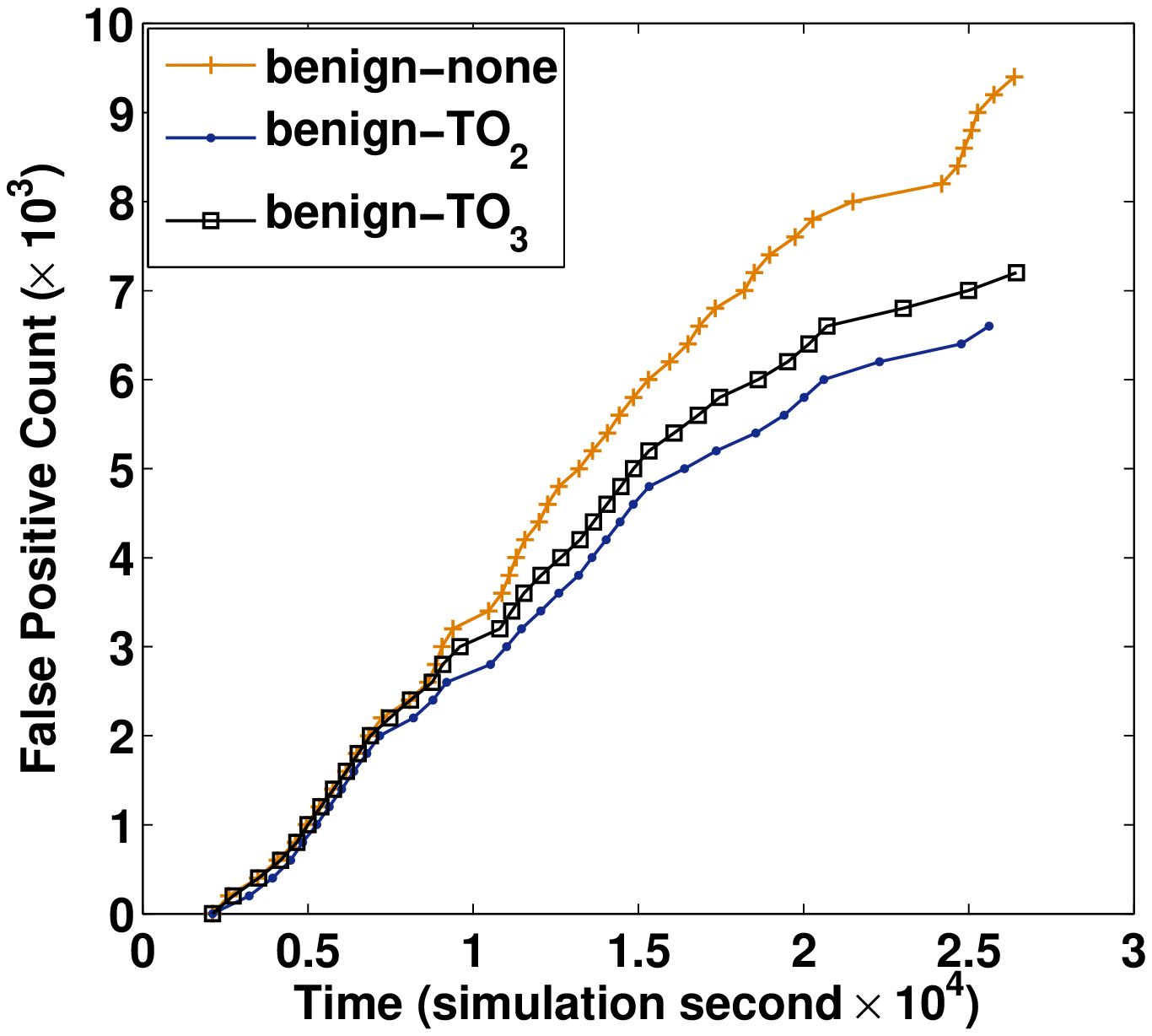}%
\label{fig13b}}
\hfil
\subfloat[Trustvalue (TopTC)]{\includegraphics[width=1.7in]{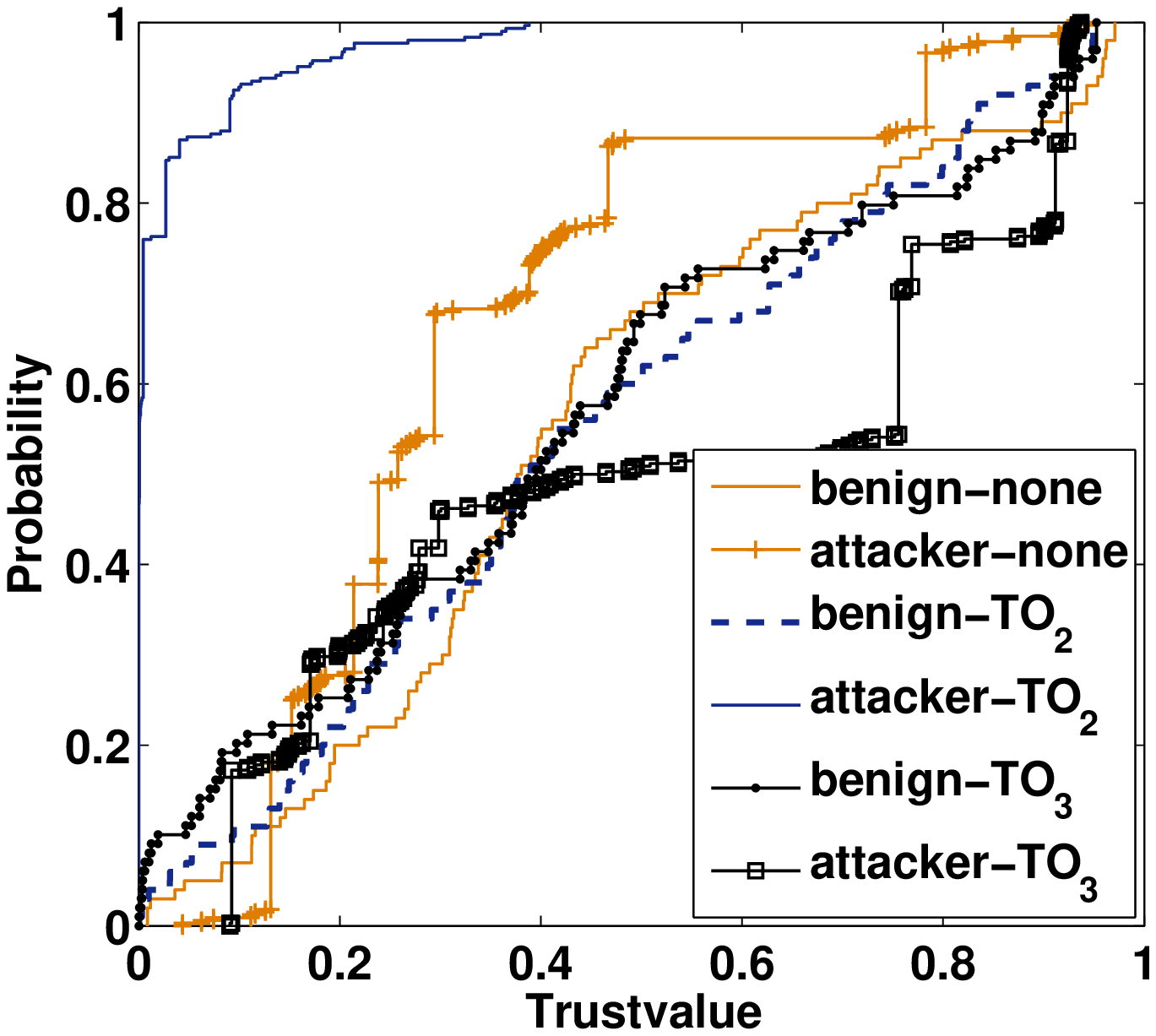}%
\label{fig13c}}
\hfil
\subfloat[False Positive Count (TopTC)]{\includegraphics[width=1.7in]{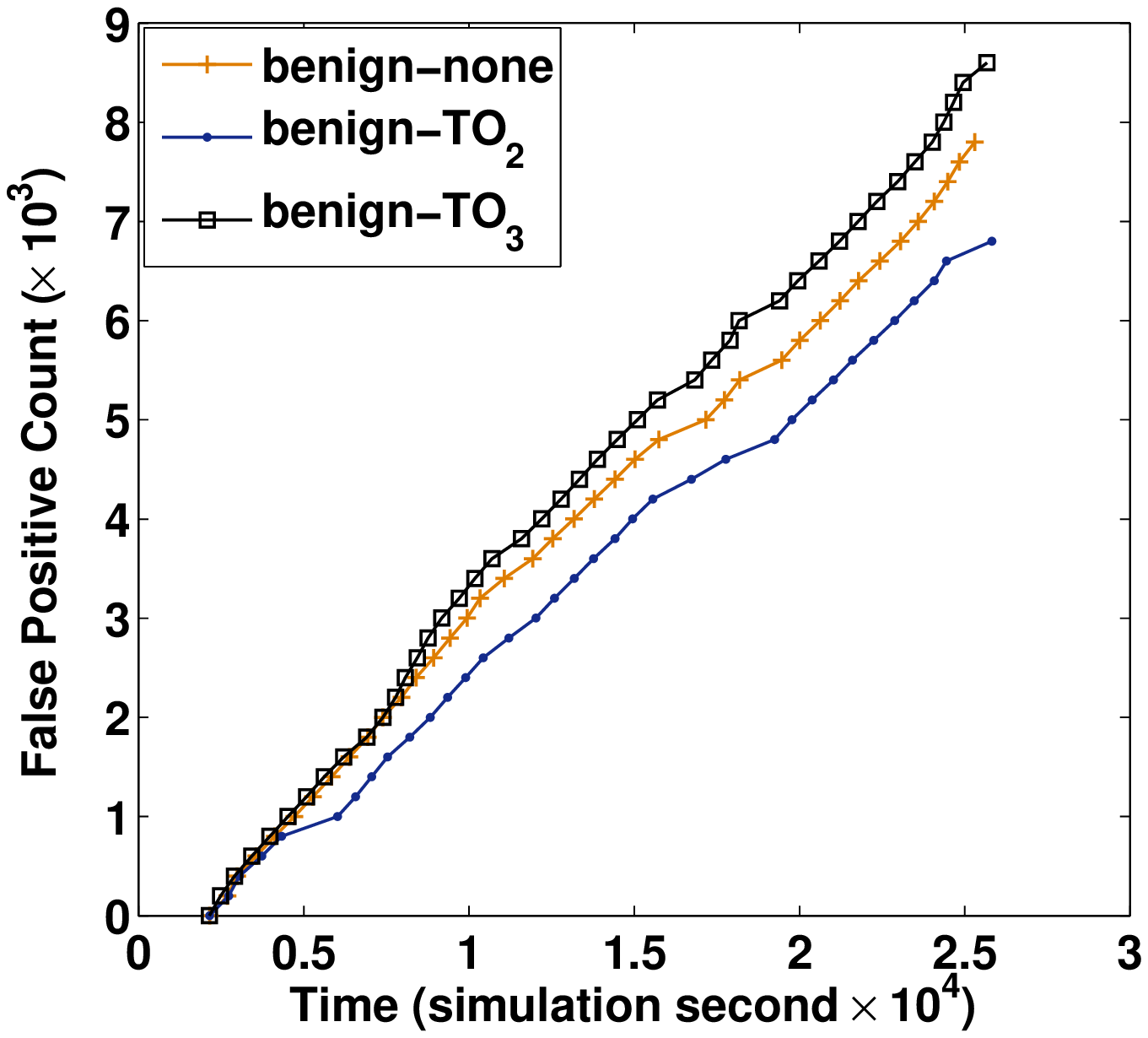}%
\label{fig13d}}
\caption{Impact of Attackers with Different Properties (Infocom).}
\label{fig13}
\end{figure*}

\begin{figure*}[!t]
\setlength{\belowcaptionskip}{-1em}
\centering
\subfloat[Trustvalue (TopTV)]{\includegraphics[width=1.7in]{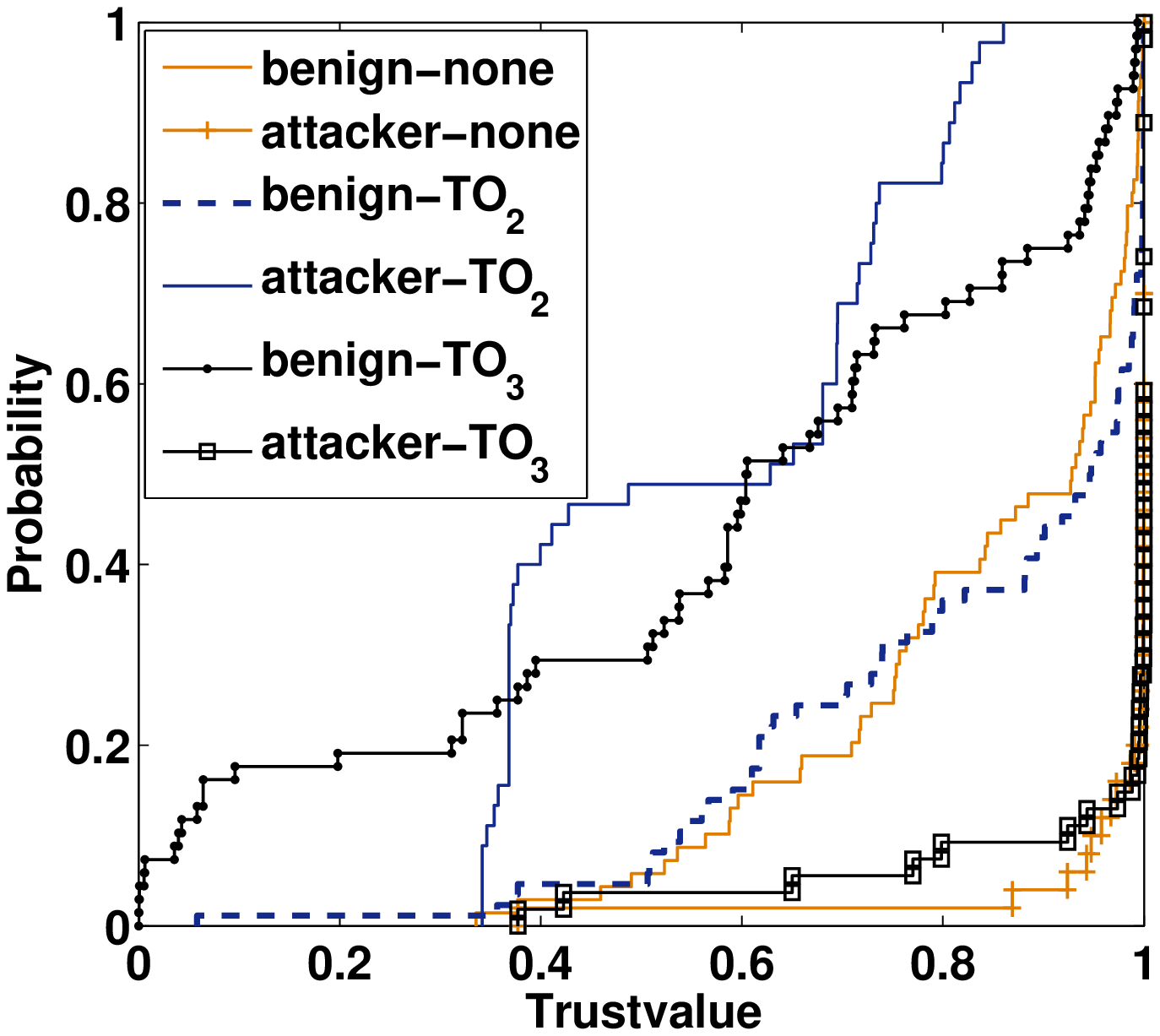}%
\label{fig14a}}
\hfil
\subfloat[False Positive Count (TopTV)]{\includegraphics[width=1.7in]{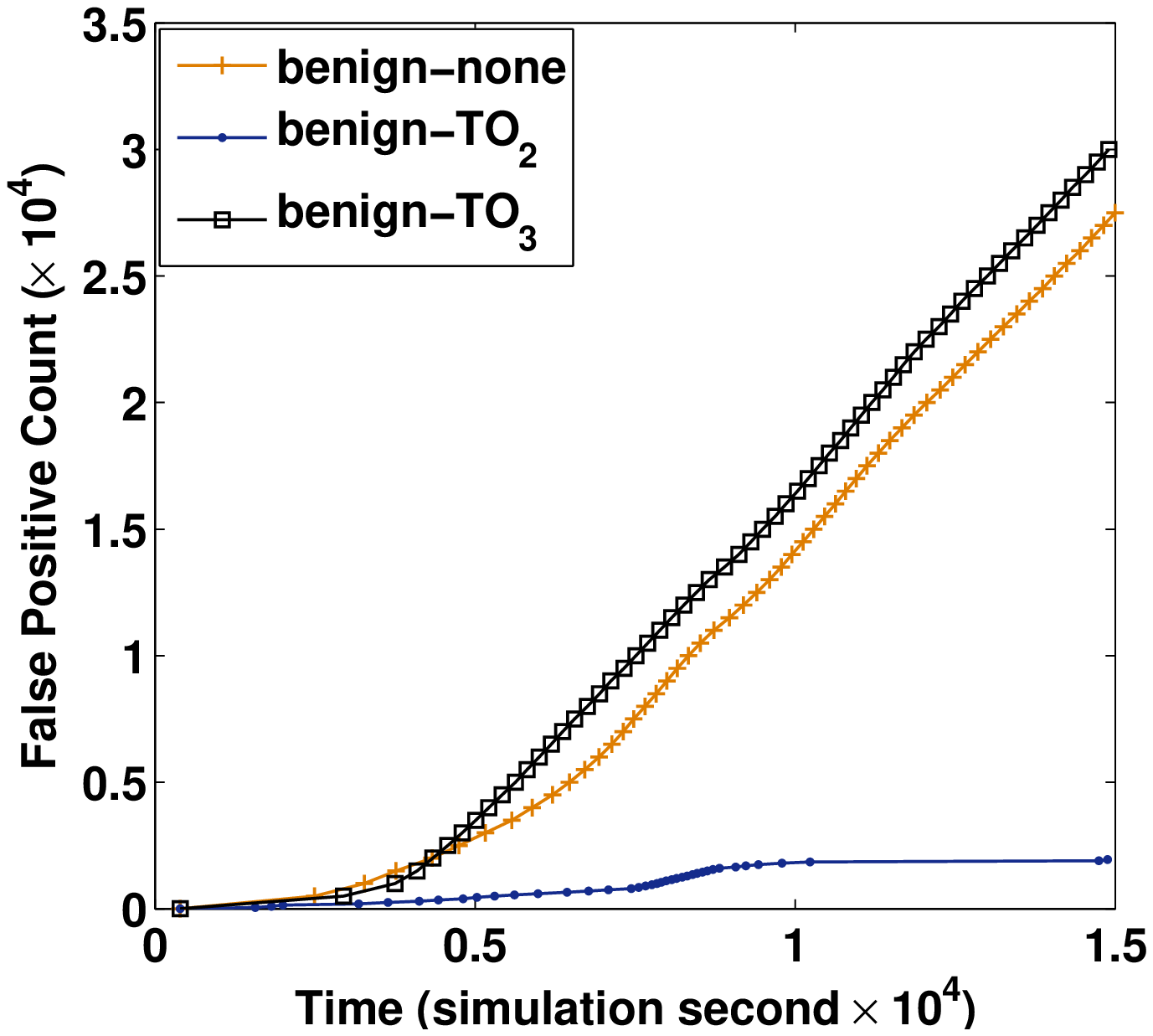}%
\label{fig14b}}
\hfil
\subfloat[Trustvalue (TopTC)]{\includegraphics[width=1.7in]{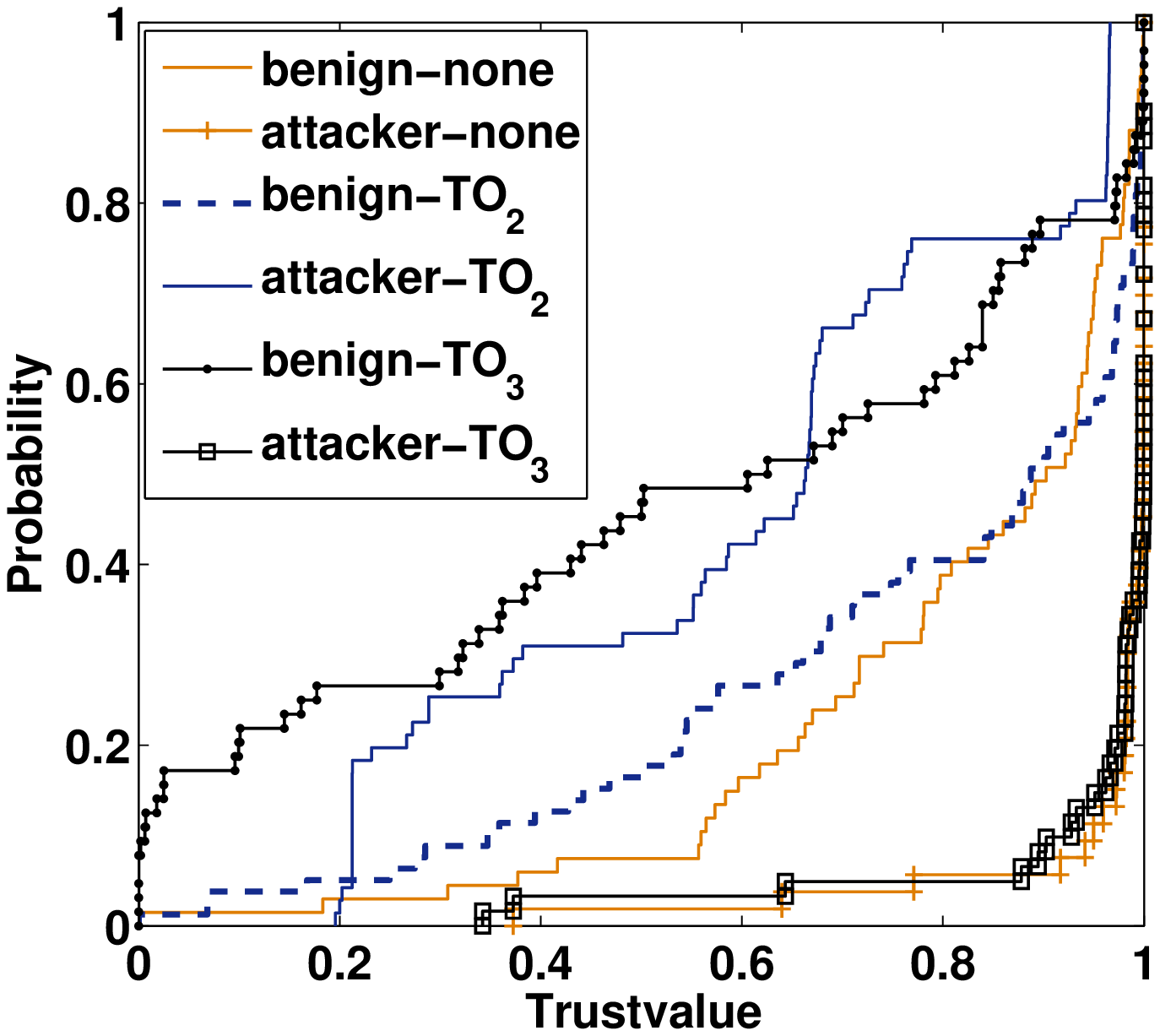}%
\label{fig14c}}
\hfil
\subfloat[False Positive Count (TopTC)]{\includegraphics[width=1.7in]{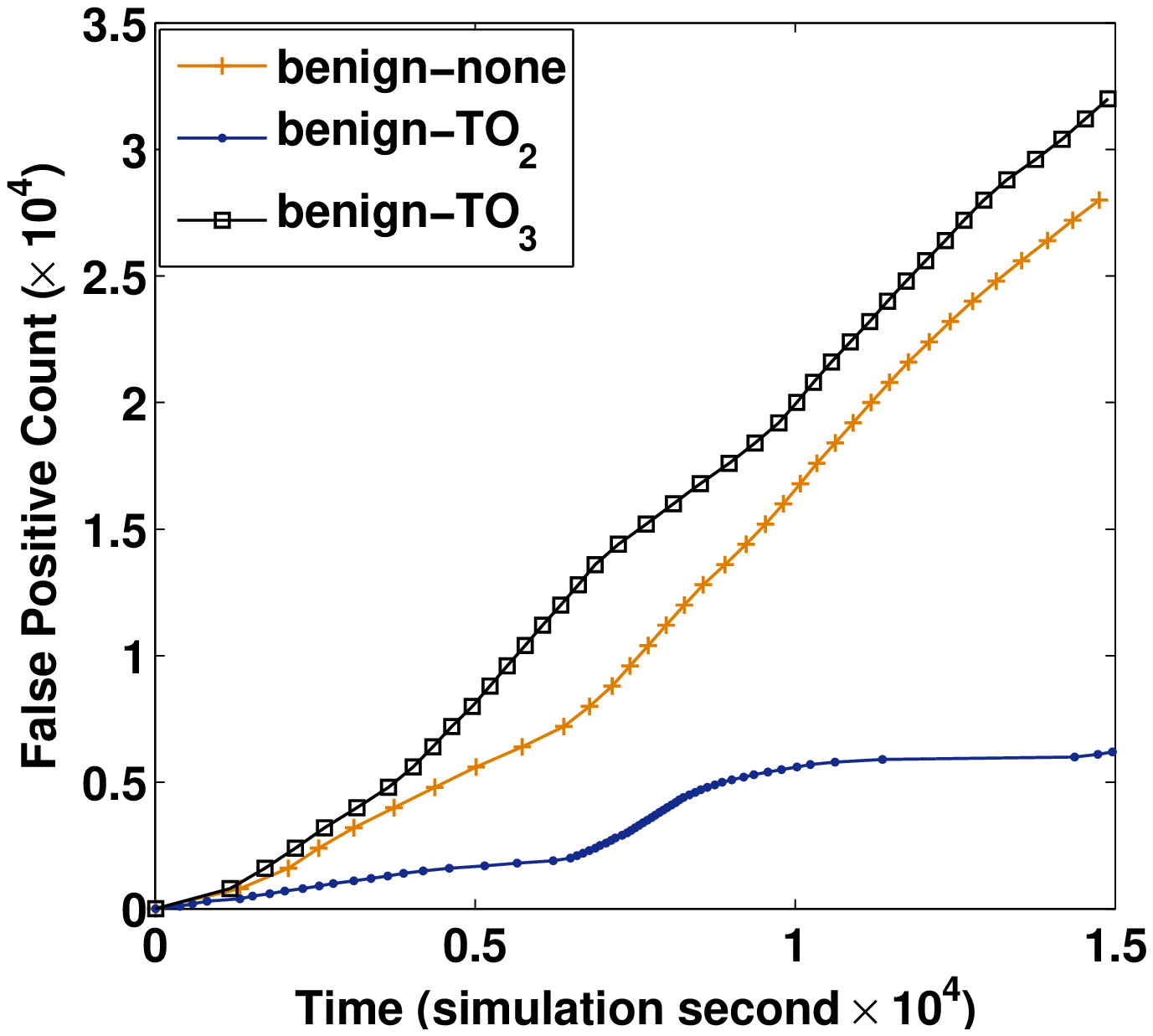}%
\label{fig14d}}
\caption{Impact of Attackers with Different Properties (Sigcomm).}
\label{fig14}
\end{figure*}

In this set of simulations, considering that TO attacks from more trustworthy or popular devices may cause deeper impacts, we specifically selected a fixed percentage of devices with different properties as continuous TO attackers, \emph{i.e.} top 7\% devices in average trustvalue (TopTV) or top 7\% devices in transaction count (TopTC) in the original scenario. The results of the Infocom and Sigcomm simulations are shown in Fig.\ref{fig13} and Fig.\ref{fig14} respectively.

For both sets of simulations, the impact of TO$_2$ attacks is well eliminated in all two scenarios: there is no obvious impact on the average trustvalue of benign devices (\emph{e.g.} benign-TO$_2$ ($0.41/9.8\%^\uparrow$) vs benign-none ($0.37$) in Fig.\ref{fig13} (a)), while the average trustvalue of attackers is severely downgraded (\emph{e.g.} attacker-TO$_2$ ($0.44/48.2\%^\downarrow$) vs attacker-none ($0.85$) in Fig.\ref{fig13} (a)), and the number of false positive transactions also decreases correspondingly (\emph{e.g.} benign-TO$_2$ ($6737/29.7\%^\downarrow$) vs benign-none ($9579$) in Fig.\ref{fig13} (b)).

For the Infocom simulations, the impact of TO$_3$ attacks from TopTV devices is well eliminated according to Fig.\ref{fig13} (a) and (b). In the TopTC scenario, attackers have a boost in terms of the average trustvalue (\emph{i.e.} attacker-TO$_3$ ($0.53/65.6\%^\uparrow$) vs attacker-none ($0.32$) in Fig.\ref{fig13} (c)) and the number of attracted transactions (\emph{i.e.} attacker-TO$_3$ ($8758/9.9\%^\uparrow$) vs attacker-none ($7969$) in Fig.\ref{fig13} (d)). However, the impact on the average trustvalue of benign devices is negligible (\emph{i.e.} benign-TO$_3$ ($0.40/7.0\%^\downarrow$) vs benign-none ($0.43$) in Fig.\ref{fig13} (c)). For the Sigcomm simulations with TO$_3$ attacks, the average trustvalue of attackers remains similar, and the number of attracted transactions has a slightly boost in all two scenarios. In fact, the impact of TO$_3$ attacks on benign devices is difficult to be restricted only when the collusive attackers possess predominant popularities (\emph{i.e.} $\frac{6\%}{1.17\%}=512.8\%$ of the average popularity of benign devices for the Infocom TopTC scenario, $\frac{4.42\%}{1.05\%}=421\%$ and $\frac{5.66\%}{1.02\%}=555\%$ for the Sigcomm TopTV and TopTC scenarios respectively). However, considering it is irrational for the most popular devices to collude in reality because of the low cost-efficiency, TDP manages to effectively eliminate the impact of TO$_3$ attacks from devices with rationally advanced trustworthiness or popularity.

\subsubsection{Impact of Attacking Intensity} \label{subsubsec624}

\begin{figure*}[!t]
\setlength{\belowcaptionskip}{-1em}
\centering
\subfloat[Trustvalue (TO$_2$)]{\includegraphics[width=1.7in]{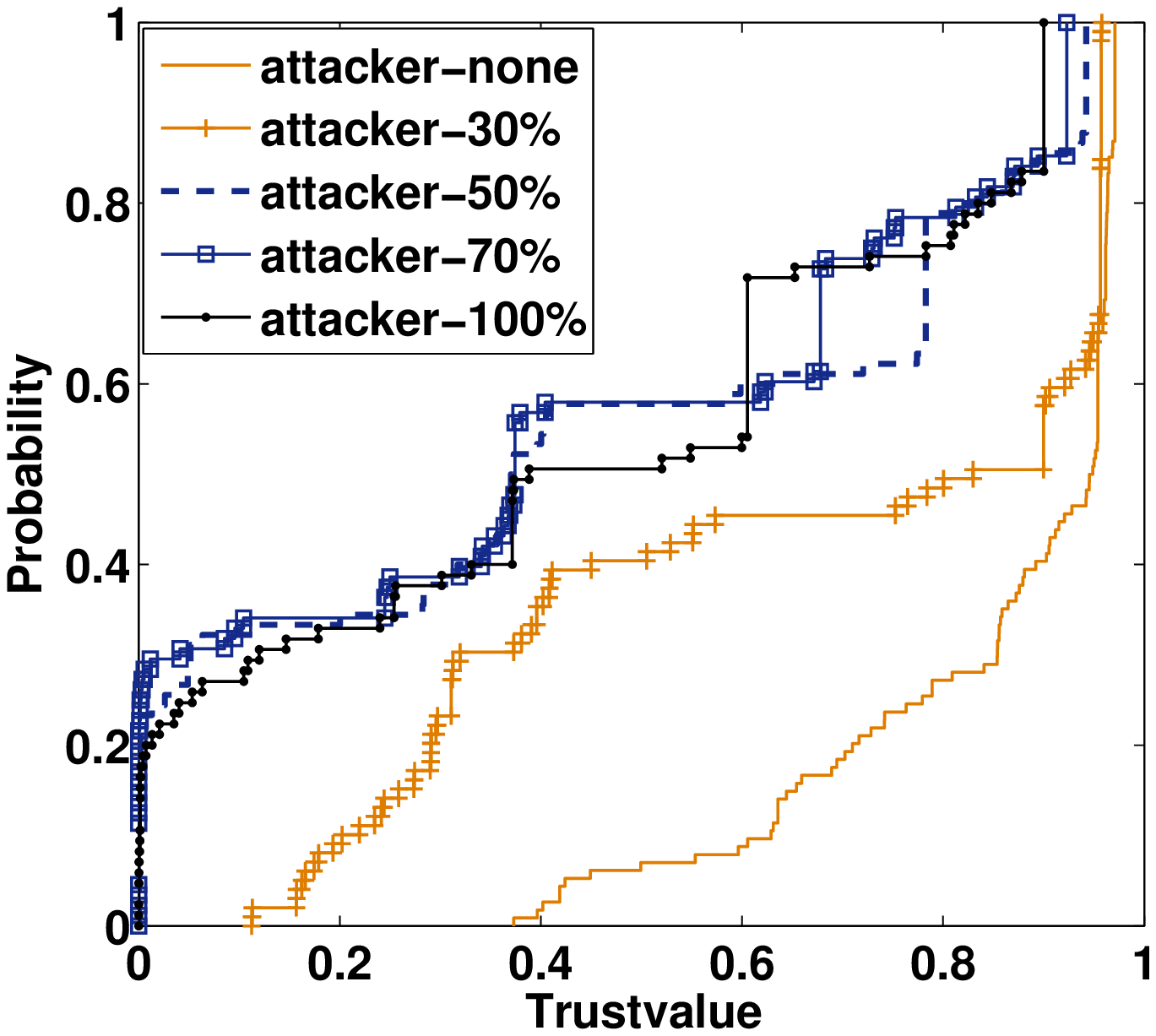}%
\label{fig15a}}
\hfil
\subfloat[False Positive Count (TO$_2$)]{\includegraphics[width=1.7in]{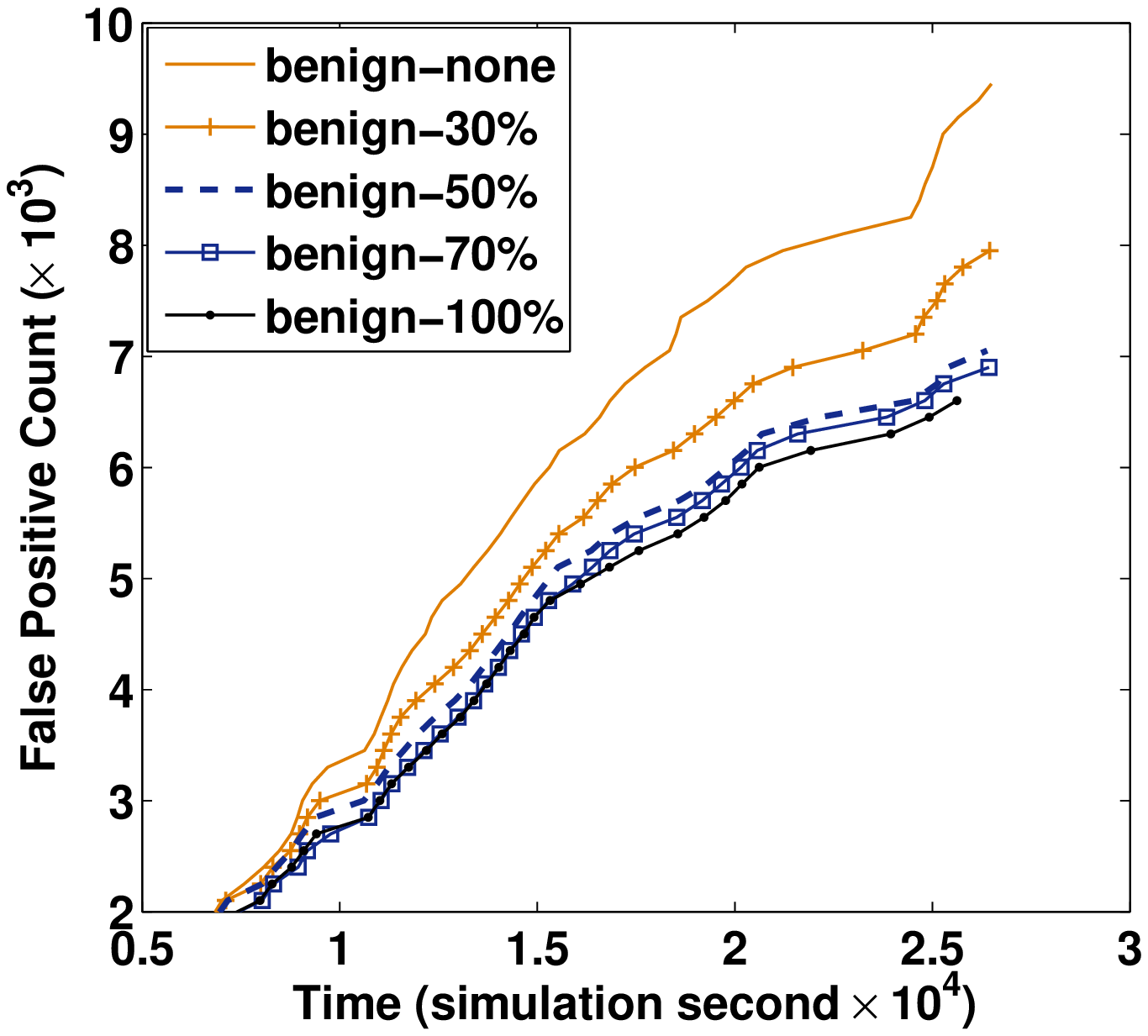}%
\label{fig15b}}
\hfil
\subfloat[Trustvalue (TO$_3$)]{\includegraphics[width=1.7in]{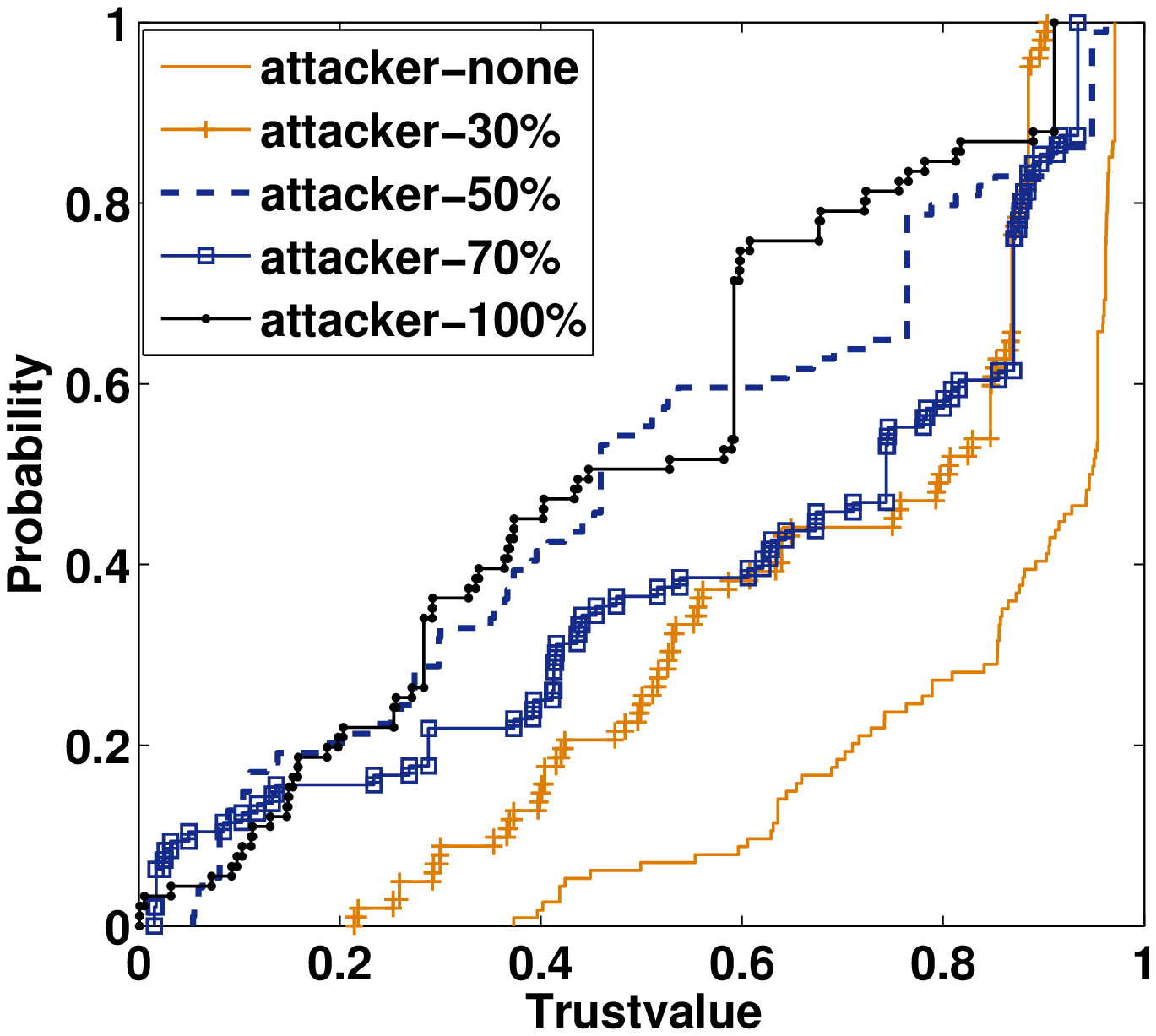}%
\label{fig15c}}
\hfil
\subfloat[False Positive Count (TO$_3$)]{\includegraphics[width=1.7in]{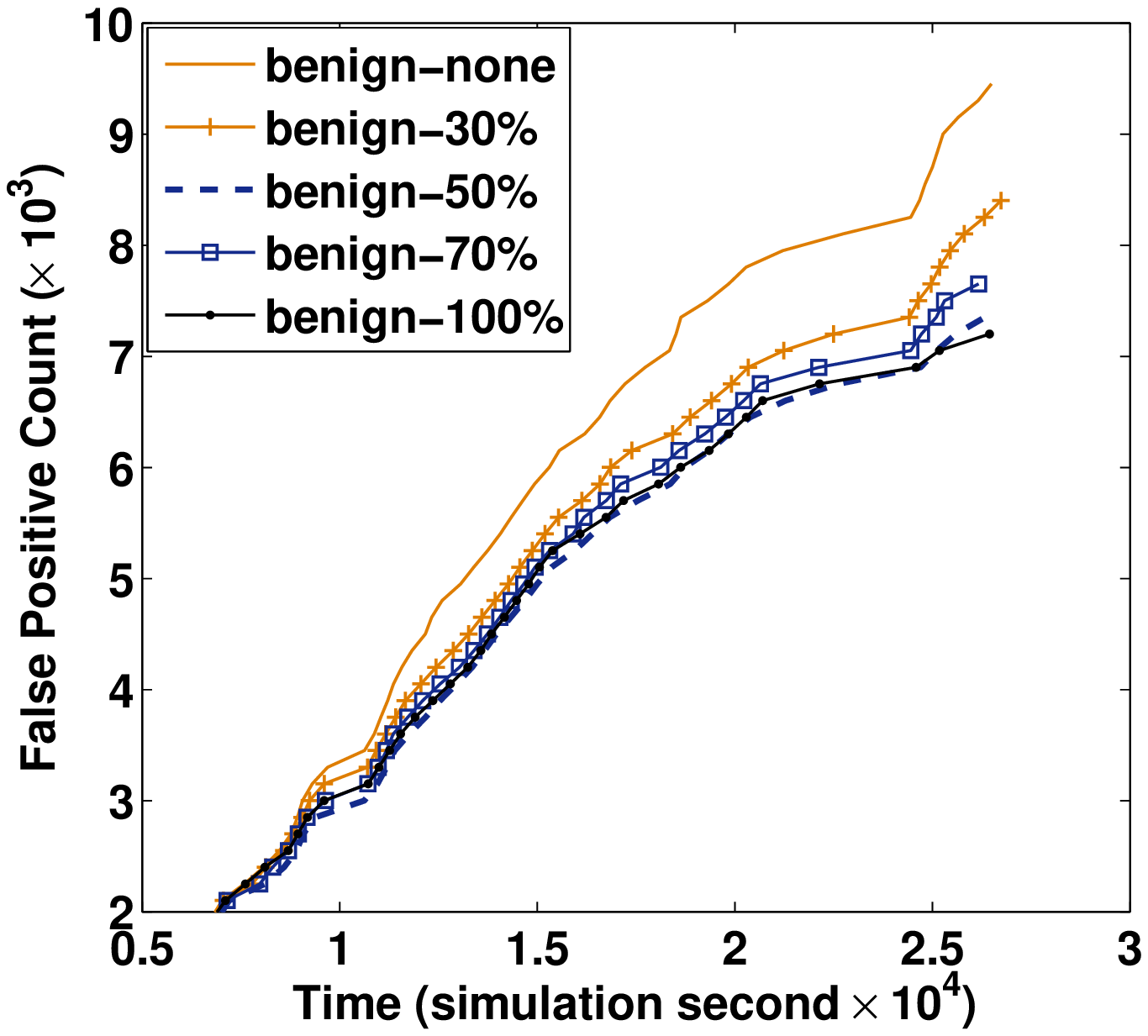}%
\label{fig15d}}
\caption{Impact of Attacks with Different Intensities (Infocom).}
\label{fig15}
\end{figure*}

\begin{figure*}[!t]
\setlength{\belowcaptionskip}{-1em}
\centering
\subfloat[Trustvalue (TO$_2$)]{\includegraphics[width=1.7in]{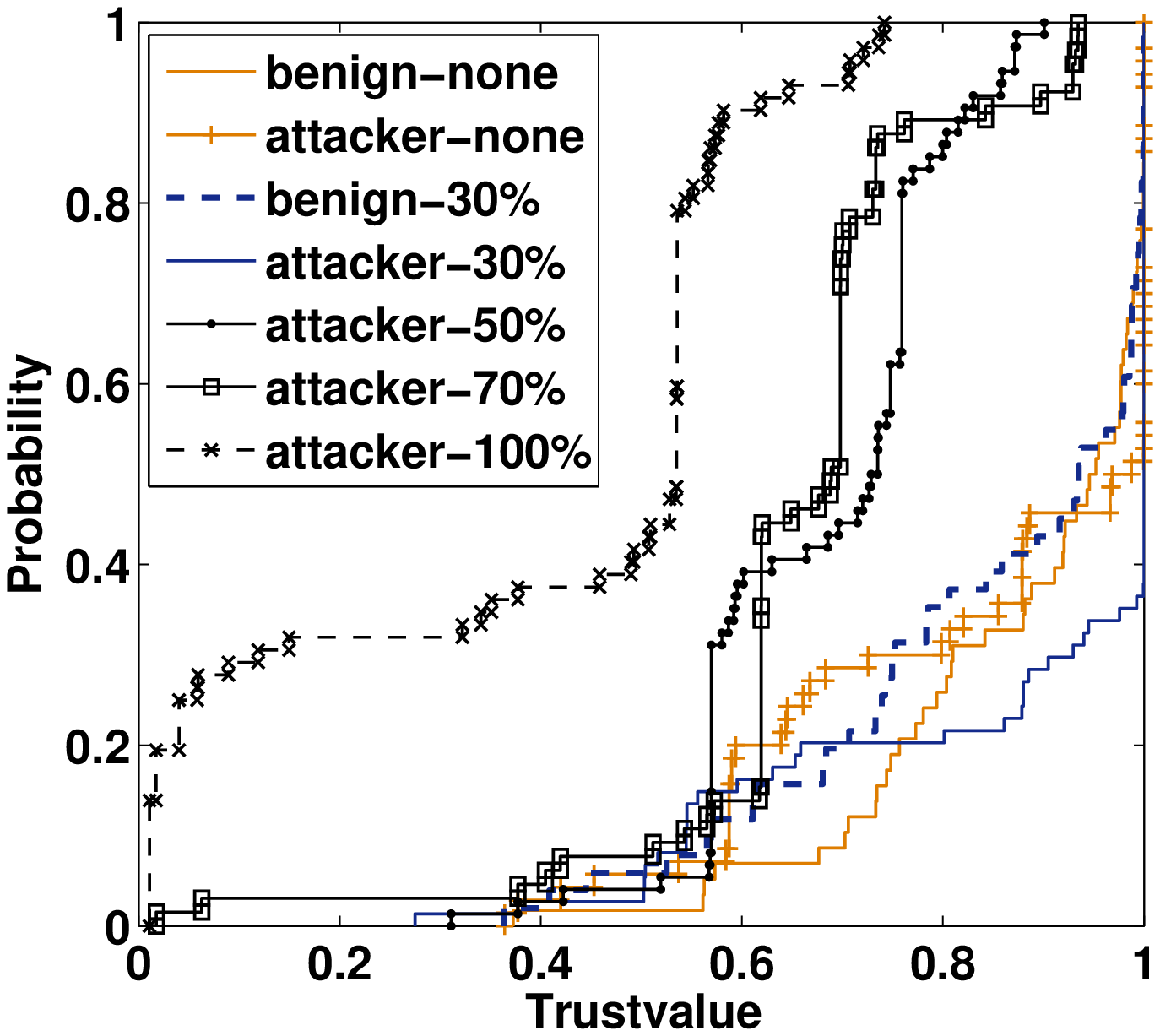}%
\label{fig16a}}
\hfil
\subfloat[False Positive Count (TO$_2$)]{\includegraphics[width=1.7in]{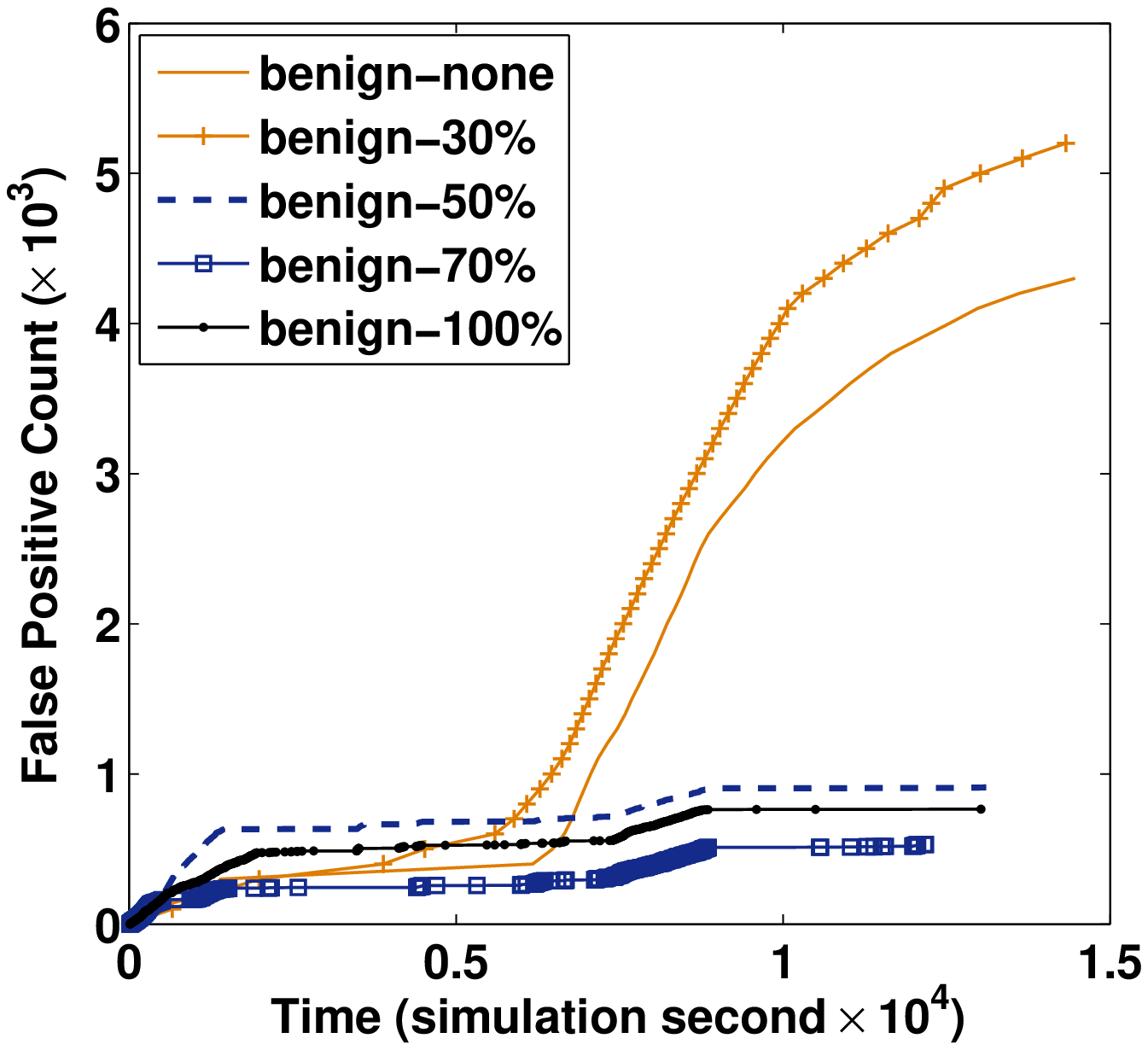}%
\label{fig16b}}
\hfil
\subfloat[Trustvalue (TO$_3$)]{\includegraphics[width=1.7in]{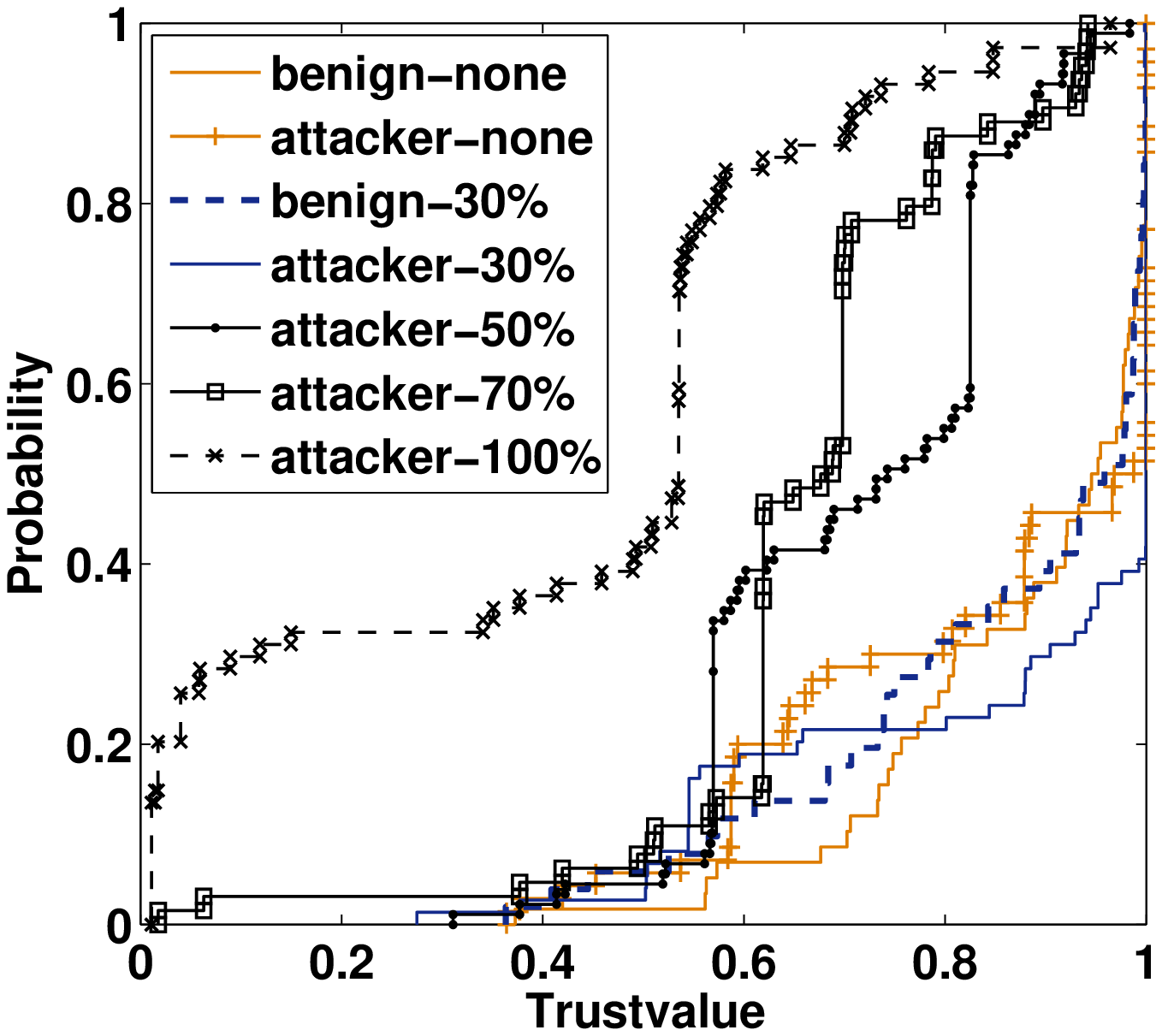}%
\label{fig16c}}
\hfil
\subfloat[False Positive Count (TO$_3$)]{\includegraphics[width=1.7in]{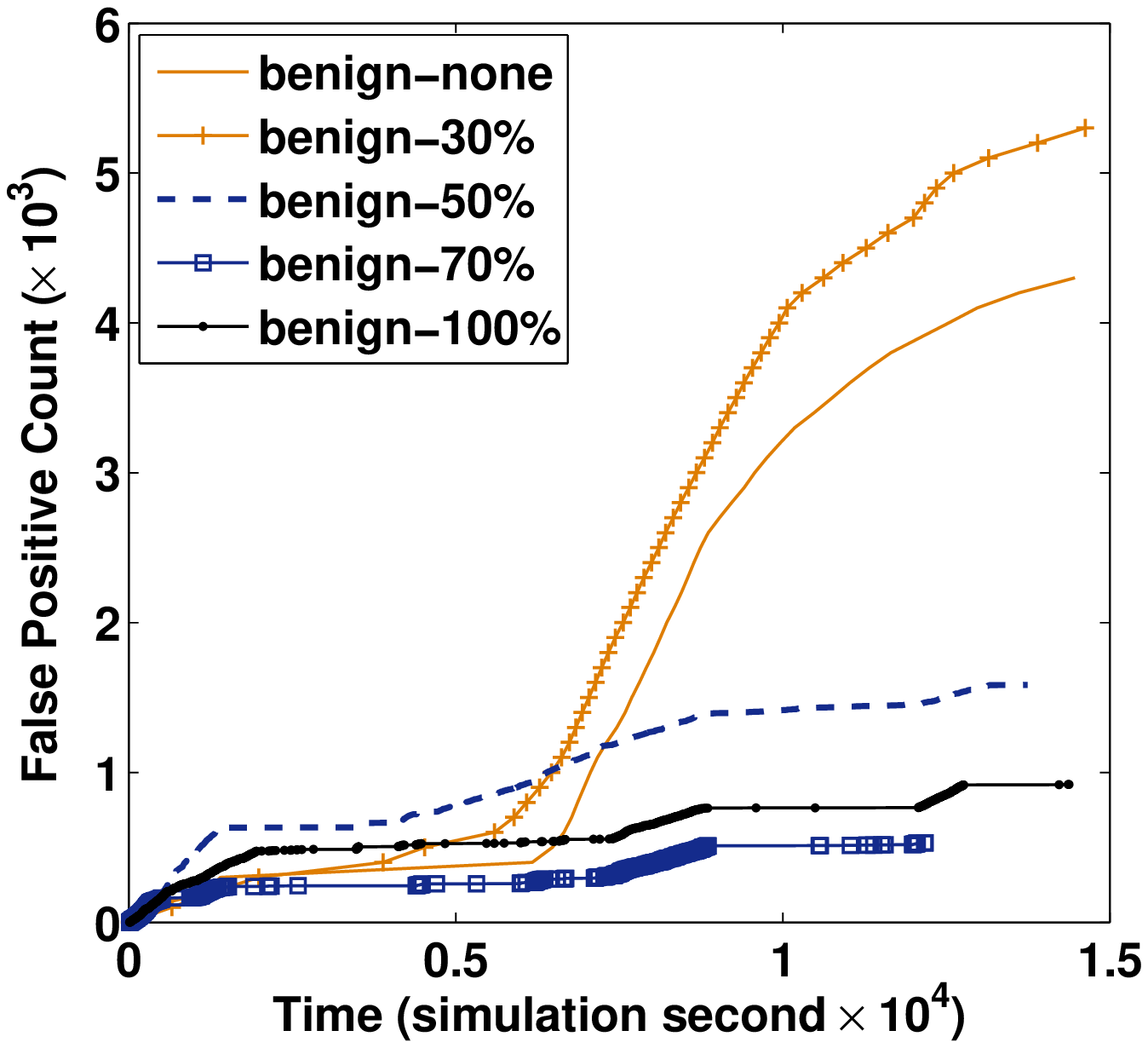}%
\label{fig16d}}
\caption{Impact of Attacks with Different Intensities (Sigcomm).}
\label{fig16}
\end{figure*}

In this set of simulations, to study the impact of TO attacks with different intensities, we randomly selected a fixed percentage (7\%) of devices as attackers that launch TO attacks in each D2D transaction with different probabilities (\emph{i.e.} 0, 30\%, 50\%, 70\%, and 100\%). The results of the Infocom and Sigcomm simulations are shown in Fig.\ref{fig15} and Fig.\ref{fig16} respectively.

For the Infocom simulations, the impacts of both TO$_2$ and TO$_3$ attacks are well eliminated: compared with the original scenario, the average trustvalue of attackers is in inverse proportion to the attack intensity (\emph{e.g.} attacker-30\% ($0.66/22.4\%^\downarrow$), attacker-50\% ($0.49/42.4\%^\downarrow$) vs attacker-none ($0.85$) in Fig.\ref{fig15} (a)). The number of false positive transactions also decreases in direct proportion to the attack intensity (\emph{e.g.} benign-30\% ($8047/16.0\%^\downarrow$), benign-50\% ($7156/25.3\%^\downarrow$) vs attacker-none ($9579$) in Fig.\ref{fig15} (b)).

For the Sigcomm simulations, the results are basically the same except for the 30\% scenarios. For both 30\% TO$_2$ and TO$_3$ attacks, attackers have a slightly boost in terms of the average trustvalue and the number of attracted transactions. In fact, since the device contact in the Sigcomm trace is relatively sparse compared to the Infocom trace (\emph{i.e.}  31 vs 110 in terms of the average transaction count of participating devices), system's reaction to TO attacks is correspondingly slower. However, the impacts of the 30\% TO attacks on the trustvalue of benign devices are negligible (\emph{e.g.} benign-30\% ($0.85/1.2\%^\uparrow$) vs benign-none ($0.86$) in Fig.\ref{fig16} (a)). Therefore, TDP manages to effectively eliminate the impact of TO attacks with different intensities.

\section{Conclusion} \label{sec7}
In this paper, we develop TDP, a rapid, user-transparent, and trustworthy device pairing scheme for Mobile Crowdsouring Systems (MCS) with opportunistic Device-to-Device (D2D) communications. We first develop a formal method to estimate the trustworthiness of participating devices at real-time. Then, we propose a CL-PKC based credential framework that allows each encountered device to spontaneously establish a secure D2D connection with the most trustworthy peer nearby. We theoretically prove that TDP manages to achieve a comparable security intensity with the off-the-shelf protocols and the immunity to the TO attacks that are not considered by existing approaches. Through real-world experiments using Android devices, we show that TDP outperforms existing approaches in terms of pairing speed and stability. In addition,  extensive trace-driven simulation results verify that TDP manages to effectively prevent the traffic manipulation of TO attackers in large-scale MCSs. Interesting future work is to apply TDP to specific MCS applications such as D2D-enabled mobile social networking and mobile cloud computing.

\ifCLASSOPTIONcaptionsoff
  \newpage
\fi

\clearpage
\renewcommand\thefigure{A\arabic{figure}}

\appendices
\section{} \label{appa}

\subsection{Determination of $c_g$} \label{appa1}

Taking the gompertz function $g(x)=e^{-e^{-c_gx}}$ as an instance, assuming that the average contact number of registered devices within a trustvalue management cycle is $50$ and there is an equal probability for a device to get a positive or a negative rating, if the average behavior estimation for a D2D transaction is $0.125$ ($q,c,\lambda=0.5$), the value of $c_g$ can be determined by solving:
\begin{equation}\label{eqa1}
\dddot{g}(50\times 0.5\times 0.125)=0, \tag{A1}
\end{equation} where $\dddot{g}(x)$ is the third derivative of $g(x)$ on $x$, and $c_g=0.308$ in this case.

\begin{figure}[H]
\setlength{\belowcaptionskip}{-1em}
\centering
\subfloat[Trustvalue Accumulation]{\includegraphics[width=1.5in]{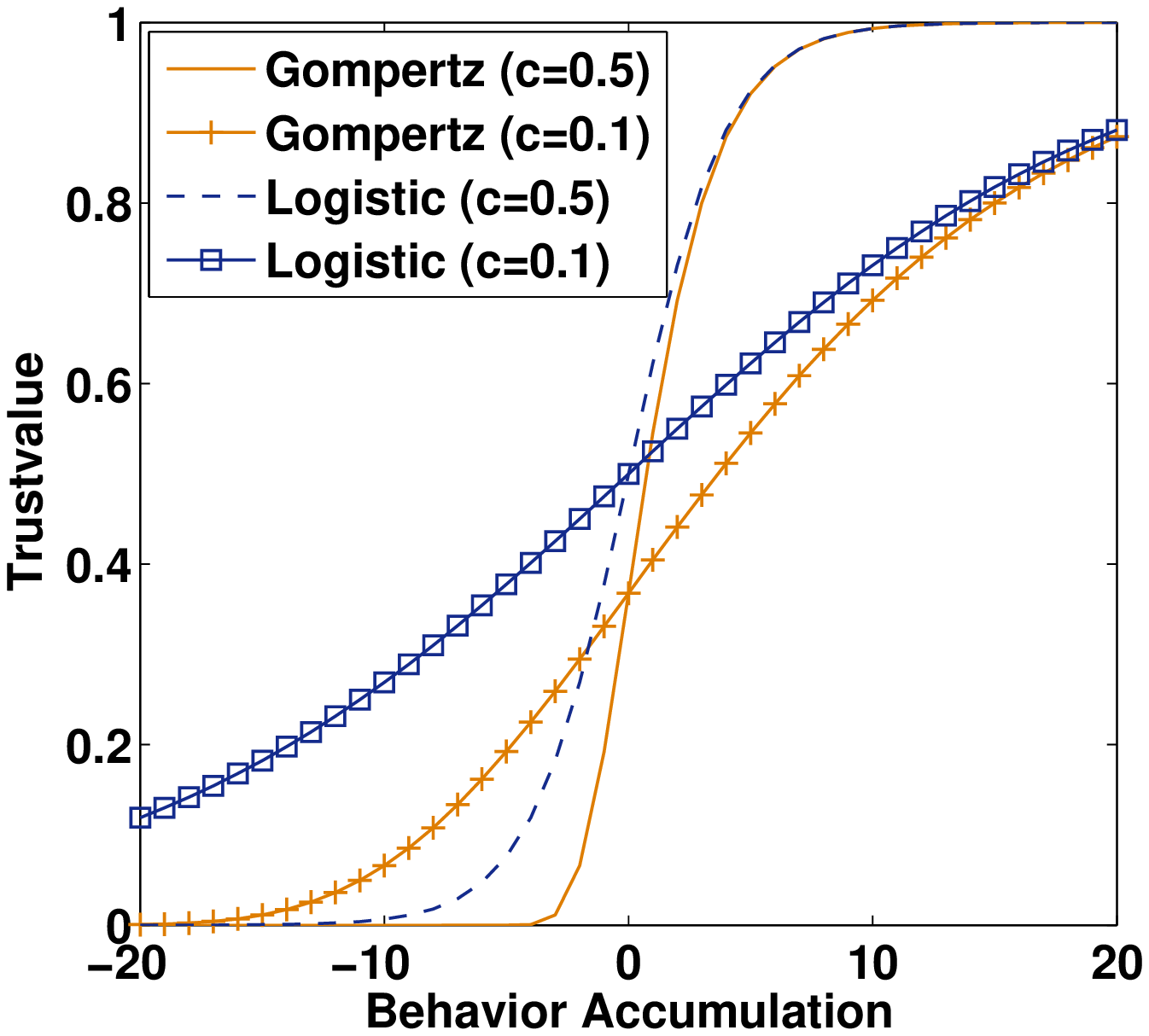}%
\label{figa11a}}
\hfil
\subfloat[Derivatives of $g(x)$]{\includegraphics[width=1.5in]{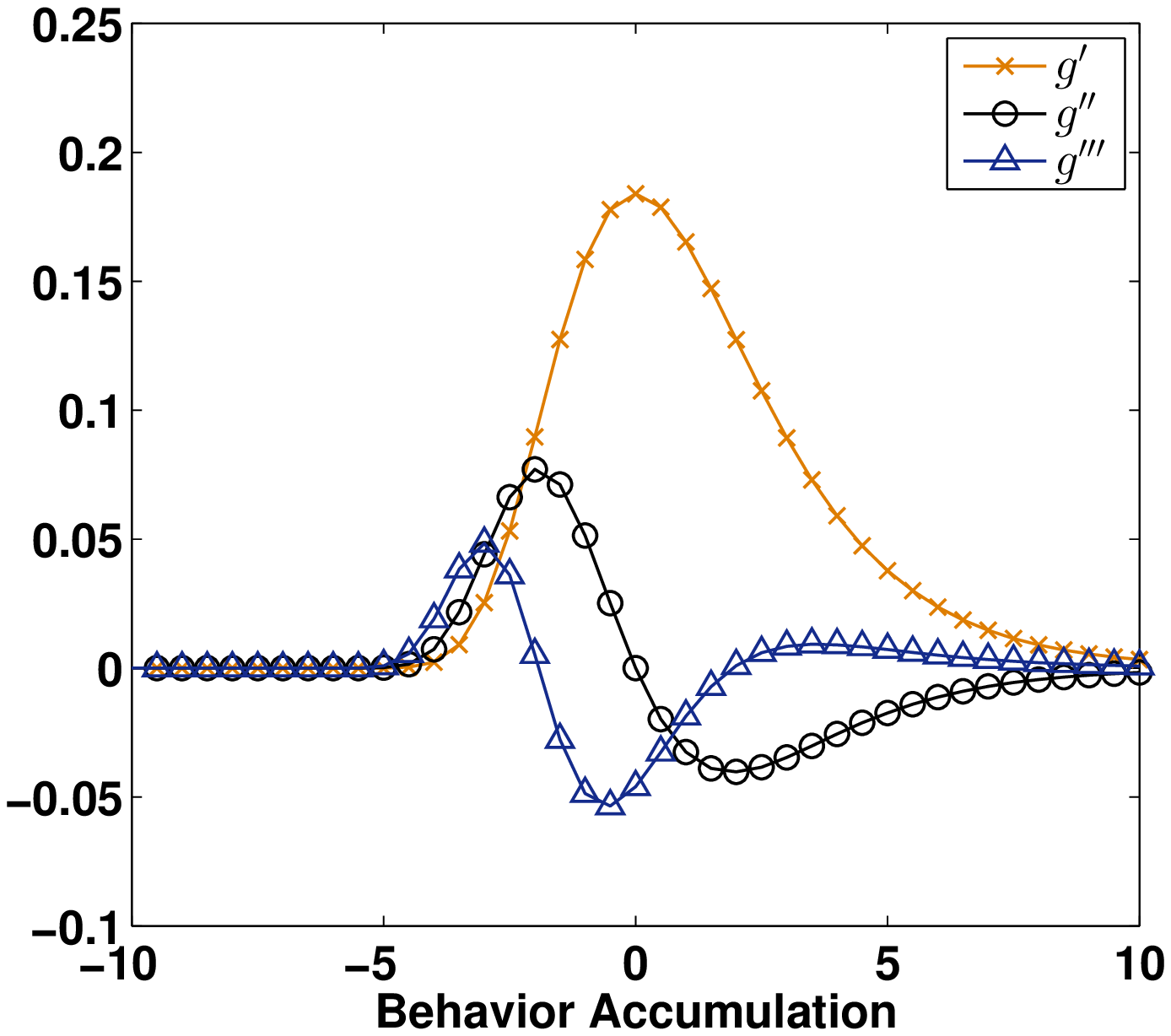}%
\label{figa11b}}
\caption{Trustvalue Accumulation Sensitivity Determination.}
\label{figa11}
\end{figure}

The purpose for this determination is to alleviate the trustvalue enhancement for devices whose contact frequencies are beyond the average. According to Fig.\ref{figa11} (b), for any registered device $a$, the climbing rate $g'$ reaches its maximum value when $x=0$. This indicates that the trustvalue adjustment for a freshly joint device is relatively sensitive. Besides, $g''$ presents the sensitivity of the variation of $g'$, which firstly has a raise and then falls to a minimum value. In fact, if device $a$ behaves well continuously, it's trustvalue will accordingly raise with three phases: a rapid enhancement, an intense damping on the climbing rate, and a relatively steady. Therefore, we set the steady point of the device trustvalue climbing at the time that the device accomplishes the average number of transactions (\emph{i.e.} Eq.(A1)).

\vspace{-1em}
\subsection{Determination of $c_w$} \label{appa2}

Taking the quadratic function $w(\sigma)=-c_w(\sigma/\bar{\sigma})^{2}+ 1$ as an instance, we assume that the average contact number of registered devices within a trustvalue management cycle is $50$. The value of $c_w$ can be determined by solving:
\begin{equation}\label{eqa2}
\dot{f}(1)=-1, \mbox{where $f(x)=-c_wx^2+1$}, \tag{A2}
\end{equation} and $c_w=0.5$ in this case.

\begin{figure}[H]
\setlength{\belowcaptionskip}{-1em}
\centering
\subfloat[Credibility Damping]{\includegraphics[width=1.5in]{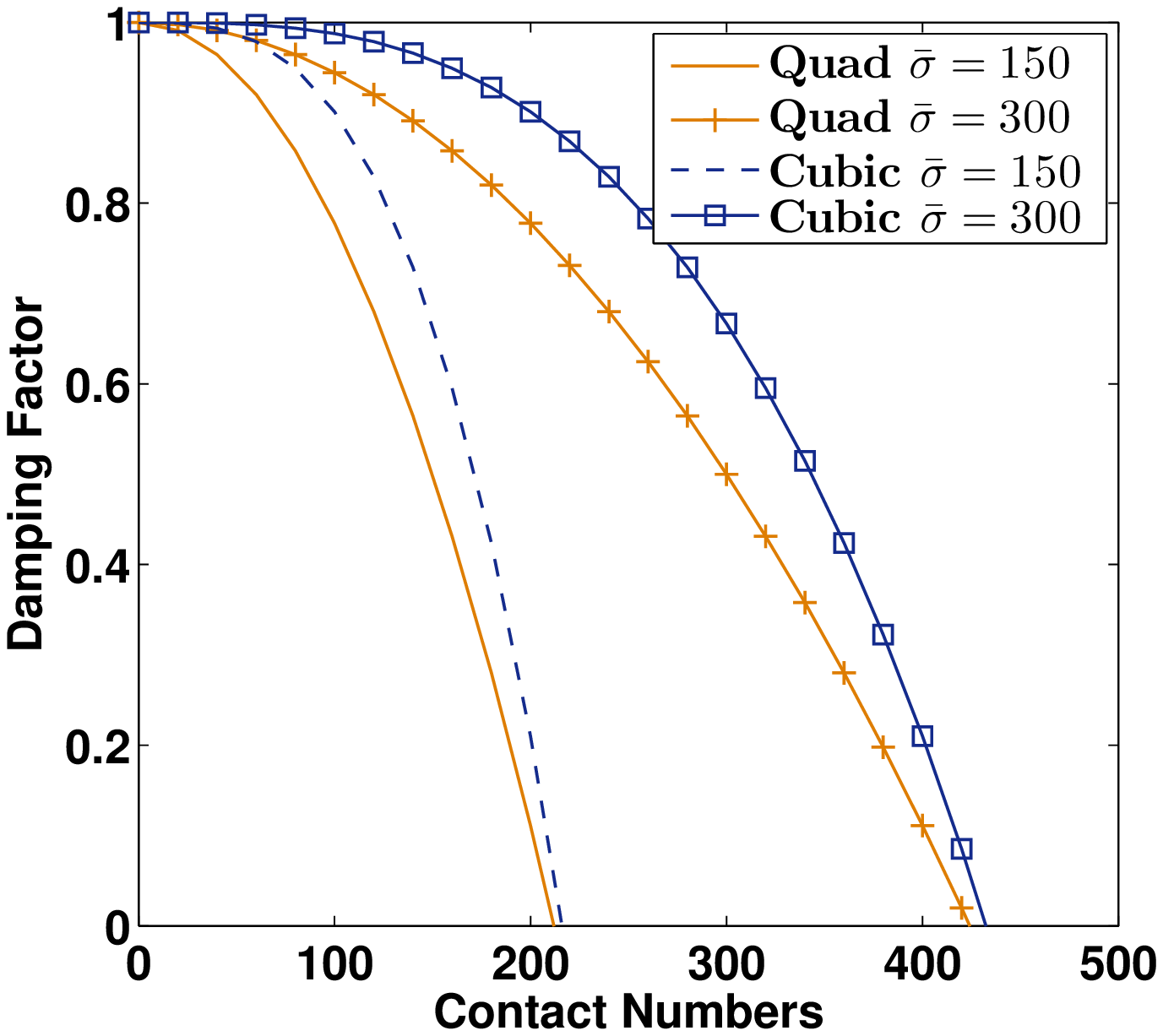}%
\label{figa21a}}
\hfil
\subfloat[Derivatives of $f(x)$]{\includegraphics[width=1.5in]{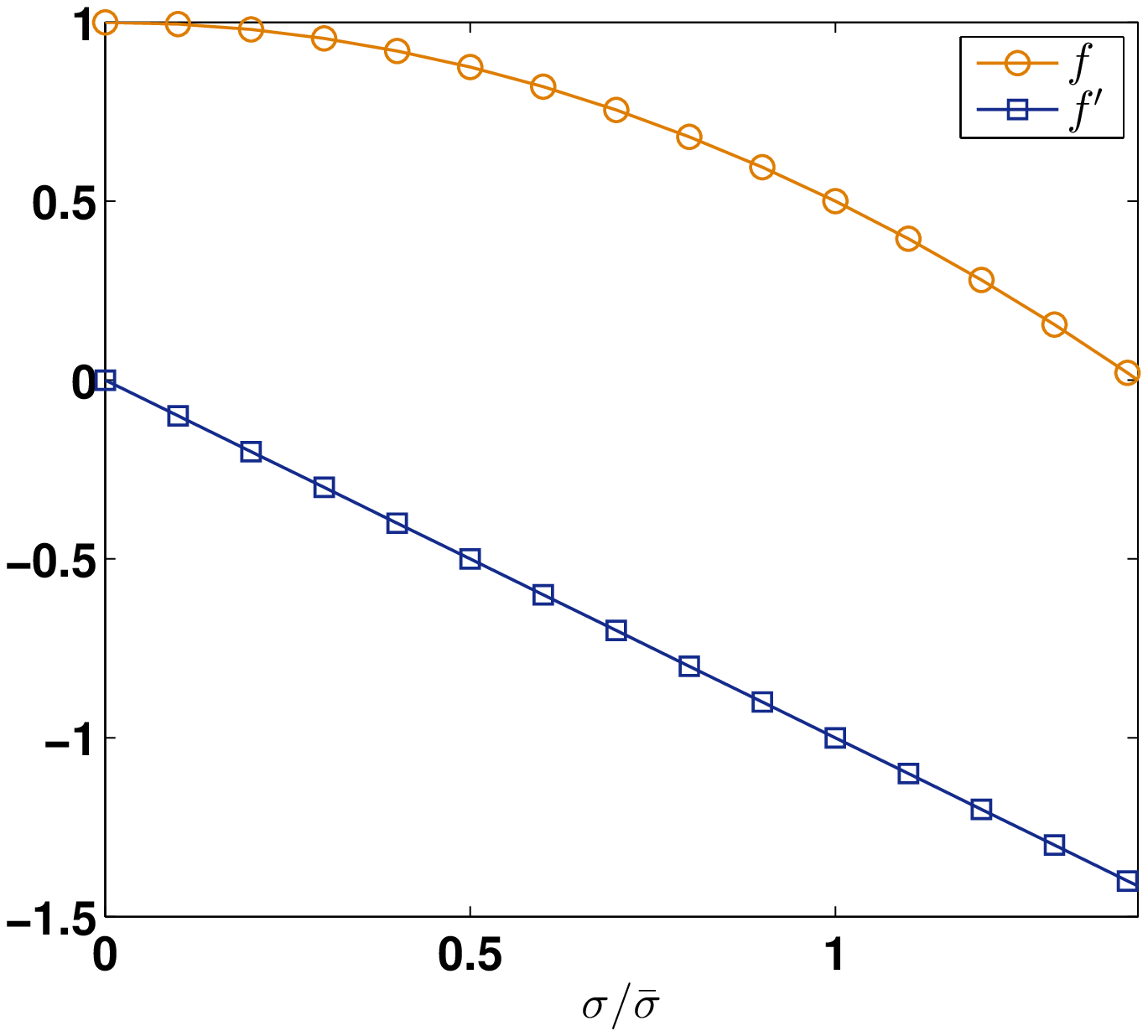}%
\label{figa21b}}
\caption{Credibility Damping Sensitivity Determination.}
\label{figa21}
\end{figure}

The purpose for this determination is to ensure a rapid credibility damping for devices with abnormal intimacy. According to Fig.\ref{figa21}, the damping factor between a fixed pair of devices, \emph{e.g.} devices $a$ and $b$, continuously drops in reverse proportion to their intimacy. Intuitively, if the contact number between $a$ and $b$ exceeds the average value, we'd like to treat them as suspicious and increase the intensity of their credibility damping rapidly. For $f(x)$, we set $x=\sigma/\bar{\sigma}$, and the rate of credibility damping is $f'=-x$. Therefore, with the determination of Eq.(A2), the credibility damping of a pair of devices whose intimacy exceeds the average value would be more intense than that of freshly encountered devices.

\section{}\label{appb}

\subsection{Proof of CO$_1$ Immunity.} \label{appb1}

\textbf{Proposition 1.}
\textit{With TDP, adversary $m$ cannot decrypt the transmission payload eavesdropped from a D2D transaction between device $a$ and $b$.}
\vspace{0.3em}
\\\textbf{Proof.}
According to Fig.5, the information that adversary $m$ can have includes $P_{pub}, a(b), pk_{a(b)}, U_{a(b)}, C_{a(b)}, F_{a(b)}$.

To decrypt the content eavesdropped, $m$ needs $k_{ab} = S_a = S_b$. The calculation of $S_a$ requires $V_a$ and $V_{b}'$. However, because of the intractability of ECDHP, $m$ cannot retrieve $l_a$. In addition, since $sk_a=(d_a,x_a)$ is privately held by $a$, $m$ cannot calculate $V_{b}'$.
Moreover, for the `benign but curious' BS (Subsection 3.2), it requires $k_{ab}$ to decrypt the content of D2D transactions. However, since $l_a$ and $l_b$ are randomly determined for each D2D transaction, it is not viable for BS to get $V_a$ and $V_b$ that are necessary to restore $k_{ab}$.
\hfill $\Box$

\vspace{-1em}
\subsection{Proof of CO$_2$ Immunity.} \label{appb2}

\textbf{Proposition 2.}
\textit{With TDP, adversary $m$ cannot establish a D2D connection with device $b$ using $a$'s public credential.}
\vspace{0.3em}
\\\textbf{Proof.}
According to Fig.5, the information that adversary $m$ can have includes $P_{pub}, a(b), pk_{a(b)}$.

To initialize a device pairing with $b$, $m$ requires $V_{b}'$ to calculate $k_{mb}$. However, the calculation of $V_{b}'$ is infeasible since $sk_a=(d_a,x_a)$ is privately held by $a$, which will lead to a failure in the successive key authentication.
\hfill $\Box$

\vspace{-1em}
\subsection{Proof of CO$_3$ Immunity.} \label{appb3}

\textbf{Proposition 3.}
\textit{With TDP, adversary $m$ cannot intercept the pairing process between device $a$ and $b$ by establishing D2D connections with them respectively without being noticed.}
\vspace{0.3em}
\\\textbf{Proof.}
Because of the infeasibility of CO${_2}$ attacks, according to Fig.5, $m$ has to choose an arbitrary public key $(R_{m}',P_{m}')$ combined with the identity of $b$ to pair with $a$. In this case, $m$ requires $V_{a}'$ to generate $k_{ma}$. Since:
\begin{center}
$V_a = l_ah_0(b\Vert R_{m}'\Vert P_{m}')P_{pub} + l_aR_{m}' + l_aP_{m}'$,
\end{center}
$m$ needs to calculate $V_{a}'$ based on $P_{pub} = xP$ and $U_a = l_aP$. Since $R_{m}' = r_{m}'P$ and $P_{m}' = x_{m}'P$ are determined by $m$, the value of $l_aR_{m}' + l_aP_{m}'$ is computable through $(r_{m}'+x_{m}')U_a$. Then, since $P_{pub} = xP$, the rest of $V_a$ can be calculated by:
\begin{center}
either $l_ah_0(b\Vert R_{m}'\Vert P_{m}')P_{pub}$ or $xh_0(b\Vert R_{m}'\Vert P_{m}')U_a$.
\end{center}
However, for the first form, $m$ cannot retrieve $l_a$ because of the intractability of ECDHP; for the second form, $x$ is intractable since it is privately held by the BS.
\hfill $\Box$

\vspace{-1em}
\subsection{Proof of TO$_1$ Immunity.} \label{appb4}

\textbf{Proposition 4.}
\textit{With TDP, adversary $m$ cannot attract unfair D2D transaction requests by fabricating its device trustvalue $\bm{t}_m$ as a higher $\bm{t}_m'$.}
\vspace{0.3em}
\\\textbf{Proof.}
To convince the peer device $a\in\mathcal{D}$ with $\bm{t}_m'$, $m$ needs to generate a verifiable signature $S_m=T_{4bs}'\Vert T_{5bs}'$. However, according to Eq.(14), $m$ requires the value of $x+r_m$ to generate $S_m$ that can be verified using the public credentials of both the BS and $m$. Since $x$ and $r_m$ are privately held by the BS, it is infeasible for $m$ to fabricate a verifiable trustvalue.
\hfill $\Box$

\vspace{-1em}
\subsection{Proof of TO$_2$ Resistance.} \label{appb5}

\textbf{Proposition 5.}
\textit{With TDP, adversary $m$ cannot attract unfair D2D transaction requests effectively by replying negative ratings to any encountered device $a\in\mathcal{D}$.}
\vspace{0.3em}
\\\textbf{Proof.}
In D2D transactions between $m$ and $a$, according to Eq.(3), the credibility is determined by $\mathcal{E}_m$ and $\mathcal{E}_a$ (together wiht $\bm{\theta}_m$ and $\bm{\theta}_a$). Since $m$ keeps replying negative ratings to others, according to Eq.(4) and (5), both $s_{ma}$ and $d_{ma}$ will be lower than normal. Therefore, $m$ will continuously receive negative ratings that downgrade its trustvalue.
\hfill $\Box$

\vspace{-1em}
\subsection{Proof of TO$_3$ Resistance.} \label{appb6}

\textbf{Proposition 6.}
\textit{With TDP, collusive adversaries $m_1,m_2\in\mathcal{M}$ cannot attract unfair D2D transaction requests effectively by (1) replying positive ratings to each other, and (2) replying negative ratings to any encountered device $a_1,a_2\in\mathcal{D}$.}
\\\textbf{Proof.}
\vspace{0.3em}
Firstly, in D2D transactions between $m_1$ and $m_2$, according to Eq.(4) and (5), they can guarantee a higher $s_{m_1m_2}$ and $d_{m_1m_2}$. However, according to Eq.(7), the damping factor $w_{m_1m_2}$ will effectively restrict the value of $c_{m_1m_2}$ if they conduct suspiciously more transactions than normal. Therefore, the trustvalue enhancement coming from their collusion will be suppressed.

Then, in D2D transactions between $m_1$ and $a_1$, since ratings from them to other benign devices (\emph{e.g.} device $a_2$) are highly different, $s_{m_1a_1}$ will be low according to Eq.(4). Besides, considering the highly different variation trends of $\bm{\theta}_{m_1}$ and $\bm{\theta}_{a_1}$, $d_{m_1a_1}$ will be low according to Eq.(5). Therefore, $m_1$ will continuously receive negative ratings that downgrade its trustvalue while $a_1$ will not be affected.
\hfill $\Box$

\end{document}